\documentclass{jfm}
\usepackage{bm}
\usepackage{graphicx}
\usepackage{amsmath}
\usepackage{hyperref}
\usepackage{apacite}
\usepackage{amssymb}
\usepackage{mathtools}
\newcommand{\numberset}[1]{\mathbb{#1}} 

\usepackage{lineno}
\usepackage{subcaption}
\usepackage{adjustbox}
\usepackage{lipsum}
\usepackage{tikz}
\usetikzlibrary{shapes,shapes.misc,decorations.markings,chains,positioning,arrows,calc}
\usepackage{array,multirow}
\usepackage{pgfplots}
\usepackage{gensymb}
\pgfplotsset{compat=1.12}
\newcolumntype{L}[1]{>{\raggedright\let\newline\\\arraybackslash\hspace{0pt}}m{#1}}
\newcolumntype{C}[1]{>{\centering\let\newline\\\arraybackslash\hspace{0pt}}m{#1}}
\newcolumntype{R}[1]{>{\raggedleft\let\newline\\\arraybackslash\hspace{0pt}}m{#1}}
\def\Otf{\Omega^\mathrm{f}(t)}
\def\bn{\mbox{\boldmath $n$}}
\newcommand{\xx}{\mbox{\boldmath $x$}}
\def\b{\mathbf}
\def\bs{\boldsymbol}

\begin{document}

\shorttitle{Decomposition of synchronized wake dynamics} 
\shortauthor{Miyanawala and Jaiman} 
\title{Decomposition of wake dynamics in fluid-structure interaction via low-dimensional models}

\author{T. P. Miyanawala \aff{1} \and R. K. Jaiman\aff{1}\corresp{\email{mperkj@nus.edu.sg}}}
\affiliation{Department of Mechanical Engineering, National University of Singapore.}
\maketitle
\begin{abstract}
We present a dynamic decomposition analysis of the wake flow in fluid-structure interaction (FSI) systems under both laminar and turbulent flow conditions. 
Of particular interest is to provide the significance of low-dimensional wake flow features and their interaction dynamics to sustain the free vibration of a square cylinder at a relatively low mass ratio. To obtain the high-dimensional data, we employ a body-conforming variational fluid-structure interaction solver based on the recently developed partitioned iterative scheme and the dynamic subgrid-scale turbulence model for a moderate Reynolds number ($Re$). The snapshot data from high-dimensional FSI simulations are projected to a low-dimensional subspace using the proper orthogonal decomposition (POD). We utilize each corresponding POD mode for detecting features of the organized motions namely the vortex street, the shear layer and the near-wake bubble.
%
We find that the vortex shedding modes contribute solely to the lift force, while the near-wake and shear layer modes play a dominating role to the drag force. We further examine the fundamental mechanism of this dynamical behavior and propose a force decomposition technique via low-dimensional approximation.
To elucidate the frequency lock-in, we systematically analyze the decomposed modes and their dynamical contributions to the force fluctuations for a range of reduced velocity at low $Re$ laminar flow.
We ascertain quantitatively that the shear layer feeds the vorticity flux to the wake vortices and the near-wake bubble during the wake-body synchronization. Based on the decomposition of wake dynamics, we suggest an interaction cycle for the frequency lock-in during the wake-body interaction, which provides the inter-relationship between the high amplitude motion and the dominating wake features. 
Through our investigation of wake-body synchronization below critical $Re$ range, we discover that the bluff body can undergo a synchronized high-amplitude vibration due to flexibility-induced unsteadiness. 
Owing to the wake turbulence at a moderate Reynolds number of $Re=22,000$, a distorted set of POD modes and the broadband energy distribution are observed, while the interaction cycle for the wake-synchronization is found to be valid for the turbulent wake flow.
\end{abstract}

\section{Introduction}
  \label{BBFlow}

\subsection{Resonant dynamics of coupled fluid-structure system}
Unsteady flows involving fluid-structure interactions are widespread in numerous engineering applications and their fundamental understanding poses serious challenges due to the richness and complexity of nonlinear coupled physics. 
Even a simple configuration of a coupled fluid-structure system can exhibit complex spatial-temporal dynamics and synchronization as functions of physical parameters and geometric variations. Synchronization is a general nonlinear physical phenomenon in fluid-structure systems whereby the coupled system has an intrinsic ability to lock to a preferred frequency and amplitude.
%
For example, when a bluff body immersed in a cross-flow is flexible or mounted elastically, there exists a strong coupling between the bluff body and the vortices forming in its wake. 
In particular, as the natural frequency ($f_n$) of the bluff body approaches to the frequency of the wake system, typically the frequency of vortex shedding ($f_{vs}$), the wake-body frequency lock-in behavior is observed which plays a crucial role in establishing the synchronization. 
During this frequency lock-in, the bluff body experiences a large self-limiting vibrations \citep{khalak1999motions} and a dynamical equilibrium between the energy transfer and dissipation exists.
This wake-body synchronization has been a major topic of research to understand the mechanism of this energy transfer and the sustenance of self-excited vibrations. In the present study, we consider a prismatic square geometry to understand the wake-body synchronization and to perform the decomposition of wake dynamics during the fluid-structure interaction.  

The phenomenon of frequency lock-in is a major concern in offshore, marine and aeronautical engineering, whereby structures are designed to avoid the large-amplitude vibrations by selecting optimal system parameters (e.g., geometric dimensions, stiffness, damping)  and/or installing active and passive devices to control the intensity of fluid-structure interaction.
In particular, several studies have been conducted with the purpose of controlling the wake-body interaction via passive and active devices \citep{law2017wake,guan2017control,narendran2018control} with the physical insight based on the reliance of frequency lock-in on the large-scale features of the wake. In fact, these studies were found to be remarkably successful in suppressing large-amplitude motion of the body by avoiding the interaction between the major organized features of the wake. However, the mechanism of the interactions among the wake features and their impact on the free motion of the bluff body is not properly explained. 
Moreover, the available experimental and numerical data can be used to provide a deeper understanding and a new insight into the kinematics and dynamics of synchronized wake-body interaction.
This paper aims at explaining how different organized flow features (i.e., near-wake structures) amplify the bluff body motion and sustain the energy transfer from the fluid flow to the vibrating body. Specifically, we examine the formation of the dominant coherent structures and their nonlinear interactions during the wake-body synchronization.

The vortex shedding pattern is undoubtedly the most prominent wake feature behind a bluff body. It is omnipresent in almost all of the separated wake flows and has been studied extensively in the literature. This primary wake feature begins at a much lower $Re$: for example in a circular cylinder wake, at $Re \approx 49$, exhibits a classical K\'arm\'an vortex street and develops the three-dimensional vorticity patterns when $Re \gtrsim 190$. 
In addition to the vortex street, a free shear layer (not attached to a solid surface) is an important dynamical feature that represents a  separating high-gradient layer behind a bluff body and it arises between the higher free stream velocity and the smaller velocity occurring in the wake region.  The shear layer behaves likes a perturbed vortex sheet and is highly sensitive and unstable to small disturbances, giving rise to alternating thickening and thinning of the vortex sheet.
The characteristic vortex structures develop when the thickening of shear layer occurs.
For the unsteady 2D regime ($49\leq Re \leq 190$ for a circular cylinder) the roll-up of the shear layers with the formation of the vortex street can be observed \citep{williamson1996vortex}. These shear layers are predominantly elongated in the streamwise direction and have a high-gradient in the cross-flow direction. Behind a moving or stationary bluff body, the region of a recirculating region with the rotational flow is present due to the fluid viscosity.  Owing to nonlinear flow separation and turbulence, complex interactions occur in the mean recirculation region, which is also referred to as the near-wake bubble. Several previous studies have explored the dynamical features inside the wake region (with the vortex shedding and the shear layer) using experimental \citep{cantwell1983experimental}, numerical \citep{braza1986numerical} and both \citep{bearman1997near, dong2006combined} techniques. 
In our present analysis, we consider the near-wake bubble as a distinct feature from the vortex street and the shear layer. The near-wake region accounts for the complex interactions of the mean circulation region, which can be considered as a general feature and can be identified separately from the other two features. Hence, we divide the wake into three dominant organized coherent structures: the vortex street, the shear layer, and the near-wake bubble. These organized features have an intrinsic dynamics of their own and influence each other in a nonlinear manner over a wide range of space and time scales. A primary goal of this paper is to employ low-dimensional models to extract the organized wake features and to examine their roles during the wake-body synchronization.

\subsection{Low-dimensional models for wake features}
To extract the large-scale organized/coherent wake features, it is required to decompose the dynamic flow fields by scales into different constituent kinematical regions. The concept of decomposition by scale has been prevalent in many fluid dynamics research ranging from a low-dimensional projection of flow field to the turbulence modeling by ensemble averaging, temporal or spatial averaging. 
A general decomposition technique can be considered to separate the space-time data for representing different characteristics of the field. For example, the proper orthogonal decomposition (POD) extracts the most energetic modes in an optimal way and provides structural information from the wake data. The POD is a popular method for constructing low-order modeling from the data \cite{holmes2012turbulence}, and it is often referred to as the Karhunen-Lo$\mathrm{\grave{e}}$ve expansion or the principal component analysis.
The key idea behind the Karhunen-Lo$\grave{e}$ve expansion is to determine a low-dimensional affine subspace from the high-dimensional data while retaining the important dynamics of the full-order model. 
After the determination of the best approximating low-dimensional subspace, a Galerkin projection is employed to project the dynamics onto it. In this work, we will employ this low-dimensional subspace projection procedure for extracting the large-scale wake features from the high-dimensional flow dynamics data. 

In the context of the present study, the POD-Galerkin projection method is quite attractive to capture the synchronized dynamics such as the vortex shedding and the near-wake interactions \citep{rempfer2003low,noack2003hierarchy}. 
In addition, it has been the prominent empirical model reduction technique incorporated for the standard flow around a stationary circular cylinder for the past few decades. For example, in the one of pioneering study, \cite{deane1991low} reproduced the flow dynamics of the laminar wake by employing merely an eight-dimensional model, which was further generalized to generate reduced spaces for 3D velocity field by \cite{ma2002low} using direct numerical simulation (DNS) data. In general, the empirical POD-Galerkin models are capable of reconstructing the reference dynamics with higher accuracy than the standalone mathematical or physical Galerkin methods, while capturing the physically most significant modes \citep{noack2003hierarchy}. With regard to the applications of the POD-Galerkin to bluff body wake flows, these modes correspond to the organized wake features such as the vortex street, the shear layer and the near-wake bubble. 
Although there exist a significant body of works on the wake modes for a stationary cylinder, they have not been examined
in the context of wake-body interaction and the lock-in process. One of the contributions of the present
study is to build some connections between the wake features and the lock-in process. 
During the lock-in/synchronization, the vibrating body undergoes
a highly nonlinear-wake interaction with self-sustained oscillations.

In the early studies of POD application to fluid flows \citep{lumley1967structure,sirovich1987turbulence}, the dynamic flow field is reconstructed by a linear combination of the most significant modes. Hence, it has a considerable local error in the highly nonlinear regions of the organized wake motions and the evaluation of the projected nonlinear term has a direct dependence on the large dimension
of the original system. This problem is mitigated to a certain extent by increasing the sampling frequency and/or refining the spatial discretization of the reference data. However, these temporal and spatial refinements increase the cost of model reduction without directly addressing the nonlinear nature of the flow. To introduce the nonlinearity,  Petrov-Galerkin projections to the Navier-Stokes formulation or Koopman operators are incorporated in some studies \citep{rowley2017model}. Instead of such explicit models, we employ the recently developed discrete empirical interpolation method (DEIM) \citep{chaturantabut2009discrete} for dynamical systems, which reconstructs the fields as a nonlinear combination of the POD modes. Apart from the POD basis subspace, the method relies on the additional POD basis to enrich the low-rank approximation of the nonlinear terms.
In the POD-DEIM, a set of best points are selected using a greedy selection and the reconstruction is based on the time history of the field data of those points. This reduces the computational cost of the technique and further allows to capture nonlinearities during the reconstruction of highly nonlinear dynamic wake fields \citep{rowley2017model}.

\subsection{Contributions and organization}
For the past few decades, the studies on the low-dimensional decomposition of wake features have been primarily focused on flow past stationary bodies, particularly on a circular cylinder \citep{deane1991low,noack2003hierarchy,taira2017modal,rowley2017model}. This may be due to the fact that the flow exhibits a diverse set of complex phenomena despite its simple geometry.
However, very few studies \citep{liberge2010reduced,yao2017model} are found on unsteady fluid-structure interaction systems. Here, we provide a modal reduction study on the flow past a freely vibrating sharp-cornered square cylinder with two-degree-of-freedom motions. 
We consider a configuration of a square cylinder for our numerical study of wake-body synchronization because: (i) this configuration has fixed and perfectly symmetric separation points at the leading sharp corners, (ii) entirely resonance-induced lock-in exists \citep{yao2017model}. The physical investigation is general for any fluid-structure system involving the interaction dynamics of flexible structures with an unsteady wake-vortex system. We hypothesize that the solution space of wake-body interaction attracts a low-dimensional manifold, which allows building a set of basis vectors for a low-dimensional representation of the high-dimensional space. The low-dimensional subspace is constructed by means of the samples collected from the high-dimensional solutions via projection-based model order reduction.
We utilize the linear and nonlinear POD-based reduced order reconstructions to understand the most significant features in the wake flow. 

To extract the modes for the dynamics of wake-body synchronization, the POD in conjunction with the nonlinear POD-DEIM is applied to a set of samples collected from the full-order simulations. We exploit the obtained POD modes to answer the following interesting questions that are prevalent in the field of fluid mechanics: (i) How does the each of large-scale features contribute to the unsteady forces acting on the bluff body? (ii) How do the wake features interact when the structural frequency and vortex shedding frequency are locked-in, such that the vortices remain very energetic even the fluid has transferred energy to the structure? (iii) Will the wake and bluff body undergo synchronized motion below the critical $Re$ due to the structural flexibility? (iv) What role does the wake turbulence play when we attempt to decompose the wake into its large-scale features?
In relation to the first question, we quantify the force contribution from each wake feature mode to the streamwise (drag) and transverse (lift) forces and explain the observed variation.
We further investigate the modal contribution of different wake features in the pre-lock-in, lock-in and post-lock-in regimes and propose a cycle explaining the sustenance of lock-in phenomena of the wake-body synchronization. 
We then explore the below critical $Re$ flows to examine whether the bluff body and the wake can undergo synchronization via flexibility-induced unsteadiness. 
Finally, we apply POD decomposition to the three-dimensional flow at moderate $Re=22,000$, whereas the wake is fully turbulent after flow separation. A well-established dynamic large-eddy simulation (LES) is
employed for generating full-order data for the turbulent wake. At this sub-critical Reynolds number, we explore the role of turbulence during the reconstruction of flow-field data and extend the wake-body synchronization cycle to the turbulent flow.

The paper is structured as follows. In Section 2, we briefly review the full-order model (FOM) for the coupled fluid-structure system, which follows by the formulation of modal reduction via linear POD and nonlinear POD-DEIM. Section 3 discusses the problem setup and the mesh convergence study performed for the full-order analysis. In Section 4, the reduced order reconstruction of fluid fields using the linear and nonlinear POD methods are presented together with the analysis on the role of wake features in generating the forces. In Section 5, the mode energy contributions from different flow features under lock-in conditions are investigated and a self-sustaining cycle is proposed to explain the wake interaction with the bluff body. Section 6 investigates the wake-body synchronization phenomenon at below critical $Re$. Section 7 explores the application of modal decomposition for moderate $Re$ flows and extends the proposed wake interaction cycle to the turbulent flow. Concluding remarks and the main results of the present study are provided in Section 8. 

\section{Numerical methodology}
We first briefly summarize our high-dimensional full-order model 
to simulate the coupled fluid-body interaction using the incompressible Navier-Stokes equations and the rigid body dynamics.
\subsection{Full-order model for fluid-body interaction}
We employ a variational formulation based on the arbitrary Lagrangian-Eulerian (ALE) to solve 
the following coupled fluid-body system
\begin{align}
\rho^\mathrm{f}\frac{\partial \mathbf{u}^\mathrm{f}}{\partial t} + \rho^\mathrm{f}\left(\mathbf{u}^\mathrm{f}-\mathbf{w}\right)\cdot \boldsymbol{\nabla} \mathbf{u}^\mathrm{f} = \boldsymbol{\nabla} \cdot {\boldsymbol{\sigma}}^\mathrm{f} 
+ \mathbf{b}^\mathrm{f}  \mbox{ on } \Otf 
\label{eq:NS} 
\\ 
\boldsymbol{\nabla} \cdot \mathbf{u}^\mathrm{f} = 0  \mbox{ on }   \Otf \label{eq:continuity}
\\
\b{M} \frac{\partial {\mathbf{u}}^\mathrm{s}}{\partial t}+ \b{C} {\mathbf{u}}^\mathrm{s}+ \b{K} \left(\boldsymbol{\varphi}^\mathrm{s}
\left(\mathbf{z}_0,t\right)-\mathbf{z}_0\right)=\mathbf{F}^\mathrm{s} \;\mbox{on}\; \Omega^\mathrm{s}
\label{eq:trans_rot}
\end{align}
where subscripts $\mathrm{f}$ and $\mathrm{s}$ denotes the fluid and structural domains, and $\Otf$ and $\Omega^\mathrm{s}$ represent the fluid and solid domains, respectively. Here $\rho^\mathrm{f}$ is the fluid density,  $\mathbf{u}^\mathrm{f}$ and $\mathbf{w}$ are the fluid and mesh velocities at a spatial point $\xx \in  \Otf$, and $\mathbf{b}^\mathrm{f}$ denotes the body force in the fluid domain. For the structural system, $\b{M}$, $\b{C}$ and $\b{K}$ are the mass, damping and stiffness matrices of the bluff body and $\mathbf{F}^\mathrm{s}$ is the external force acting on the body. The function 
$\boldsymbol{\varphi}^\mathrm{s}\left(\mathbf{z}_0,t\right)$
maps the initial position vector of the center of mass ($\mathbf{z}_0$) 
to its position at time $t$, and
${\boldsymbol{\sigma}}^\mathrm{f}$ is the Cauchy stress tensor for a Newtonian fluid given by:
\begin{equation}
{\boldsymbol{\sigma}}^\mathrm{f} = -{p} \mathbf{I} + \mu^\mathrm{f}\left(\boldsymbol{\nabla} \mathbf{u}^\mathrm{f} + \left(\boldsymbol{\nabla} \mathbf{u}^\mathrm{f}\right)^T\right),
\end{equation}
where $p$ is the fluid pressure. 
In addition to the initial conditions and the standard Neumann/Dirichlet conditions, the coupled system incorporates the velocity and traction continuity conditions at 
the fluid-body interface $\Gamma$ as follows:
\begin{align}
\mathbf{u}^\mathrm{f}(t) = \mathbf{u}^\mathrm{s}(t),
\label{eq:bcsVelocity} \\
\int_{\Gamma(t)} \boldsymbol{\sigma}^\mathrm{f}(\xx,t)\cdot\bn \: \mathrm{d} \Gamma + \mathbf{F}^\mathrm{s} = \b{0},
\label{eq:bcsTraction}
\end{align}
where $\bn$ is the outer normal to the fluid-body interface.
The above fluid-body interface conditions are satisfied by the body-conforming Eulerian-Lagrangian treatment, which provides accurate modeling of 
the boundary layer and the vorticity generation over a moving body.
While Eqs. (\ref{eq:NS}-\ref{eq:trans_rot}) of the coupled fluid-body system are directly solved for  low $Re$ flows,  we consider the well-established dynamic subgrid-scale model for high $Re$ turbulent flow. The spatially-filtered Navier-Stokes and continuity equations are solved in the variational form. Details of the dynamic subgrid-scale model are provided in \cite{jaiman_caf2016}.

The weak variational form of Eq.~(\ref{eq:NS}) is discretized in space 
using equal-order iso-parametric finite elements for the fluid velocity and pressure.
In the present study, we utilize the nonlinear partitioned staggered procedure for the full-order simulations of fluid-structure interaction \citep{nifc2016}.
The motion of structure is driven by the traction forces exerted 
by the fluid flow, whereby 
the structural motion predicts the new interface position and the geometry changes 
for the moving fluid domain at each time step. The movements of the internal ALE fluid nodes are updated such that the mesh quality 
does not deteriorate as the motion of solid structure becomes large.
To extract the transient flow characteristics, we solve the Navier-Stokes equations at discrete timesteps which lead to a sequence of linear systems of equations via
Newton-Raphson type iterations. 
We employ the Conjugate Gradient (CG), with a diagonal
preconditioner for the symmetric matrix arising from the pressure projection and  the standard Generalized Minimal Residual (GMRES)  solver  based  on  the  modified  Gram-Schmidt  orthogonalization  for the non-symmetric velocity-pressure matrix.
The above coupled variational formulation completes the presentation 
of the full-order model for the fluid-structure interaction. 

From a model reduction viewpoint, the coupled system of the nonlinear differential equations for the fluid-body interaction can be written in the following form:
\begin{equation}
\frac{d\mathbf{y}}{dt} = \b{F}(\mathbf{y}),
\label{eq:systemEq}
\end{equation}
where $\mathbf{y}$ is the column state vector describing the unknown degrees of freedom and $\b{F}$ is a vector-valued function describing the spatially discretized governing equations. 
In the present fluid-body system, the state vector comprises of the fluid velocity and the pressure as $\mathbf{y} = \{\mathbf{u}^\mathrm{f}, p \}$ and the structural velocity involves the three translational degrees-of-freedom. For a discretized domain of $m$ elements and $n$ timesteps, the full-order simulation outputs a high-fidelity data set $\mathbf{y}\in \numberset{R}^{m\times n \times q}$, where $q$ is the number of variables in $\mathbf{y}$. This dataset is extremely valuable to determine the instantaneous physics of the fluid-body system and to construct a low-order representation that preserves the behavior of the original system.
\subsection{Low-order models}
We now turn to the data-driven model reduction technique whose goal is to decompose the aforementioned high-dimensional data set into a set of low-dimensional modes. For that purpose, we can consider the decomposition of the nonlinear mapping $F$ of Eq. (\ref{eq:systemEq}) as
\begin{equation}
\b{F}(\mathbf{y}) = \b{f} + \b{A}\mathbf{y} + \b{F}'(\mathbf{y}),
\label{eq:decompEq}
\end{equation}
where $\b{f}$ denotes a constant column vector with $m$ rows, $\b{A}$  and $\b{F}'$ are the linear and nonlinear terms.
For the ease of explanation, consider the solution vector $\mathbf{y}(\xx, t) \in \numberset{R}^{m\times n}$ comprising a single quantity of interest which have been determined at discretized locations $\xx$ of the spatial domain and for a particular time $t$ and the matrix operator $\b{A}$ is an $ m\times m$ matrix which captures the linear dynamics while $\b{F}'(\mathbf{y})$ is a nonlinear function of $\mathbf{y}$.
Using the projection-based model reduction, we can represent the state vector $\mathbf{y}$ by an element in a low-rank vector subspace spanned by the column vectors of an $m \times k$ matrix $\bs{{\mathcal{V}}}=[v_1 \: v_2 ...\: v_k]$, where $k \ll m$. The state vector $\mathbf{y}$ can be approximated by $\bs{{\mathcal{V}}}\hat{\mathbf{y}}$, where $\hat{\mathbf{y}}$ is a reduced column vector with $k$ entries. Since the columns of $\bs{{\mathcal{V}}}$ are orthonormal (i.e., $\bs{{\mathcal{V}}}^T\bs{{\mathcal{V}}} = \b{I}$), via the Galerkin projection onto the basis $\bs{{\mathcal{V}}}$, we get the following reduced dynamics:
\begin{equation}
\frac{d\hat{\mathbf{y}}}{dt} = \bs{\mathcal{V}}^T \b{f} + \bs{\mathcal{V}}^T\b{A}\bs{\mathcal{V}} \hat{\mathbf{y}} + \bs{\mathcal{V}}^T \b{F'}(\bs{\mathcal{V}}\hat{\mathbf{y}}).
\end{equation}

Next, we have to choose a suitable subspace for the mode decomposition. Using the reduced singular value decomposition (SVD), the above state vector $\b{y}$ can be expressed as
\begin{equation}
\mathbf{Y} = \bs{\mathcal{V}} \bs{\Sigma} \bs{\mathcal{W}}^T = \sum_{j=1}^{k} \sigma_j \b{v}_j \b{w}_j^T,
\end{equation}
where the vectors $\b{v}_j$ are the POD modes of the matrix $\mathbf{Y}$ with rank $k$, $\bs{\mathcal{W}}$ is an orthonormal matrix with $n \times k$ , and $\bs{\Sigma}$ is a $k \times k$  diagonal matrix with diagonal entries $\sigma_1 \geq \sigma_2 \geq ... \geq \sigma_{k} \geq 0$. For any $r \leq k$, the subspace spanned by $\{\b{v}_1, ... , \b{v}_r\}$ provides an optimal representation of $\mathbf{y}$ in the subspace of dimension $r$ using the SVD process. The total energy contained in each POD mode $\b{v}_j$ can be computed by the singular value $\sigma_j^2$. Note that $\bs{\mathcal{V}}$ and $\bs{\mathcal{W}}$ are the orthonormal eigenvectors of $\mathbf{Y}\mathbf{Y}^T$ and $\mathbf{Y}^T\mathbf{Y}$, respectively.
%

\subsubsection{Proper orthogonal decomposition}
The POD method provides an algorithm to decompose a set of data into a minimal number of modes. We give a brief outline of this projection-based model reduction for the dynamical analysis of wake-body interaction.
The general POD algorithm can be expressed as follows. Here, the eigenvectors of $\mathbf{Y}\mathbf{Y}^T$ are determined instead of performing the SVD. The algorithm adopted from \cite{taira2017modal} is summarized in Algorithm 1.
\\
\\
\underline{ALGORITHM 1: Snapshot POD}\\
\fbox{%
	\parbox{\linewidth}{%
Input: Snapshots of spatial field expressed as $\mathbf{Y}(\xx,t)$ where $\mathbf{Y} \in \mathbb{R}^{m \times k}$ ($m$ - number of spatial points, $k$ - number of snapshots.)

Output: Significant $r$ POD modes $\boldmath \bs{\mathcal{V}}$ $= [\b{v}_1, \b{v}_2,...,\b{v}_r]$
            
\begin{enumerate}
  \item Develop the fluctuation matrix by subtracting the mean: $\tilde{\mathbf{Y}}(t) = {\mathbf{Y}}(\xx,t) - \overline{\mathbf{Y}}(\xx)$
  \item Construct the covariance matrix ${\b{R}} =\tilde{\mathbf{Y}}{\tilde{\mathbf{Y}}^T} \in \mathbb{R}^{m \times m}$
  \item Find the eigenvalues and eigenvectors of $\b{R}$ by $\b{R}\bs{\mathcal{V}} = \bs{\Lambda}\bs{\mathcal{V}}$
  \item Determine the number of required POD modes ($r$) using $\sum_{j=1}^{r} \lambda_j / \sum_{j=1}^{m} \lambda_j \approx 1.0$, where $\lambda_j$ are the eigenvalues given by $\bs{\Lambda}$.
\end{enumerate}
}
}
\\
\\

The standard linear POD is almost in the same order expensive as the  full-order analysis since it is using the $\tilde{\mathbf{Y}}\tilde{\mathbf{Y}}^T$ matrix which has the size of ${m \times m}$.
In a typical time-dependent flow analysis, it is unnecessary to generate $m$ POD modes for comparison as the POD mode energy decays exponentially. Hence an alternative method the so called snapshot POD \citep{sirovich1987turbulence} is applied to extract the most significant modes. In the snapshot method, the eigenvalue decomposition is performed on $\tilde{\mathbf{Y}}^T\tilde{\mathbf{Y}} \in \numberset{R}^{k \times k}$ which is significantly smaller than $\tilde{\mathbf{Y}}\tilde{\mathbf{Y}}^T$ as $k \ll m$. Let the eigenvalues and eigenvectors of $\tilde{\mathbf{Y}}^T\tilde{\mathbf{Y}}$ be given by
\begin{equation}
\tilde{\mathbf{Y}}^T\tilde{\mathbf{Y}}\bs{\mathcal{W}} = \bs{\Lambda}\bs{\mathcal{W}},
\end{equation}
then using the relationship between the eigenvectors of $\tilde{\mathbf{Y}}\tilde{\mathbf{Y}}^T$ and $\tilde{\mathbf{Y}}^T\tilde{\mathbf{Y}}$, a maximum of $k$ significant POD modes can be extracted by
\begin{equation}
\bs{\mathcal{V}} = \tilde{\mathbf{Y}}\bs{\mathcal{W}}\bs{\Lambda}^{-1/2}.
\end{equation}
Throughout the study, every POD decomposition will be performed via the snapshot POD method due to its low computational cost and memory usage.
After extracting the significant POD modes, the constant and linear components of the instantaneous state vector can be recovered as a linear combination of the identified significant modes
\begin{equation}
{\mathbf{Y}}(\xx,t) \approx \overline{{\mathbf{Y}}}(\xx) + \sum_{j=1}^{r} \hat{{y}}_j(t)\b{v}_j,
\label{eq:linearPOD}
\end{equation}
where $r$ is the number of significant POD modes.
The temporal coefficients of the linear combination are determined by the $L^2$ inner product $\langle \: . \: , \: .  \: \rangle$ between the fluctuation matrix and the modes as follows
\begin{equation}
\hat{\mathbf{Y}}(t) = \langle \mathbf{Y}-\overline{\mathbf{Y}},\bs{\mathcal{V}} \rangle.
\label{eq:innerProduct}
\end{equation}
This summarizes the process of POD by performing the SVD on the snapshots of the sampled solutions at certain timesteps.

While the above POD-Galerkin process can reconstruct the linear term to the expected error threshold, the nonlinear term will not be reconstructed properly in the context of nonlinear incompressible flow which involves quadratic nonlinearity. The linear POD reconstruction requires a higher number of modes and/or a smaller sampling interval for snapshots to obtain the required local domain accuracy. In other words, the spatial and/or temporal discretizations of the POD method have to be so small such that the nonlinearities behave almost linearly. Subsequently, the POD reconstruction may result in a similar order of computational expense as full-order simulation. This issue can be handled by employing discrete empirical interpolation method (DEIM).
The DEIM introduces the nonlinearity by supplementing an additional basis for a low-order representation of nonlinear terms. This gives rise to the reduction in the requirement of POD modes, hence decreasing the computational cost while capturing the nonlinear regions properly.

\subsubsection{Discrete empirical interpolation method}
To overcome the difficulty in the linear POD, \cite{chaturantabut2009discrete} proposed the discrete empirical interpolation method to reconstruct the full-order variable as a nonlinear combination of the POD modes.
The aim of DEIM is to design a low-order representation for the nonlinear terms by introducing an additional basis. Consider $\bs{\mathcal{U}}$ as a basis generated from the leading $l$ modes of the POD, which is attracted to a low-dimensional subspace. We can approximate the nonlinear term in Eq. (\ref{eq:systemEq}) by the sequence of nonlinear snapshots as $\b{F}'(\bs{\mathcal{V}}\hat{\mathbf{y}}(t)) \approx \bs{\mathcal{U}}\hat{\b{c}}$. The coefficients $\hat{\b{c}}$ can be selected based on Algorithm 2, which relies on a greedy approximation of nonlinear function. 
In Algorithm 2, $\hat{\rho}$ and $\wp_1$ denote the assigned value and  the assigned index of $\max \{|\b{v}_1|\}$, and $\boldsymbol{e}_{\wp_i}=[0,...,0,1,0,...0]^T \in \numberset{R}^{m}$ is the $\wp_i$th column of the identity matrix of size $m \times m$.
The accuracy of DEIM approximation depends on the error induced by the POD projection and the estimation of $\|(\bs{\mathcal{P}}^T\bs{\mathcal{U}})^{-1}\|$.  
Further details of the DEIM process can be found in  \cite{chaturantabut2009discrete}.
\\
\\
\underline{ALGORITHM 2: POD-DEIM}\\
\fbox{%
	\parbox{\linewidth}{%
			Output: Indices of $l$ best points $\bs{\wp} = [\wp_1,\wp_2,....,\wp_l]^T$ \\
            Input: Most significant $l$ POD modes $ [\b{v}_1, \b{v}_2,...,\b{v}_l]$
            
\begin{enumerate}
  \item $[\hat{\rho} \: \wp_1 ]= \max \{|\b{v}_1|\}$
  \item $\bs{\mathcal{U}} = [\b{v}_1]$, $\bs{\mathcal{P}} = \boldsymbol{e}_{\wp_1}$, $\bs{\wp} = [\wp_1]$
  \item for i=2 to $l$ do
  	\begin{enumerate}
    	\item Solve $(\bs{\mathcal{P}}^T\bs{\mathcal{U}})\hat{\b{c}} = \bs{\mathcal{P}}^T\b{v}_i$ for $\hat{\b{c}}$
        \item Compute residual $\hat{\b{r}} = \b{v}_i - \bs{\mathcal{U}}\hat{\b{c}}$
        \item Assign $[\hat{\rho} \: \wp_i] = \max \{|\hat{\b{r}}|\}$
        \item Augment $\bs{\mathcal{U}} \gets [\bs{\mathcal{U}} \: \b{v}_i]$, $\bs{\mathcal{P}} \gets [\bs{\mathcal{P}} \: \boldsymbol{e}_{\wp_i}]$, 
        $\bs{\wp} \gets [\bs{\wp} \: \wp_i]^T$
    \end{enumerate}
  \item end for
\end{enumerate}
            }
} 
\\

Here, a set of entries $\bs{\wp} \subset \{1,2,...,l\}$ often called optimal (best) points are selected to determine $\hat{\b{c}}$ by the following relation
\begin{equation}
\hat{\b{c}} = (\bs{\mathcal{P}}^T\bs{\mathcal{U}})^{-1}\bs{\mathcal{P}}^T \b{F}'(\bs{\mathcal{V}}\hat{\mathbf{y}}(t)).
\end{equation}
Assuming that $\b{F}'$ is a component-wise function $\bs{\mathcal{P}}^T \b{F}'(\bs{\mathcal{V}}\hat{\mathbf{y}}(t)) = \b{F}'(\bs{\mathcal{P}}^T \bs{\mathcal{V}} \hat{\mathbf{y}}(t))$, we can rewrite Eq. (\ref{eq:systemEq}) as
\begin{equation}
\frac{d}{dt}\hat{\mathbf{y}}(t) = (\bs{\mathcal{V}}^T\b{A}\bs{\mathcal{V}})\hat{\mathbf{y}}(t) + \bs{\mathcal{V}}^T \bs{\mathcal{U}} (\bs{\mathcal{P}}^T\bs{\mathcal{U}})^{-1} \b{F}'(\bs{\mathcal{P}}^T\bs{\mathcal{V}}\hat{\mathbf{y}}(t)).
\label{eq:LinNLin}
\end{equation}
In the present work, we perform the nonlinear POD on the same fluctuation matrix $\tilde{\mathbf{y}}_{m \times k}$ without separating the linear and nonlinear components. Consider the approximation of $\tilde{\mathbf{y}}$ as a nonlinear combination of the POD modes:
\begin{equation}
\tilde{\mathbf{y}}(t) \approx \bs{\mathcal{V}}\bs{\theta}(t).
\end{equation}
The coefficients $\bs{\theta}(t)$ are calculated by the conditions imposed by the POD-DEIM. 
While the POD modes are linearly independent, we can obtain a unique number of DEIM points if $\bs{\mathcal{P}}^T\bs{\mathcal{U}}$ matrix is invertible. 
By using just the $\bs{\wp}$ rows of $\bs{\mathcal{V}}$ and $\bs{\mathcal{U}}$, we can establish the following relationship:
\begin{equation}
\bs{\mathcal{V}}_{\wp} \bs{\theta}(t) = \tilde{\mathbf{y}}_{\wp}(t)
\end{equation}
which further gives
\begin{equation}
\tilde{\mathbf{y}}(t) \approx \bs{\mathcal{V}}\bs{\mathcal{V}}_{\wp}^{-1}\tilde{\mathbf{y}}_{\wp}.
\end{equation}
If the number of points used is higher than the number of significant modes, i.e. $l>r$, which is often the case, $\bs{\mathcal{V}}_{\wp}$ becomes a rectangular matrix. This makes the coefficients $\bs{\theta}(t)$ given by the gappy POD reconstruction
\begin{equation}
\bs{\theta}(t) = arg \; min ||\tilde{\mathbf{y}}_{\wp}(t) - \bs{\mathcal{V}}_{\wp}\b{\hat{a}}||_2, \: \b{\hat{a}} \in \numberset{R}^r.
\label{eq:GappyPOD}
\end{equation}
The solution to the least square problem (Eq.~\ref{eq:GappyPOD}) gives the result
\begin{equation}
\tilde{\mathbf{y}}(t) \approx \bs{\mathcal{V}} \bs{\mathcal{V}}_{\wp}^{+}\tilde{\mathbf{y}}_{\wp},
\end{equation}
where $\bs{\mathcal{V}}_{\wp}^{+}$ is the {Moore-Penrose pseudoinverse} 
of $\bs{\mathcal{V}}_{\wp}$. \\

The POD-DEIM provides a way to introduce nonlinearity to the POD reconstructions, however, due to this nonlinear behavior, it is not guaranteed to converge to the full-order results. In other words, the use of more POD modes or DEIM points does not assure an improvement in the result. Therefore, determining the optimal sizing of the low-dimensional representation is critical when using POD-DEIM for reconstruction. In the next section, we present the full-order model for generating high-dimensional data.
%

\section{Full-order simulations}
\subsection{Problem set-up}
\begin{figure}
    \centering 
\begin{subfigure}{0.49\textwidth}
   \centering 
\includegraphics[scale=0.3]{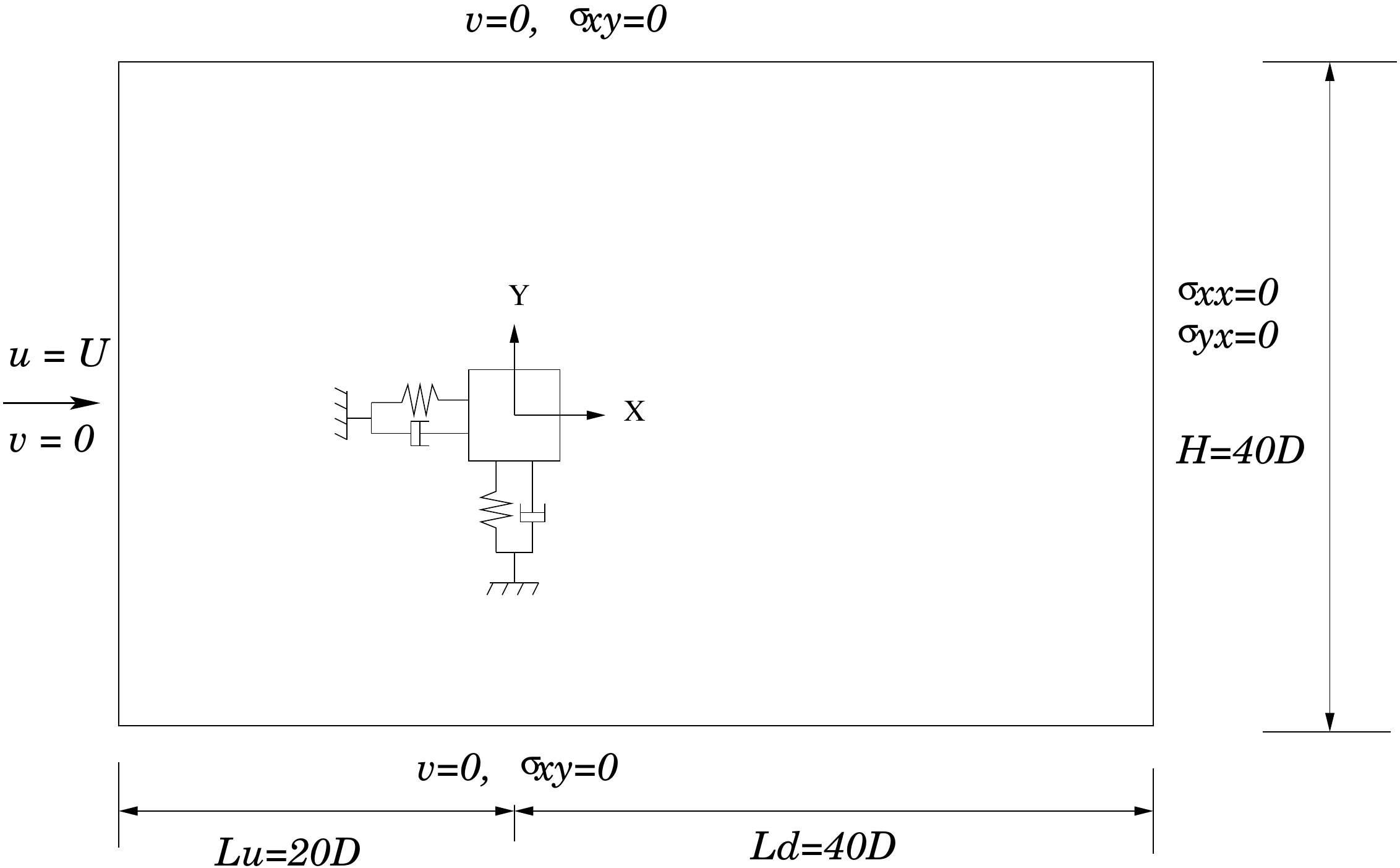}
\caption{}
\label{fig:Domain_2D_SC}
\end{subfigure}~
\begin{subfigure}{0.49\textwidth}
		\centering
\includegraphics[scale=0.3]{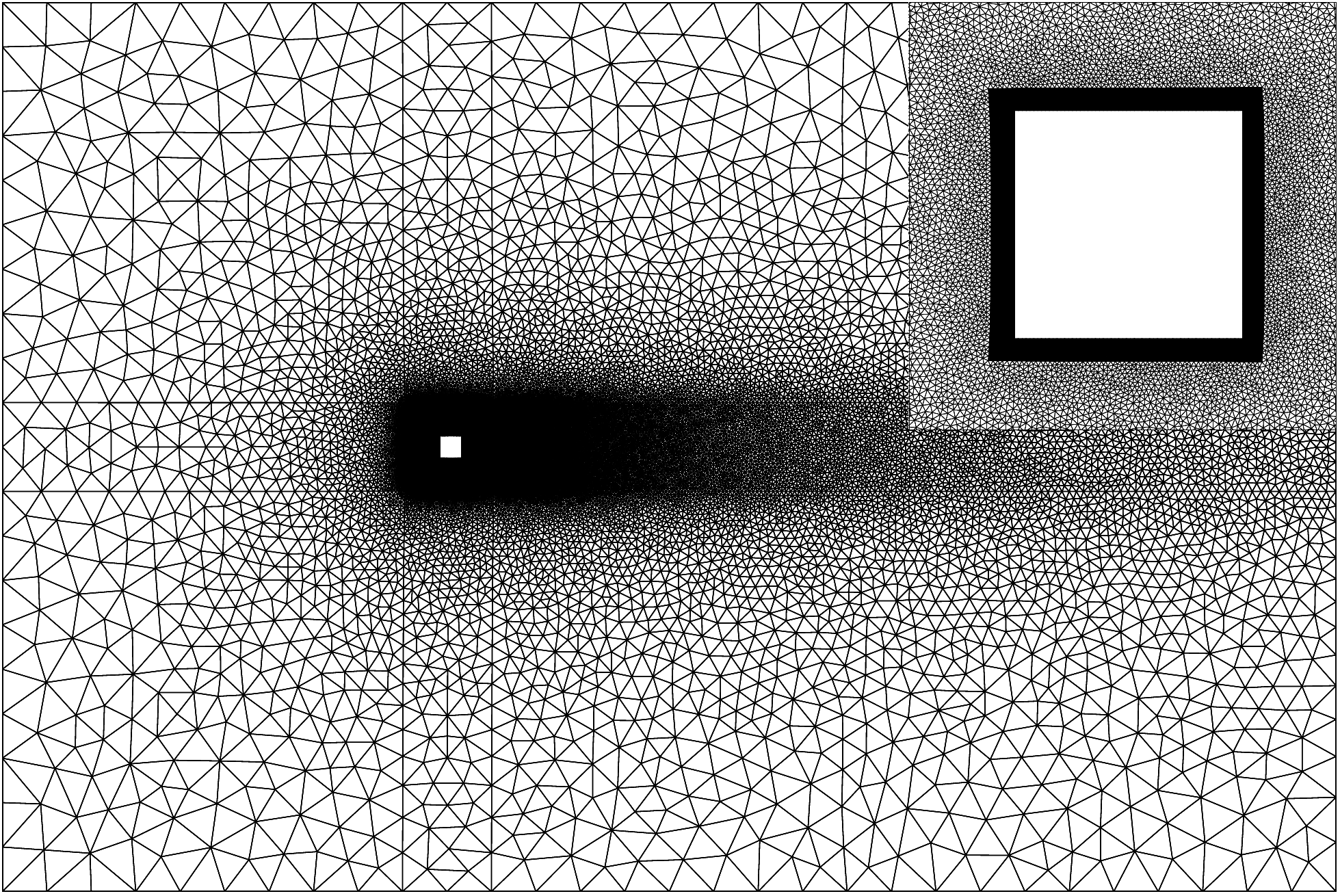}
\caption{}
  \label{fig:mesh_2D_SC} 
\end{subfigure}
\caption{Full-order problem setup for fluid-structure interaction: (a) schematic diagram of the computational domain and boundary conditions, and (b) representative $Z$- plane slice of the unstructured mesh. The top right inset displays the near cylinder mesh.}
  \end{figure}

In this section, we give an overview of full-order simulations for a freely vibrating structure immersed in a viscous incompressible fluid flow.
Specifically, the focus of this section is to present numerical results on the flow past an elastically mounted 
square cylinder, whereby the cylinder is free to oscillate 
in the streamwise ($X$) and the transverse ($Y$) directions.  
The mass and natural frequencies are identical in both $X$- and $Y$-directions.
The translational flow-induced vibration of a cylinder is strongly influenced by 
the four key non-dimensional parameters, 
namely mass-ratio $\left(m^*\right)$, 
Reynolds number $\left(Re\right)$, reduced velocity $\left(U_r \right)$,
and critical damping ratio $(\zeta)$
defined as
\begin{align}
m^{*}= \frac{M}{m_f}, \ \ \ \ Re = \frac{\rho^{f} U_{\infty} D}{\mu^\mathrm{f}}, \ \  \ \ U_r = \frac{U_{\infty}}{f_{n} D}, \ \
 \ \  \zeta = \frac{C}{2\sqrt{K M}},
\label{eq:DampingRatio}
\end{align}
where $M$ is the mass per unit length of the body, $C$ and $K$ are the damping and stiffness 
coefficients, respectively for an equivalent spring-mass-damper system of a vibrating structure, 
$U_{\infty}$ and $D$ denote the free-stream speed and the diameter of cylinder, respectively. 
The natural frequency of the body 
is given by $f_n=(1/2\pi) \sqrt{K/M}$ and  the mass of displaced fluid by the structure is
 $m_f = \rho^f D^2 L_c$ for a square cross-section, 
and $L_c$ denotes the span of the cylinder. In the above definitions, 
we make the isotropic assumption for the translational motion of the rigid body, i.e., the mass vector
$\mathbf{M}=(m_x,m_y)$ with $m_x=m_y=M$, the damping vector $\mathbf{C}=(c_x,c_y)$ with $c_x=c_y=C$,
the stiffness vector $\mathbf{K}=(k_x,k_y)$ with $k_x=k_y=K$. 
The fluid loading is computed by integrating the surface traction considering the first layer of elements located on the cylinder surface. The instantaneous lift and drag force coefficients are evaluated as
\begin{align}
C_L = \frac{1}{\frac{1}{2}\rho^\mathrm{f}U_{\infty}^2D L_c}\int_{\Gamma}(\mathbf{\sigma}^{\mathrm{f}} .\bn).\bn_y d\Gamma, \\
C_D = \frac{1}{\frac{1}{2}\rho^\mathrm{f}U_{\infty}^2D L_c}\int_{\Gamma}
(\mathbf{\sigma}^{\mathrm{f}} .\bn).\bn_x d\Gamma. 
\end{align}
Here ${\bn_x}$ and ${\bn_y}$ are the Cartesian components of the unit outward normal {$\bn$}. 
Figure \ref{fig:Domain_2D_SC} illustrates a schematic of the two-dimensional simulation domain used for the fluid-body interaction problem.
The center of the square column is located at the origin of the Cartesian coordinate system. The side length of the square 
column is denoted as $D$. The 
distances to the upstream and the downstream boundaries are $20D$ and $40D$, respectively. The distance between the side walls is $40D$, which corresponds to a 
blockage of 2.5\%.  The flow velocity $U_{\infty}$ is set to unity at the inlet and a no-slip wall is implemented at the surface of 
the square column. While the top and bottom boundaries are defined as slip walls, the computational domain is assumed to be periodic in the spanwise direction for the 3D simulations.

\subsection{Mesh convergence study}
For the high-dimensional approximation of full-order model, the computational domain is discretized using an unstructured finite-element mesh, wherein a boundary layer mesh surrounding the body and three-node triangle (2D) and six-node wedge (3D) elements outside the boundary layer region. Three more grids are
generated where the mesh elements are successively increased by approximately a factor of 2, designated as M2, M3 and M4. The discretized domain, along with a close-up view of the corners of the square column is illustrated in figure \ref{fig:mesh_2D_SC}. 
Results of grid convergence study are recorded in Table (\ref{table:meshcv}) for the lock-in region. All cases for the mesh convergence are simulated at $Re = 100$, $m^*=3$ and $U_r = 5.0$. The mesh convergence error is computed by considering the finest mesh M4 as the reference case. 
The force coefficients, the shedding frequency and the root mean square (rms) of the transverse amplitude are analyzed.
It can be seen
that values recorded for mesh M3 and M4 differ by less than 1\%.
Therefore, the mesh M3 is adequate for the present study. 
Furthermore, the adopted full-order solver and the numerical discretizations have been  extensively validated in several earlier studies for both low $Re$ \citep{miyanawala2016flow,jaiman_caf2016} and moderate $Re$ \citep{jaiman_caf2016,miyanawala2018self} flows.
\begin{table}
\centering
\caption{Grid convergence study at $Re = 100$, $m^{*}=3$ and $U_r=5.0$.}
\begin{tabular}{L{4cm}|C{2cm}C{2cm}C{2cm}C{1.5cm}}
\hline
 & M1 & M2 & M3 & M4\\
\hline
number of nodes&17,622&34,302&87,120&145,608\\
number of elements&17,389&34,027&86,631&145,195\\
time-step size $\Delta t$&0.025&0.025&0.025&0.025\\
shedding frequency ${f/f_n}$& 0.9798 & 0.9798 & 0.9798 & 0.9798\\
rms amplitude ${A_y^{rms}}/D$&0.097 (56.0\%)&0.200 (9.7\%)&0.220 (0.72\%)& 0.2211 \\
mean drag  $\overline{C_D}$&1.623 (24.1\%)&1.994 (6.7\%)&2.134 (0.18\%)&2.1377 \\
rms lift $C_L^{\mathrm{rms}}$ &0.485 (29.7\%)&0.604 (12.4\%)&0.687 (0.30\%)&0.6893 \\
\end{tabular}
\label{table:meshcv}
\end{table} 
In the next section, the modal decomposition of the pressure field is presented for a representative reduced velocity of $U_r=6.0$ in the lock-in region at $(Re, m^*, \zeta)  =(100, 3.0, 0)$. The snapshots of the FOM performed for the flow past a vibrating square cylinder is utilized to recover the POD modes and the DEIM points. The accuracy of the linear POD and POD-DEIM are systematically assessed with regard to their effectiveness to extract the flow features.

\section{Assessment of low-order model for wake decomposition}
\begin{figure}
\centering
\begin{subfigure}[h]{0.5\textwidth}
\includegraphics[scale=0.4]{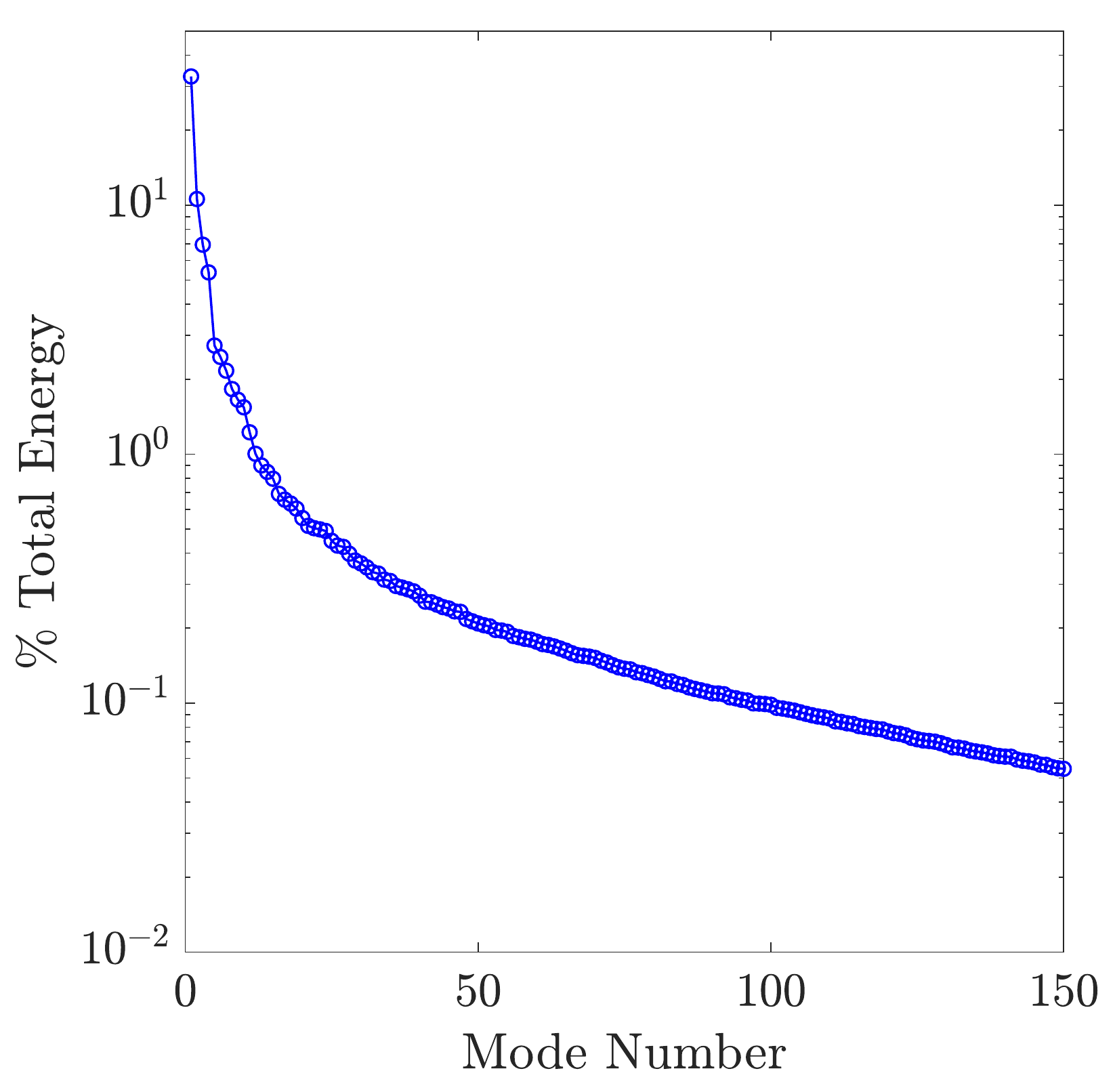}
\caption{}
\end{subfigure}~
\begin{subfigure}[h]{0.5\textwidth}
\includegraphics[scale=0.4]{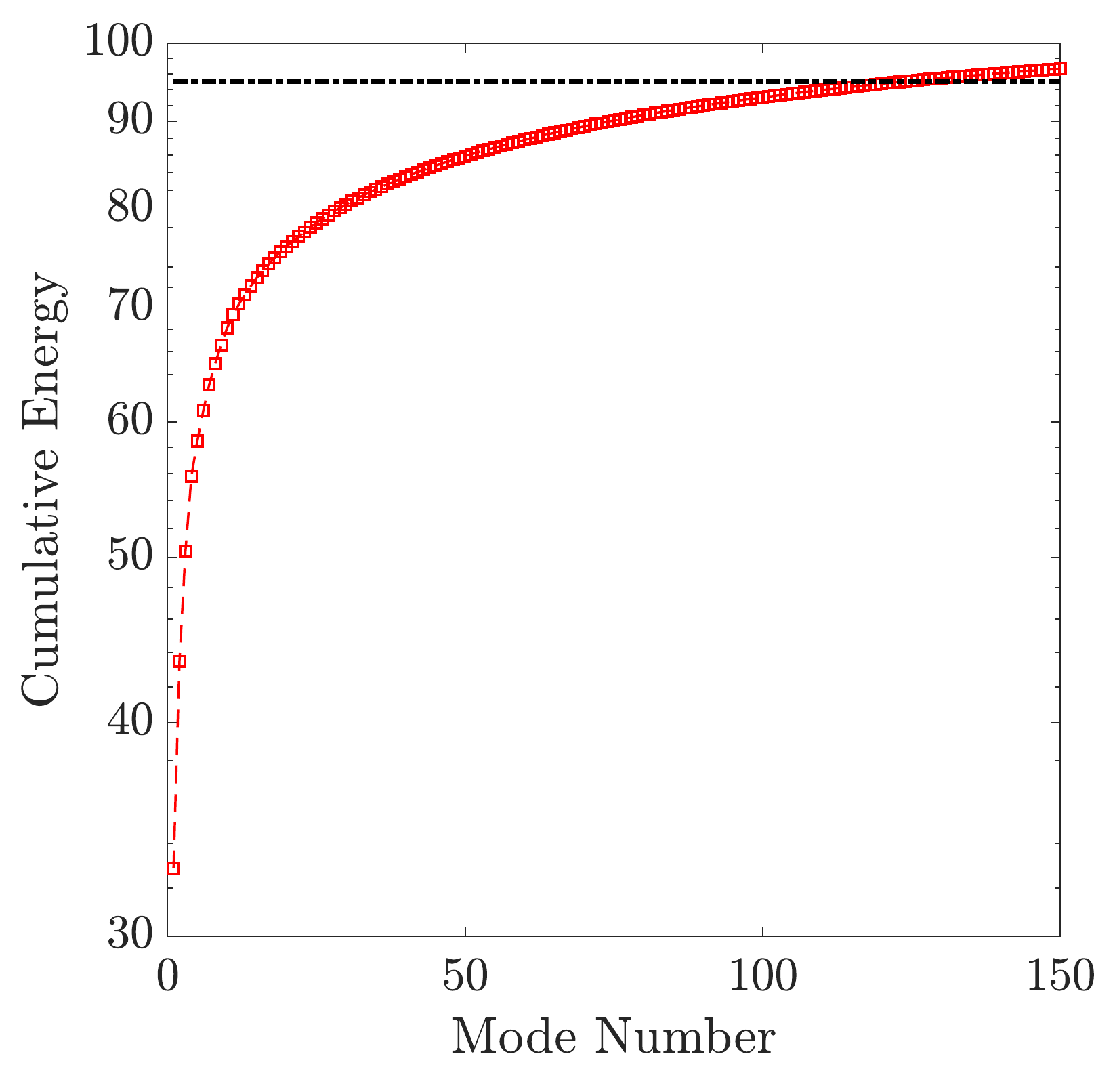}
\caption{}
\end{subfigure}
\caption{Distribution of modal energy for a laminar flow past a freely vibrating square cylinder: (a) energy decay of POD modes, and (b) cumulative energy of POD modes. 
}
\label{PODEnergy}
\end{figure}

As described earlier, we incorporate the snapshot POD method described to obtain the low-dimensional decomposition of the wake dynamics. As found in \cite{miyanawala2018self}, the laminar bluff body flow involves simply a few significant features. It will be ineffective to generate the entire set of POD modes, e.g., the order of the mesh points of 87,120 for this particular problem. Hence, we use the snapshot POD technique and obtain just the most significant POD modes, which are a few order of magnitude smaller. We reconstruct the pressure field using the linear and nonlinear techniques and compare their effectiveness to capture the organized wake features.
In the present analysis, the unsteady pressure field values for all the mesh points, are collected to a $m \times k$ matrix $\b{P}$ where $m$ (mesh count) = $87,120$ and $k$ (number of snapshots) = $320$.
The fluctuation matrix $\tilde{\mathbf{y}}_{m \times k}$ is then  generated by subtracting the mean value ($\overline{\b{P}}$) of each point over the snapshots
$\tilde{\mathbf{Y}} = \b{P} - \overline{\b{P}}$.
The POD modes are extracted using the eigenvalues
$\bs{\Lambda}_{k \times k} = diag[\lambda_1,\lambda_2,...,\lambda_{k}]$ and eigenvectors $\bs{\mathcal{W}} = [\b{w}_1 \: \b{w}_2 \: ... \: \b{w}_{k}]$ of the covariance matrix $\tilde{\mathbf{Y}}^T\tilde{\mathbf{Y}} \in \numberset{R}^{k \times k}$ given by $ \tilde{\mathbf{Y}}^T\tilde{\mathbf{Y}} \bs{\mathcal{W}} = \bs{\Lambda}\bs{\mathcal{W}}$.
As presented earlier, the POD modes $\bs{\mathcal{V}} = [\b{v}_1 \: \b{v}_2 \: ... \: \b{v}_{k}]$ are related to $\bs{\Lambda}$ and $\bs{\mathcal{W}}$ by
$\bs{\mathcal{V}} = \tilde{\mathbf{Y}} \bs{\mathcal{W}}\bs{\Lambda}^{-1/2}$.
Each eigenvalue represents the energy/strength of the POD mode. Since the mean pressure distribution is initially removed from the pressure field, the relative strength of the mode directly expresses the contribution from each mode for the pressure fluctuations. Figure \ref{PODEnergy}a displays the energy of these modes normalized by the total energy of the 320 modes obtained. It is clear that this energy decays exponentially and the most energetic mode has 56\% of the total energy. In fact, the first 9 most significant modes contain 99\% of the total energy of the modes, as shown in figure \ref{PODEnergy}b. Initially, these 9 significant modes are used to recover the pressure field in the linear POD reconstruction. We refer to these modes as mode 1, mode 2, etc. and they are in the descending order of mode energy ($\lambda_i$). We first incorporate the linear reconstruction method wherein we assume the final flow field is a linear combination of the flow features captured by the POD modes.

\subsection{Linear POD reconstruction}
\begin{figure}
\centering
\begin{subfigure}[]{0.5\textwidth}
\centering
\includegraphics[trim={2cm 3cm 2cm 3.9cm},clip,scale=0.225]{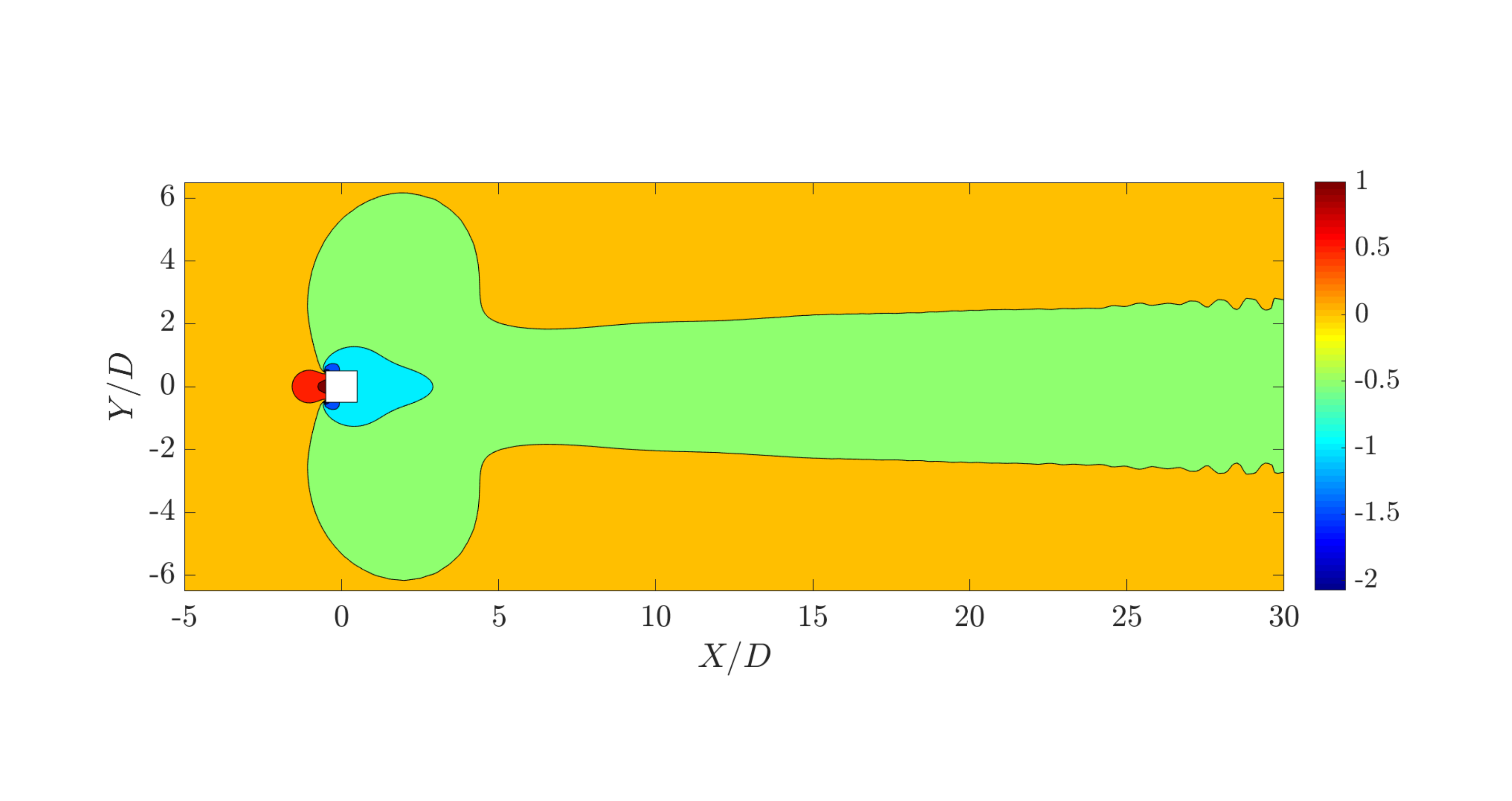}
\caption{Mean Pressure}
\end{subfigure}~
\begin{subfigure}[]{0.5\textwidth}
\centering
\includegraphics[trim={2cm 3cm 2cm 3.9cm},clip,scale=0.225]{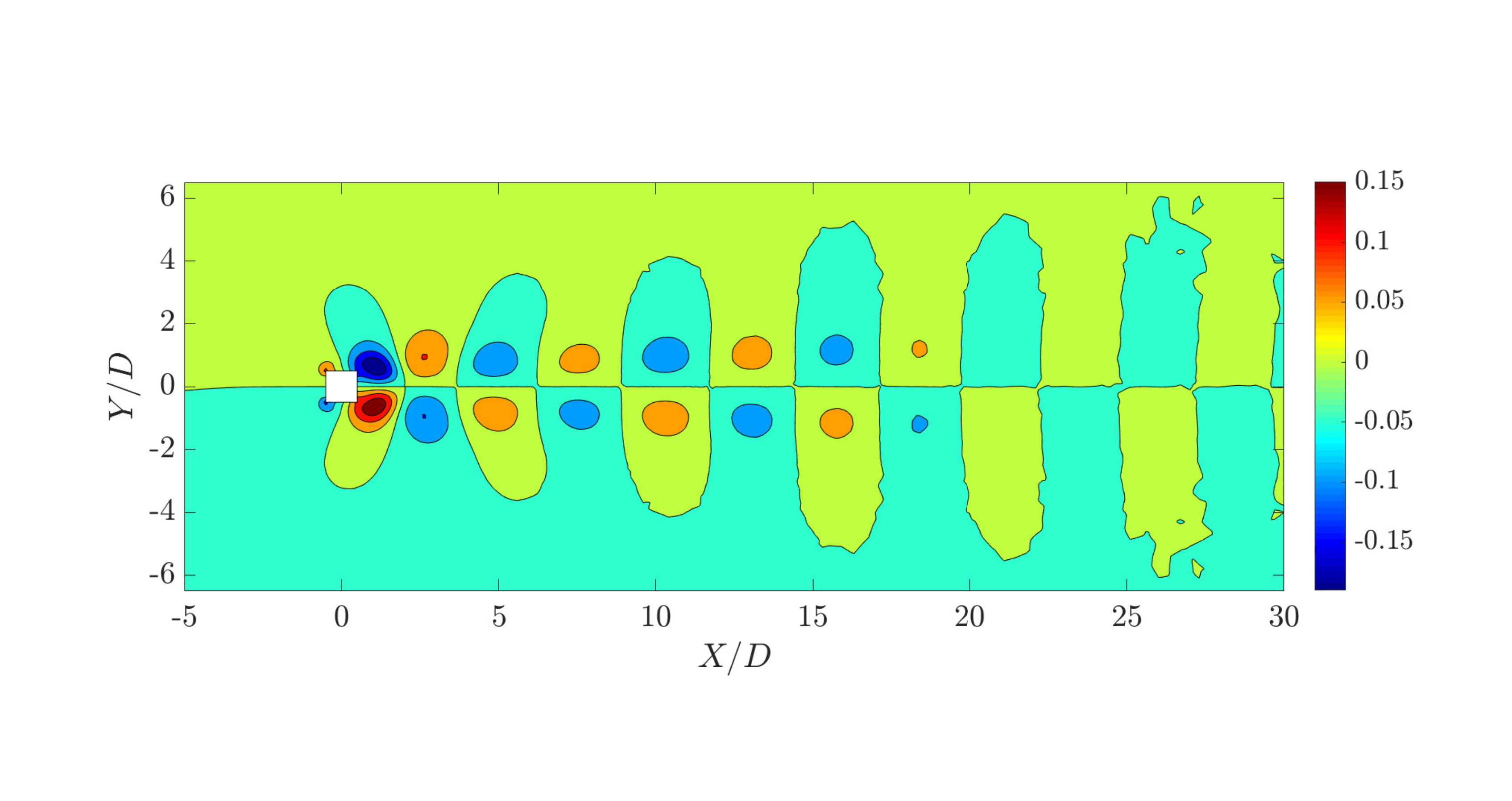}
\caption{Mode 1 (56\%)}
\end{subfigure}
\begin{subfigure}[]{0.5\textwidth}
\centering
\includegraphics[trim={2cm 3cm 2cm 3.9cm},clip,scale=0.225]{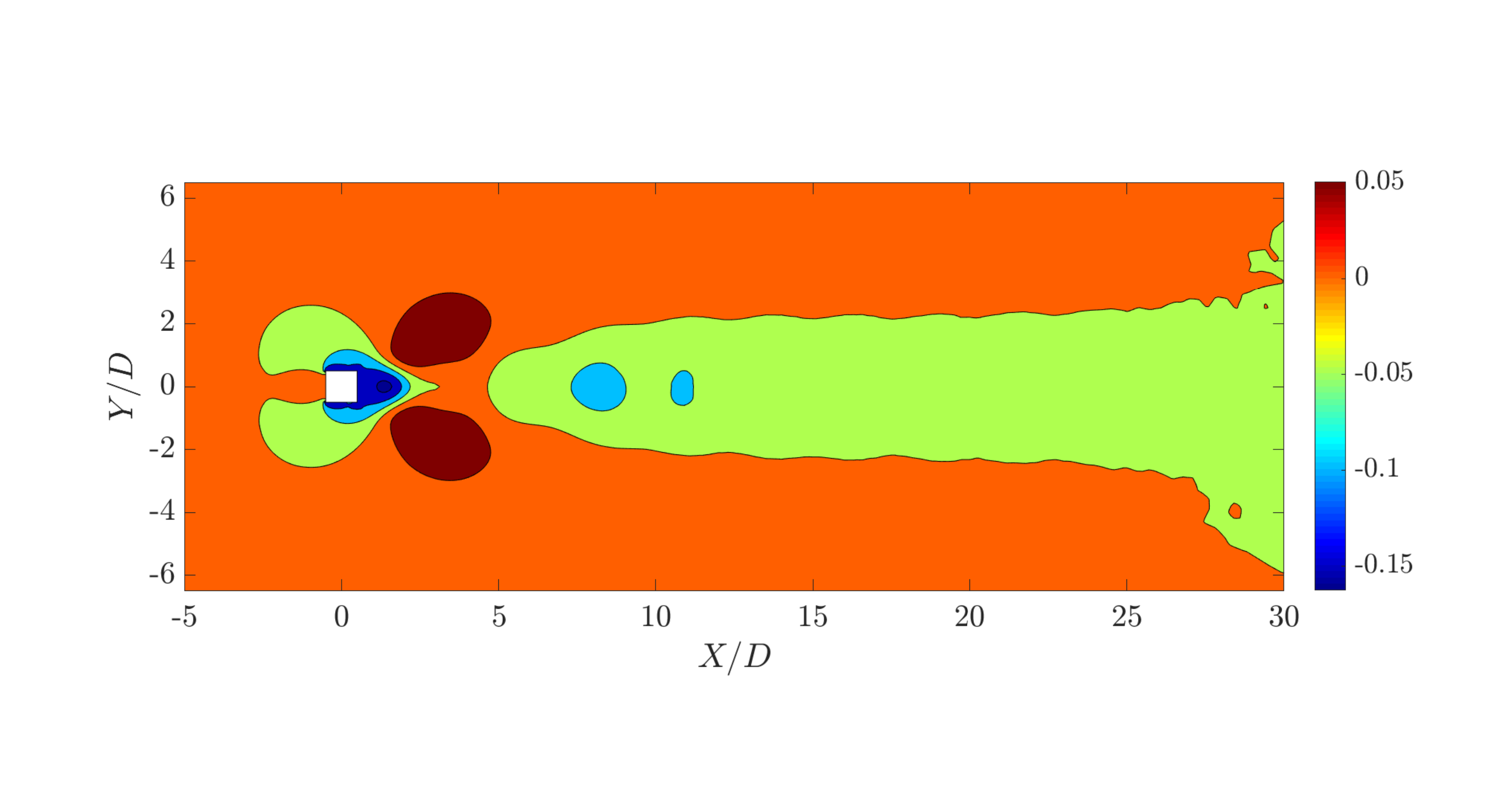}
\caption{Mode 2 (20.5\%)}
\end{subfigure}~
\begin{subfigure}[]{0.5\textwidth}
\centering
\includegraphics[trim={2cm 3cm 2cm 3.9cm},clip,scale=0.225]{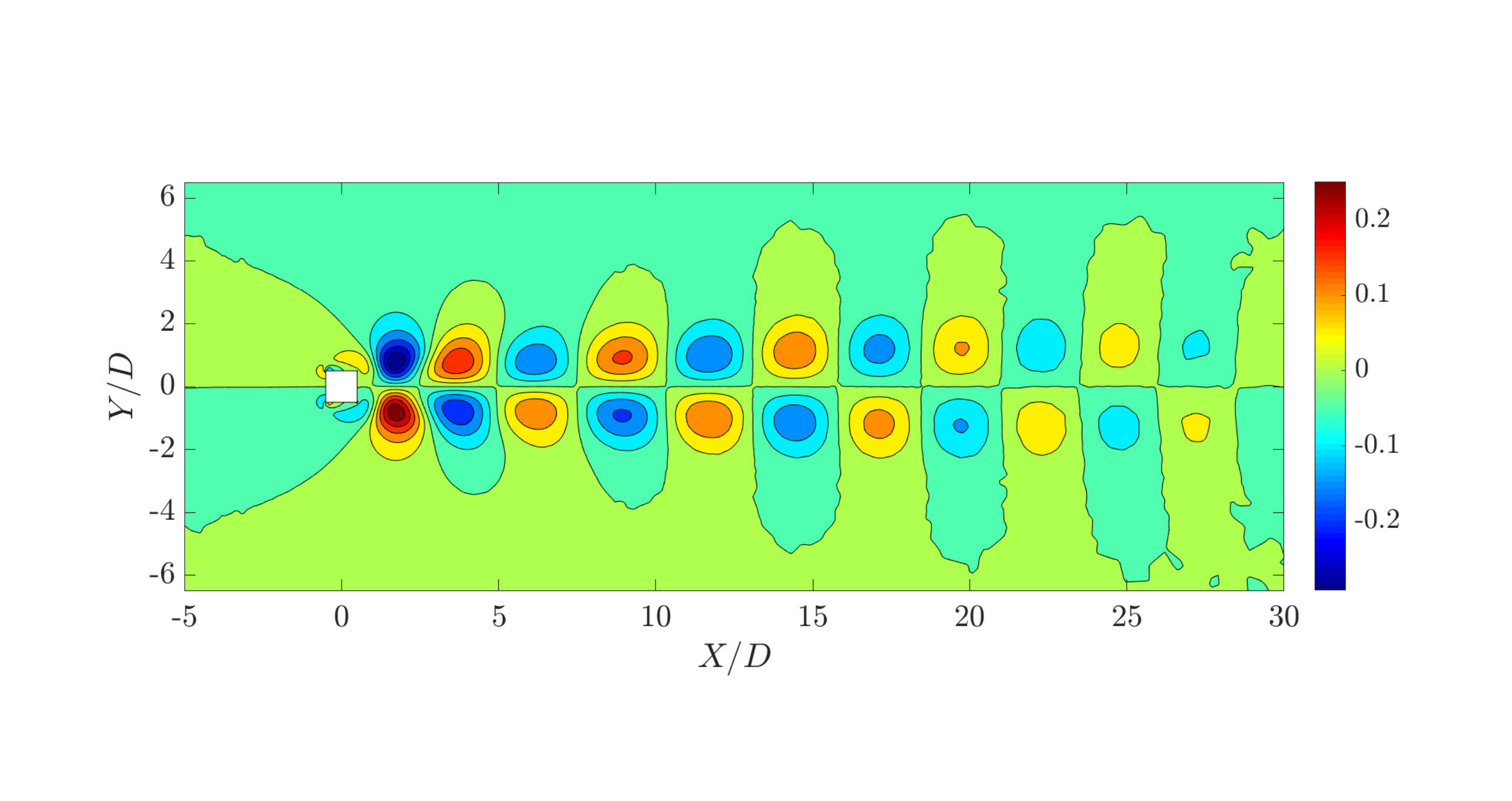}
\caption{Mode 3 (12.6\%)}
\end{subfigure}
\begin{subfigure}[]{0.5\textwidth}
\centering
\includegraphics[trim={2cm 3cm 2cm 3.9cm},clip,scale=0.225]{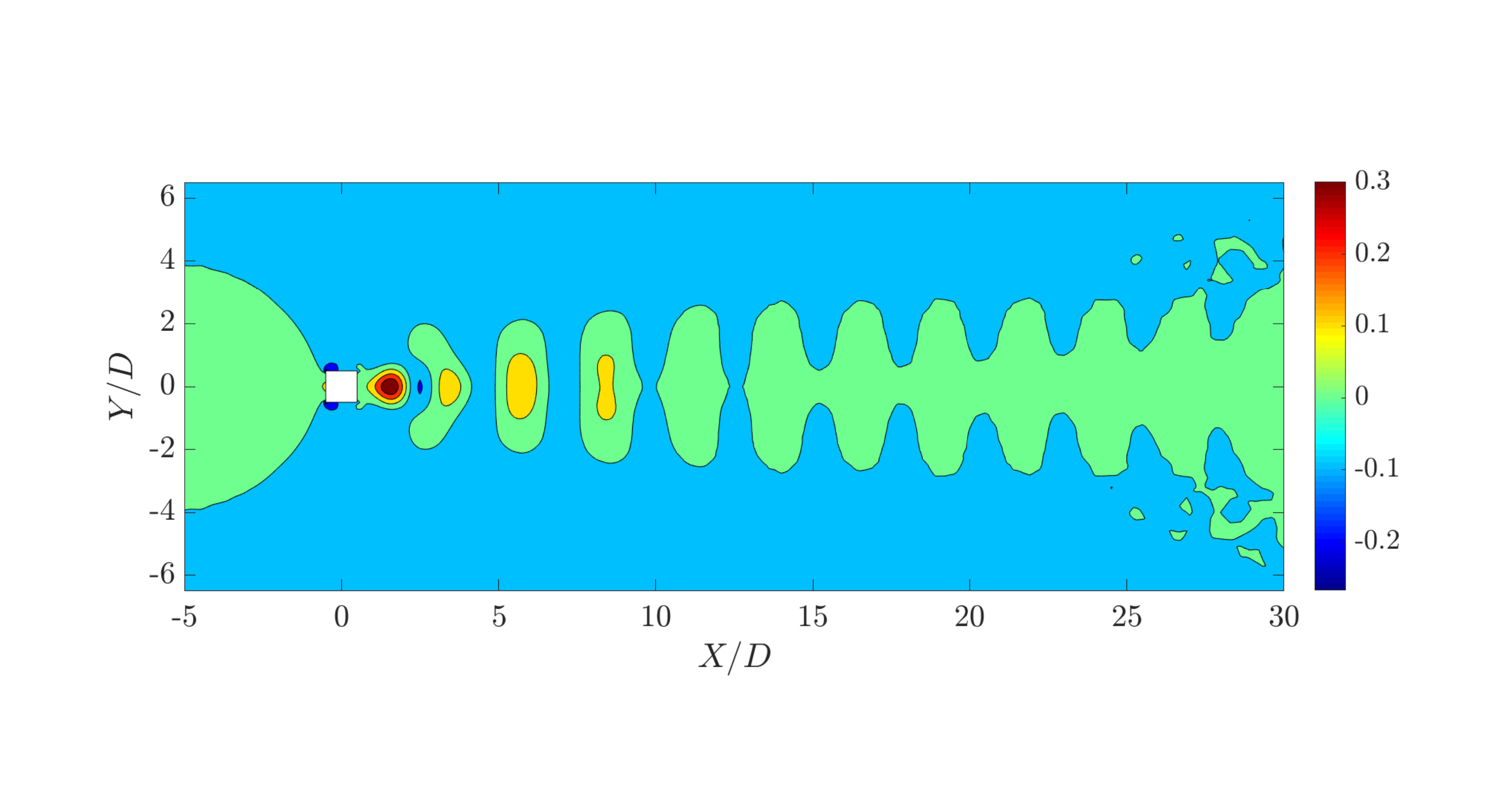}
\caption{Mode 4 (3.4\%)}
\end{subfigure}~
\begin{subfigure}[]{0.5\textwidth}
\centering
\includegraphics[trim={2cm 3cm 2cm 3.9cm},clip,scale=0.225]{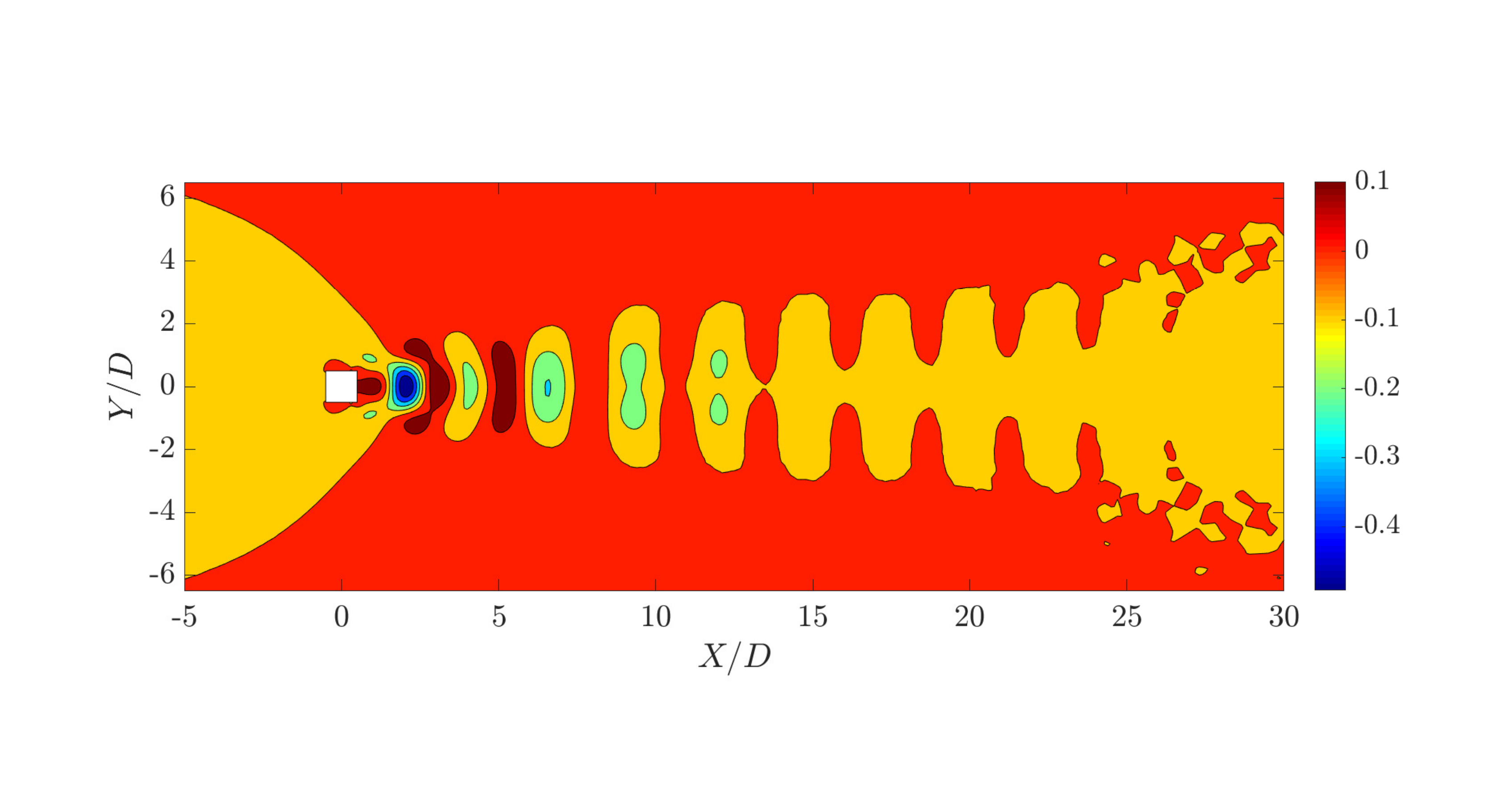}
\caption{Mode 5 (2.6\%)}
\end{subfigure}
\begin{subfigure}[]{0.5\textwidth}
\centering
\includegraphics[trim={2cm 3cm 2cm 3.9cm},clip,scale=0.225]{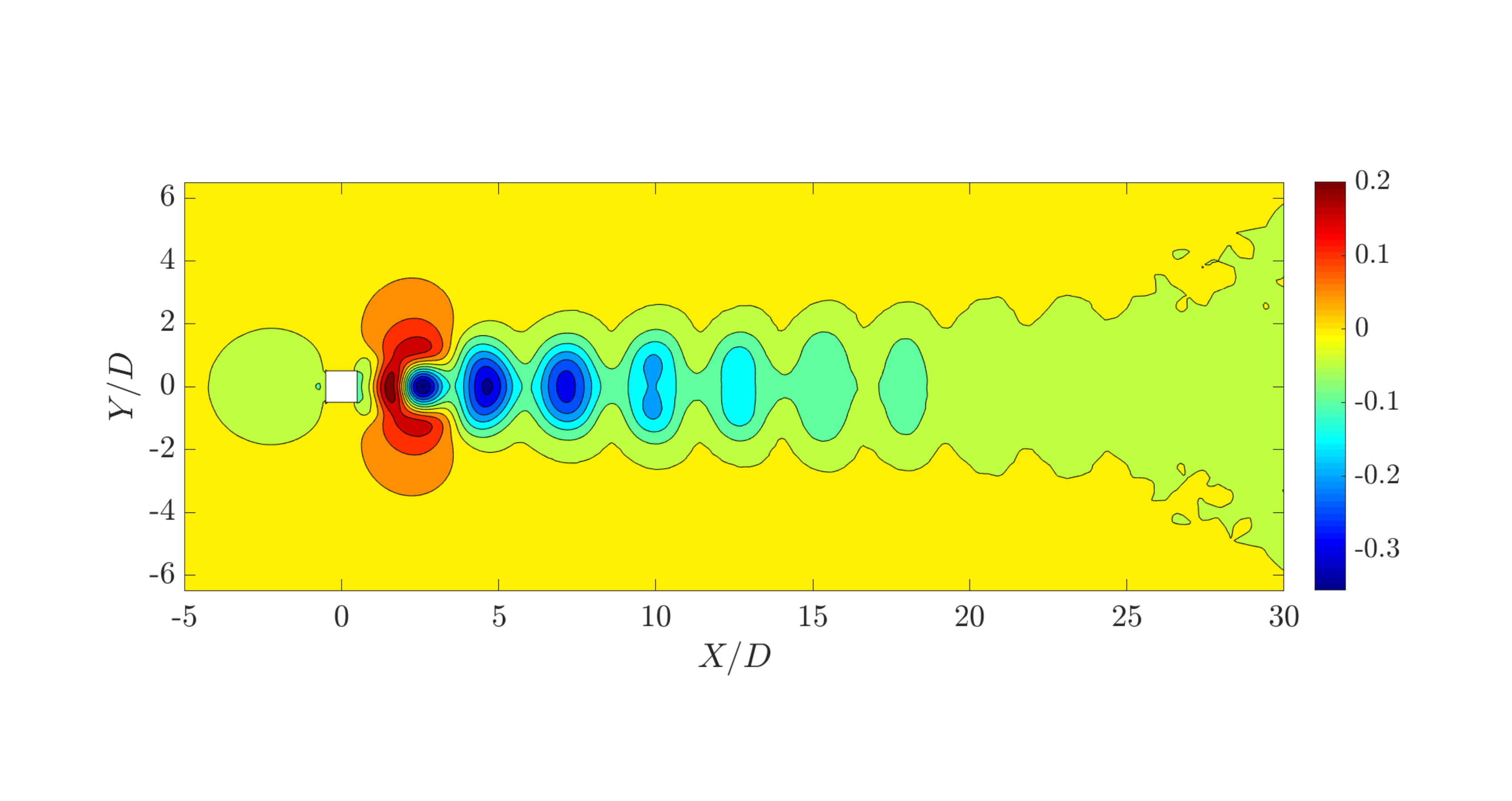}
\caption{Mode 6 (1.39\%)}
\end{subfigure}~
\begin{subfigure}[]{0.5\textwidth}
\centering
\includegraphics[trim={2cm 3cm 2cm 3.9cm},clip,scale=0.225]{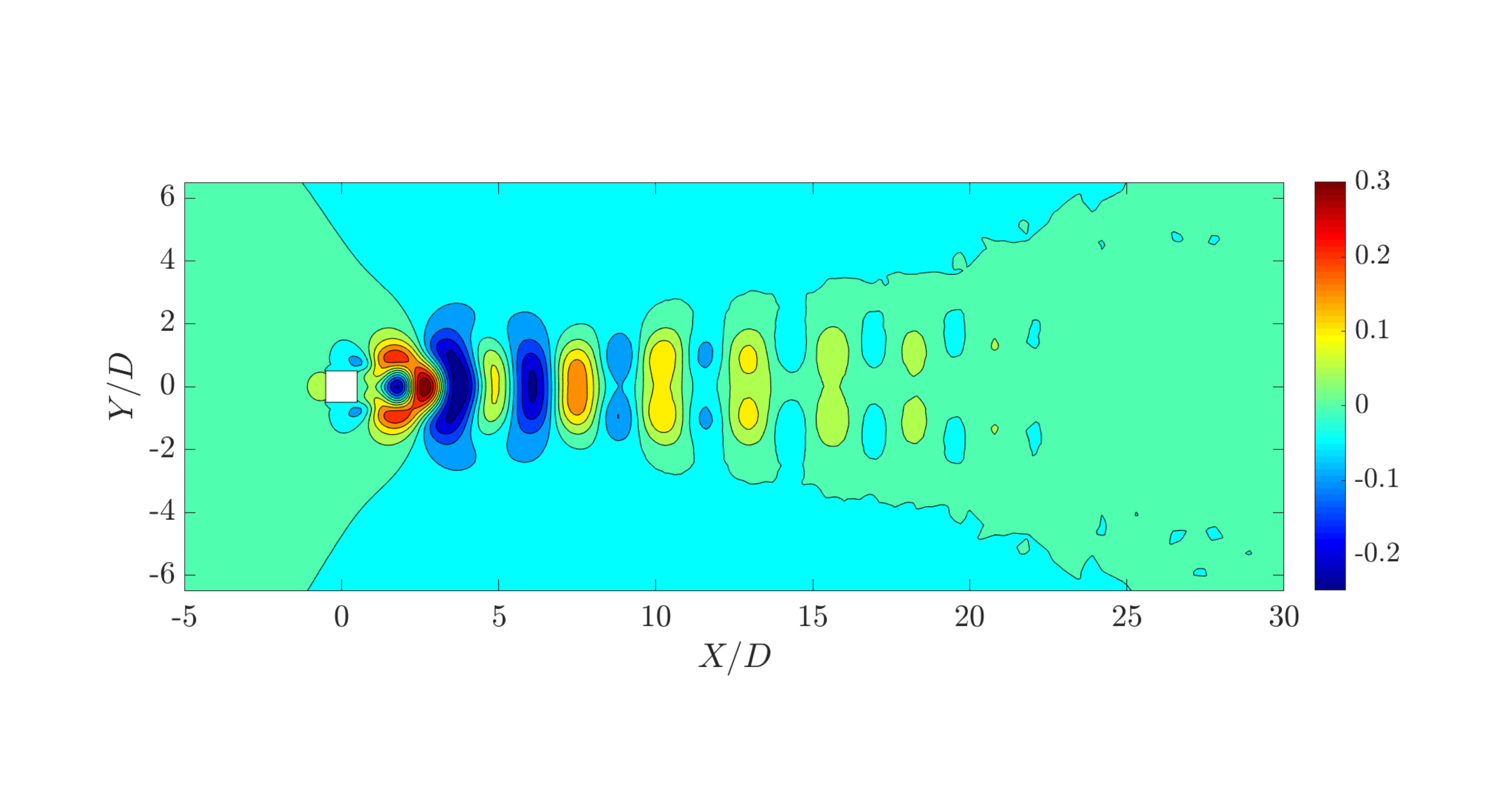}
\caption{Mode 7 (1.23\%)}
\end{subfigure}
\begin{subfigure}[]{0.5\textwidth}
\centering
\includegraphics[trim={2cm 3cm 2cm 3.9cm},clip,scale=0.225]{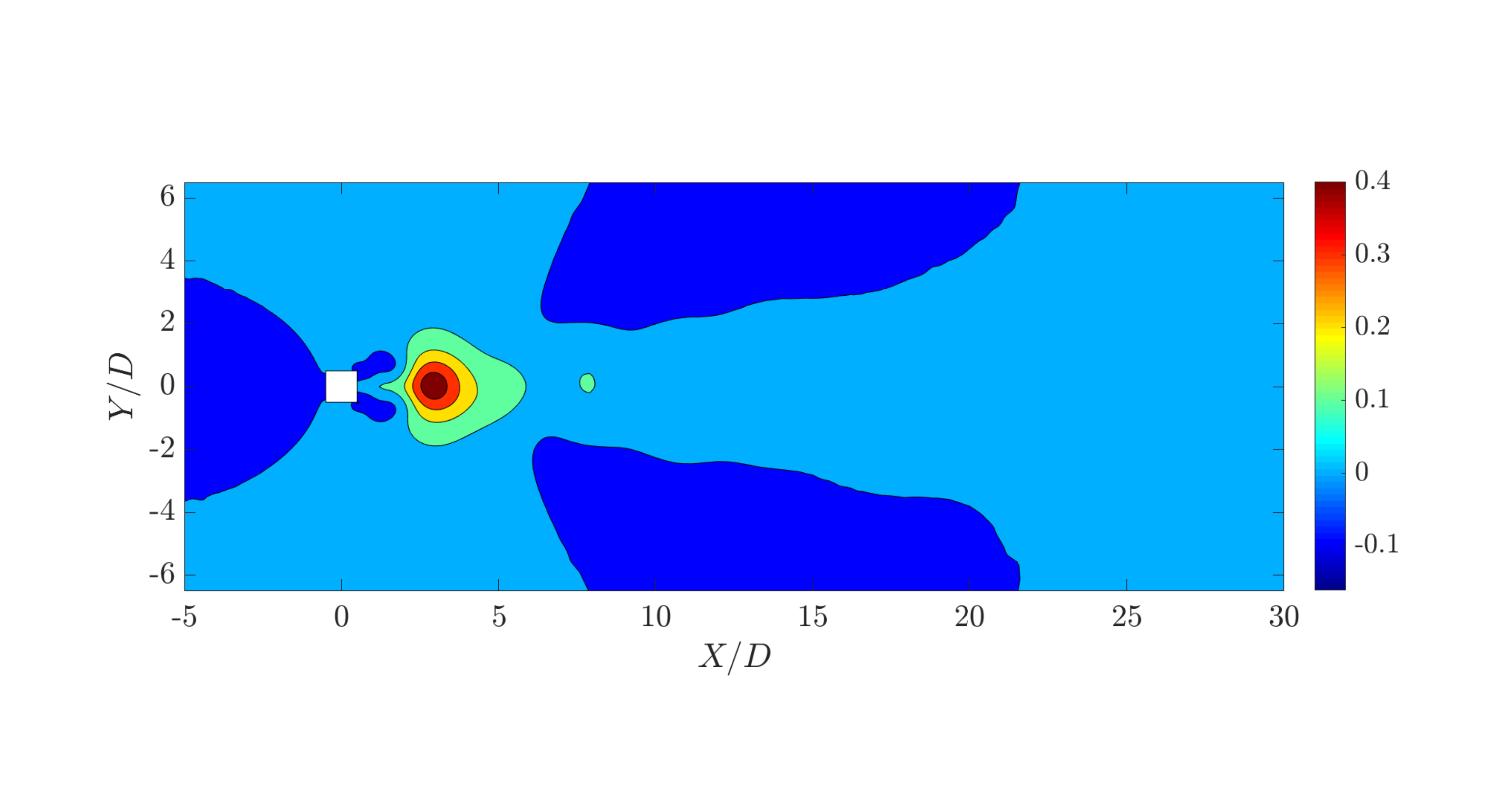}
\caption{Mode 8 (0.52\%)}
\end{subfigure}~
\begin{subfigure}[]{0.5\textwidth}
\centering
\includegraphics[trim={2cm 3cm 2cm 3.9cm},clip,scale=0.225]{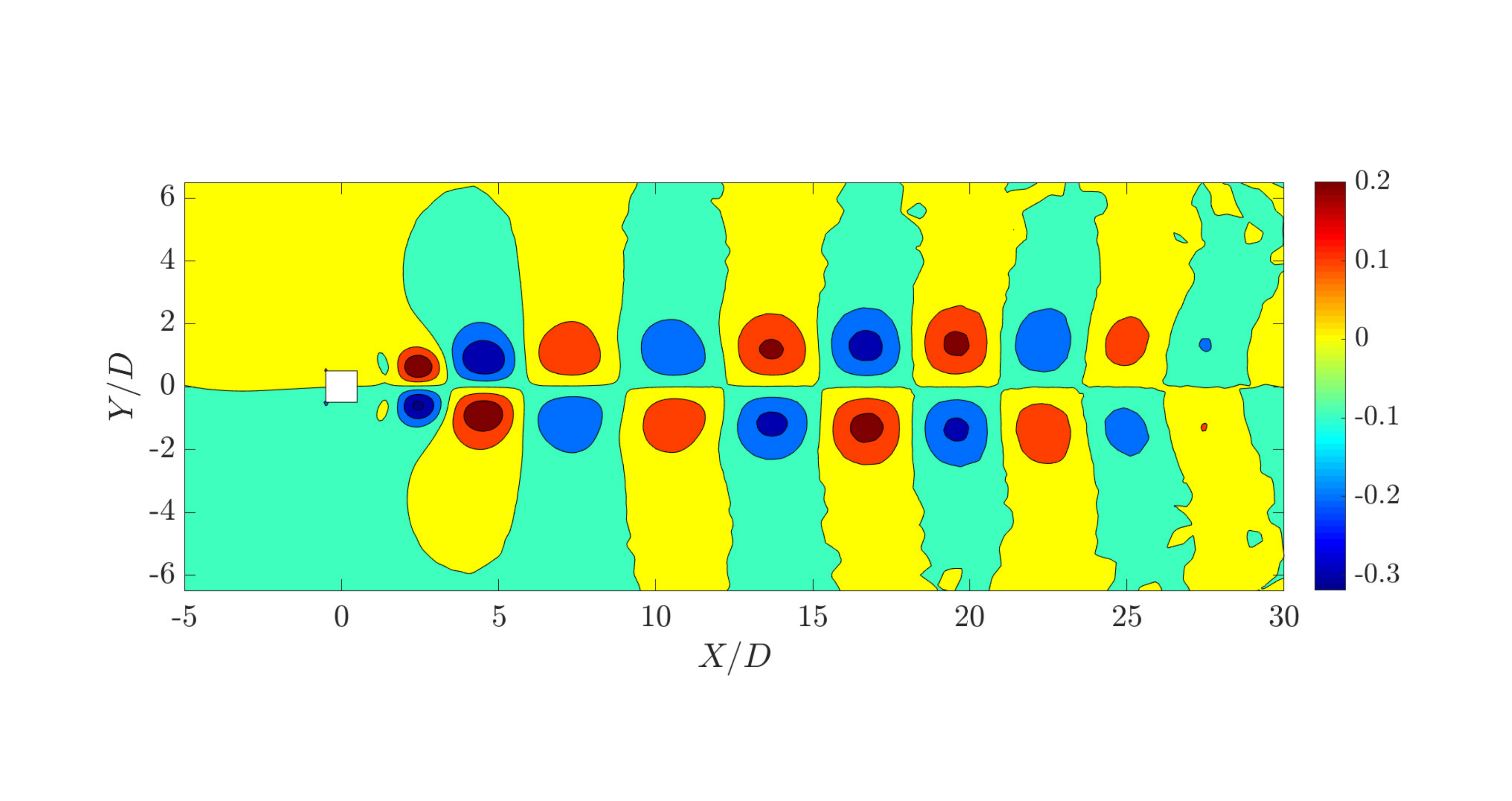}
\caption{Mode 9 (0.46\%)}
\end{subfigure}
\caption{The mean field and the first 9 significant POD modes. The energy fraction of the POD mode is mentioned in brackets. 
The values are normalized by $1/2\rho^{\mathrm{f}}U_{\infty}^2$. The flow is from left to right.}
\label{PODModes}
\end{figure}   

In the linear POD reconstruction method, the instantaneous pressure field is recovered by the mean and a linear combination of the identified significant modes. 
In this analysis, $r$ is set to $9$, which represents the most energetic modes containing $\sim 99\%$ of the total contribution to the pressure fluctuations.
The temporal coefficients $\hat{y}_j$ are determined by the $L^2$ inner product between the fluctuation matrix and the modes as expressed in Eq. (\ref{eq:innerProduct}). 
The mean pressure distribution and the first 9 POD modes are displayed in figure \ref{PODModes}. The mean field is symmetric around the $X$-axis along the wake centerline. This is expected as the time-averaged distribution of the flow past a symmetrical bluff body should be symmetrical. Furthermore, the modes 2, 4, 5, 6, 7 and 8 are symmetric around the wake centerline while modes 1, 3 and 9 are anti-symmetric with equal values and opposite signs about the wake centerline. It is evident that the first, third and ninth modes correspond to the Karman vortex street with alternating positive and negative pressure regions about the $X$-axis and the pressure contours resulting from a staggered vortex street. The POD modes 2 and 8 have a high gradient behavior in the near-wake ($0.5D-5D$) region almost parallel to the top and bottom edges of the square cylinder suggesting that this mode represents the influence from the shear layer. The modes 4, 5, 6 and 7 originate from the near-wake region and diffuse symmetrically towards the far wake. We can attribute these contributions to the near-wake bubble and its local dynamical property. For the ease of explanation, we refer to these modes as the vortex shedding, the shear layer and the near-wake.

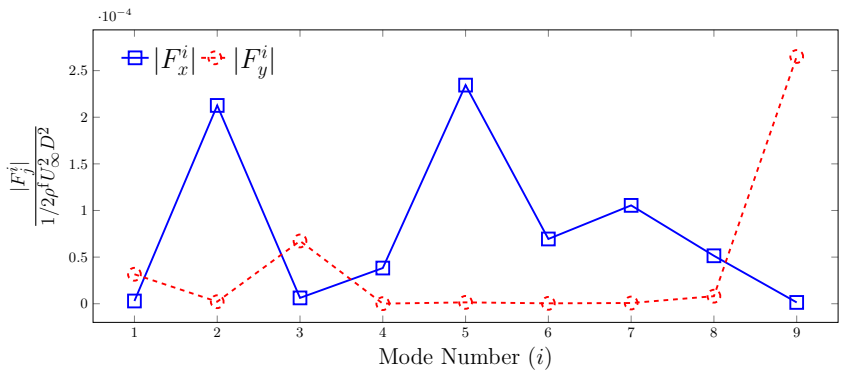
\begin{figure}
\centering
\begin{tikzpicture}[trim axis left, 
trim axis right,
scale=0.6, 
baseline]
\begin{axis}[
    xlabel={\Large Mode Number $(i)$},
    ylabel={\LARGE $\frac{|F_j^i|}{1/2 \rho^{\mathrm{f}} U_{\infty}^2 D^2}$},
    xmin=0.5, xmax=9.5,
     ymin=-0.00002,
    xtick={1,2,3,4,5,6,7,8,9},
    width =18cm,
    height = 8cm,
    legend pos=north west,
    legend style={draw=none},
    legend columns = 2,
]
\addplot[
    color=blue,
    solid,
    mark=square,
    very thick,
    mark size = 4,
    ]
    coordinates {
(1,2.9383e-06)(2,2.1266e-04)(3,6.2534e-06)(4,3.8276e-05)(5,2.3445e-04)(6,6.9448e-05)(7,1.0548e-04)(8,5.1432e-05)(9,1.4005e-06)
    };
    \addlegendentry{\LARGE $|F_x^i|$}
    \addplot[
    color=red,
    dashed,
    mark=o,
    very thick,
    mark size = 4,
    ]
    coordinates {
(1,3.1611e-05)(2,2.1805e-06)(3,6.7318e-05)(4,1.1762e-07)(5,1.4662e-06)(6,4.1714e-07)(7,7.8217e-07)(8,8.0932e-06)(9,2.6516e-04)
    };
    \addlegendentry{\LARGE $|F_y^i|$}
\end{axis}
\end{tikzpicture}
\caption{Absolute value of the time invariant contribution from each mode to the in-line($|F_x^i|$) and cross-flow($|F_y^i|$) forces. 
Modes 1, 3 and 9 capture the vortex shedding, modes 2 and 8 correspond to the effect of the shear layer and modes 4, 5, 6 and 7 capture the near-wake bubble effects. 
}
\label{fig:FjModes}
\end {figure}
Figure \ref{fig:FjModes} quantifies the time-invariant contributions ($F_j^i$) from each mode to the  drag and lift forces. For the definition of $F_j^i$, $j=(x,y)$ is the direction of the force and $i$ is the mode number. These values are calculated based on the fluid-solid boundary values of the mode fields displayed in figure \ref{PODModes}. It is clear that the vortex shedding modes (modes 1, 3 and 9) contribute entirely to the lift force, while the shear layer and near-wake modes contribute entirely to the drag force. Further details on the force decomposition procedure using the modal contributions are presented in Appendix A. Due to this directionally independent contribution of the bluff body features for the forces, the time coefficients ($\hat{y}_j(t)$) of these modes should display the same frequencies of the lift and drag forces. 

\begin{figure}
\centering
\begin{subfigure}[]{\textwidth}
\centering
\includegraphics[trim={0.5cm 0.3cm 0 0},clip,scale=0.4]{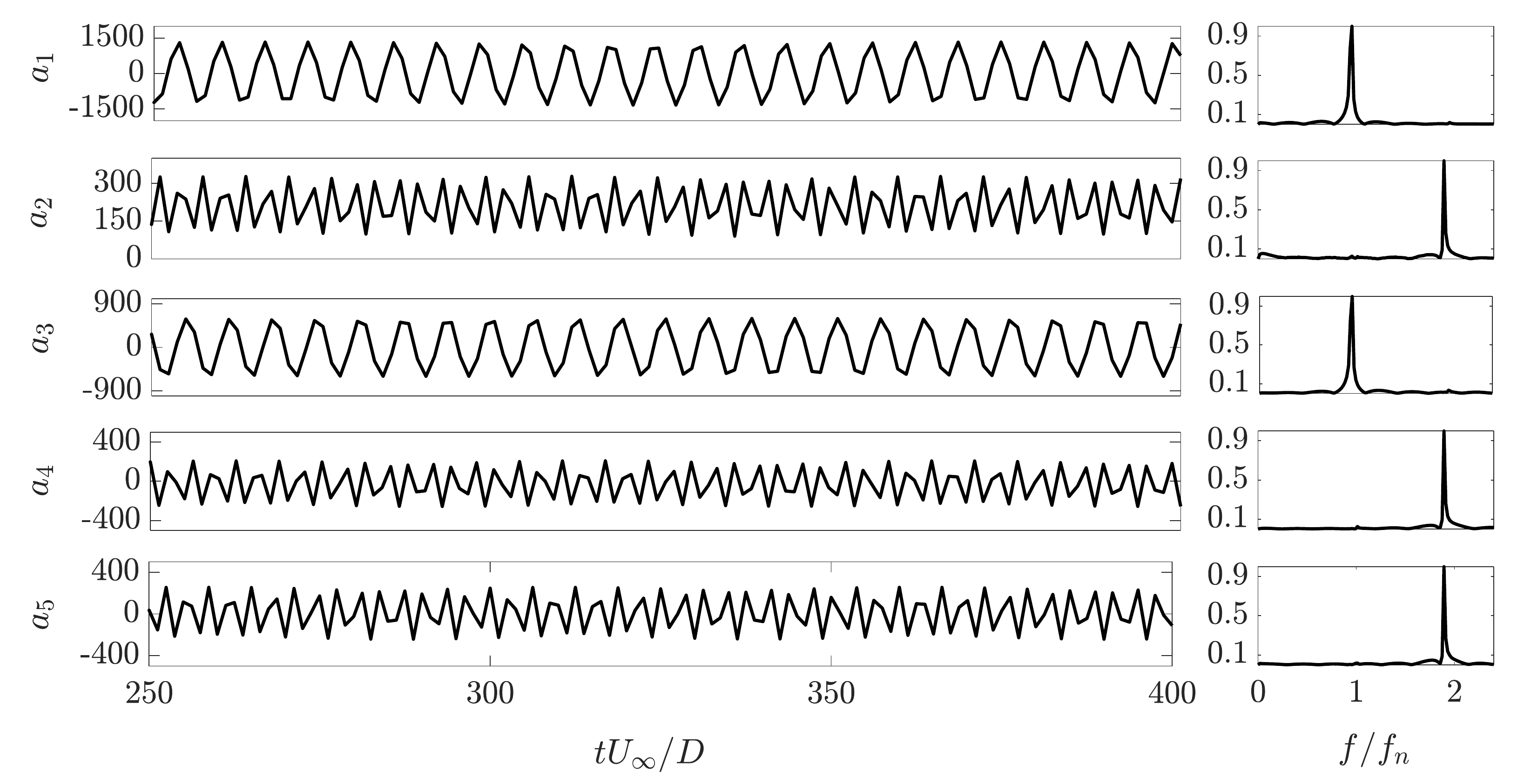}
\caption{}
\end{subfigure}
\begin{subfigure}[]{\textwidth}
\centering
\includegraphics[trim={0.6cm 0.6cm 0.6cm 0.8cm},clip,scale=0.4]{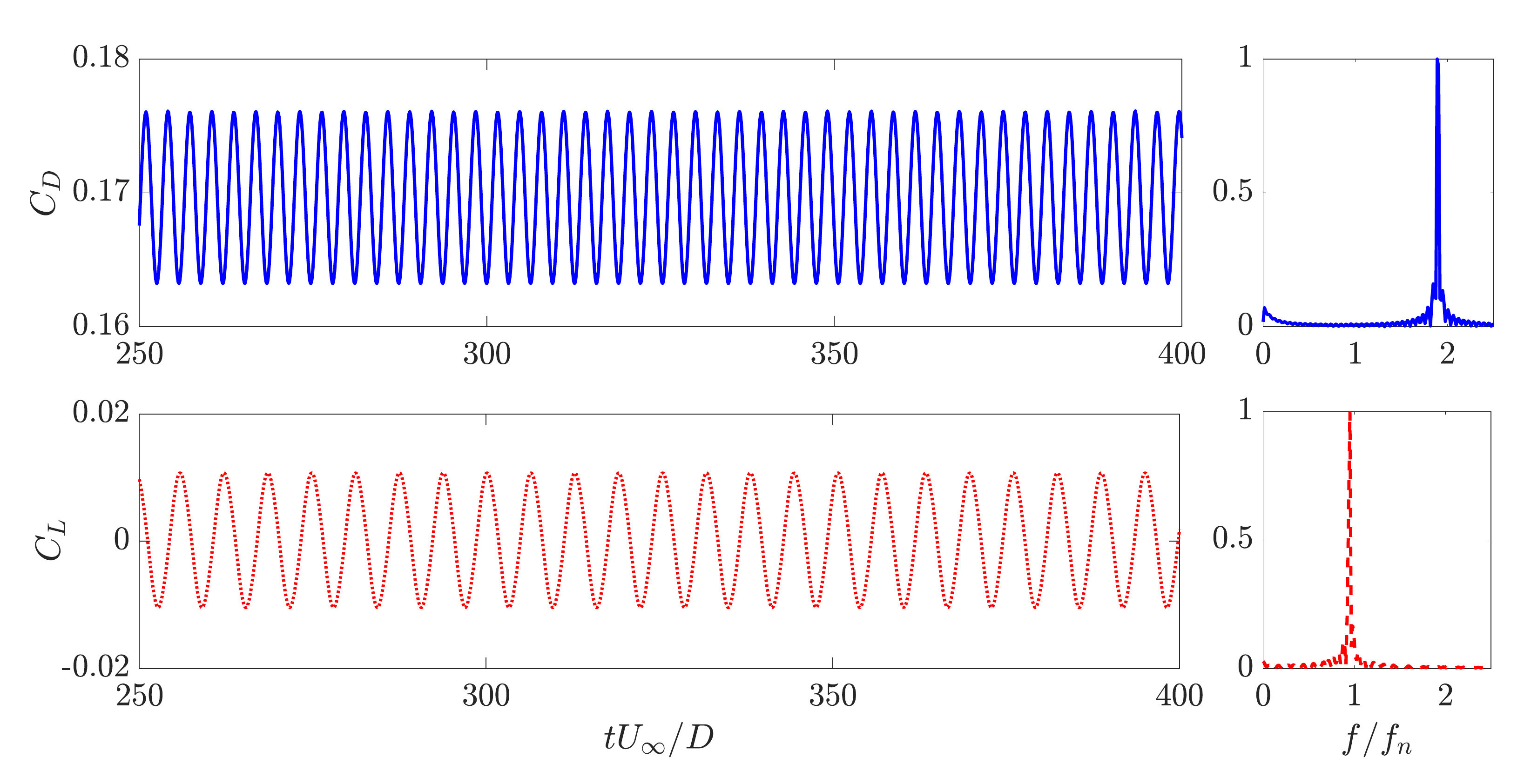}
\caption{}
\end{subfigure}
\caption{Time history and FFT comparisons: (a) temporal coefficients of first 5 POD modes, (b) force coefficients. Note that modes 2, 4 and 5 have the same frequency ($2f_n$) as the drag and modes 1 and 3 has the same frequency ($f_n$) as the lift. 
}
\label{ModeTcoeff}
\end{figure}

The time histories and the FFT spectra of the first 5 POD modes are shown in figure \ref{ModeTcoeff}a. The time histories have sharp variations, because of the snapshot sampling of the data. The first and third mode coefficients have a low-frequency sinusoidal variation with the natural frequency ($f_{n}$). The second mode coefficient has a non-zero mean with a $\approx 2\times f_{n}$ frequency. The fourth and fifth modes have a nearly zero mean variation with $\approx 2\times f_{n}$ frequency. Interestingly, as presented in figure \ref{ModeTcoeff}b, $f_n$ and $2f_n$ coincide with the frequencies of lift and drag, respectively. Hence, we can further confirm that the modes 1 and 3 make their sole contribution to the fluctuating lift while the modes 2, 4 and contribute to the drag force. Furthermore, the non-zero mean drag can be attributed to the non-zero mean of mode 2, which contains the largest component of the drag. From these observations, we can further confirm that the vortex shedding process contributes exclusively to the lift force and the near-wake and the shear layer phenomena influence the drag force.  

\begin{figure}
\begin{subfigure}[]{\textwidth}
\centering
\includegraphics[trim={2cm 2.5cm 1cm 3.5cm},clip,scale=0.45]{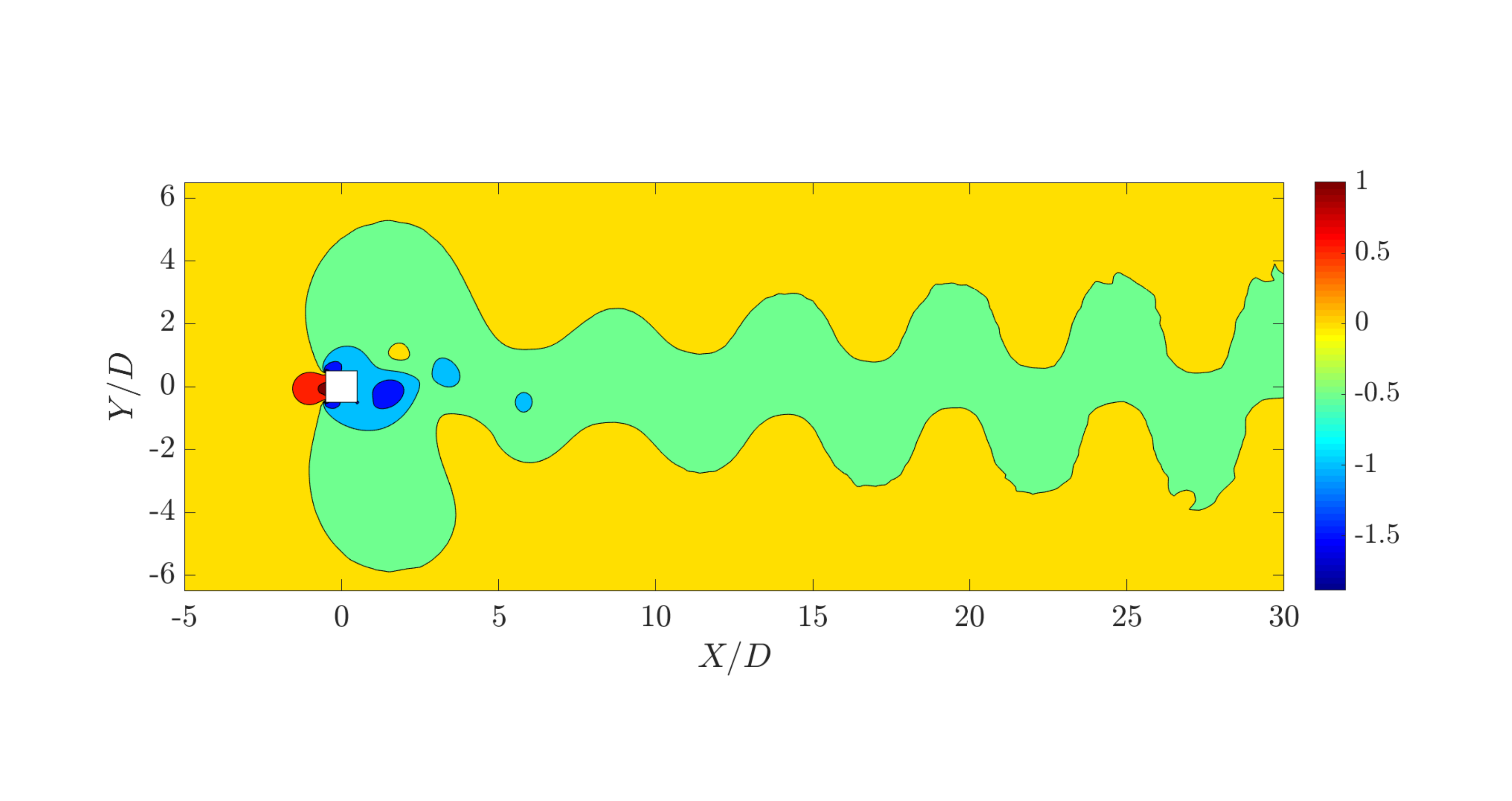}
\caption{}
\end{subfigure}
\begin{subfigure}[]{\textwidth}
\centering
\includegraphics[trim={2cm 2.5cm 1cm 3.5cm},clip,scale=0.45]{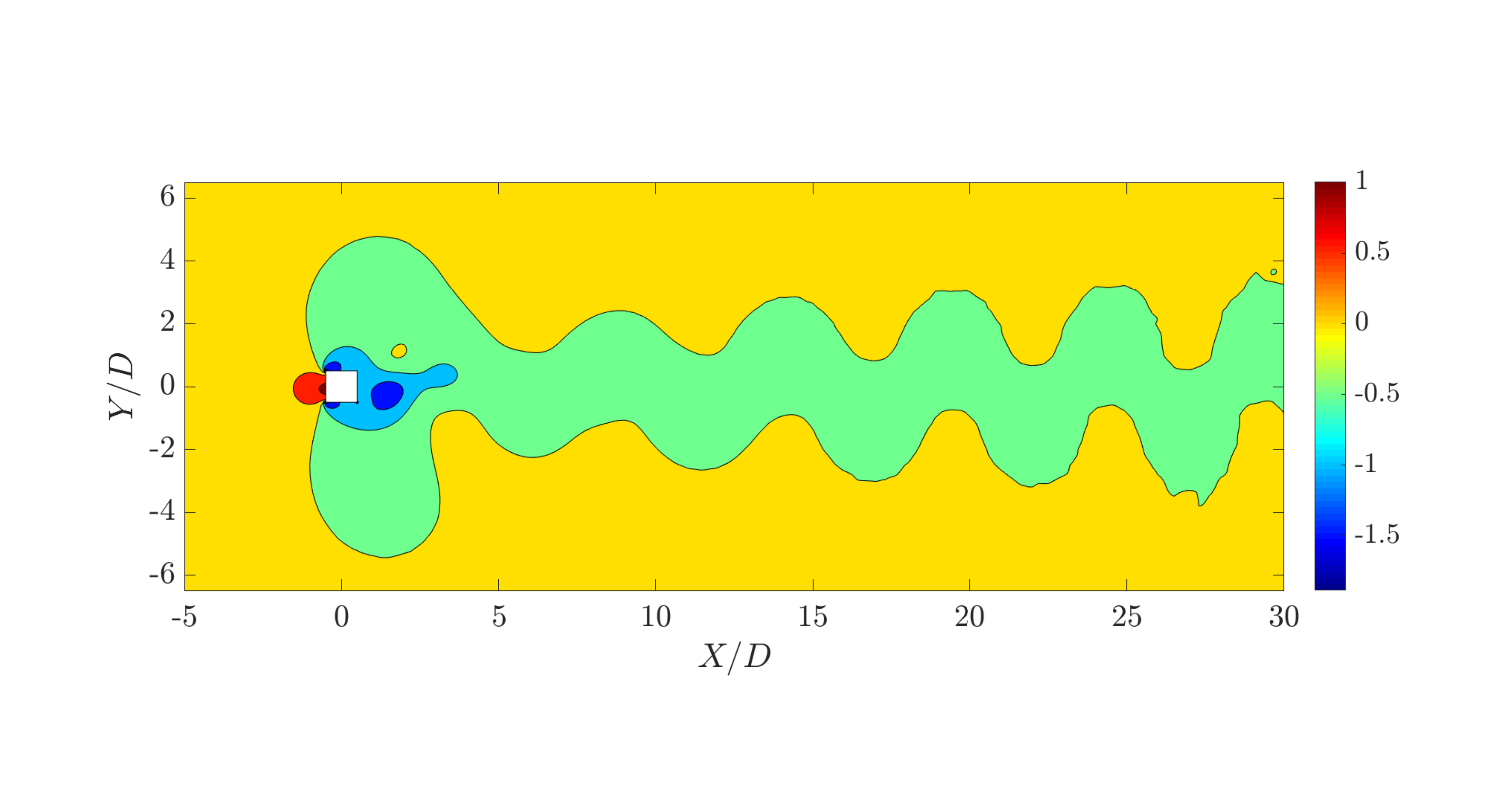}
\caption{}
\end{subfigure}
\begin{subfigure}[]{\textwidth}
\centering
\includegraphics[trim={0.5cm 2cm 0 1.8cm},clip,scale=0.4]{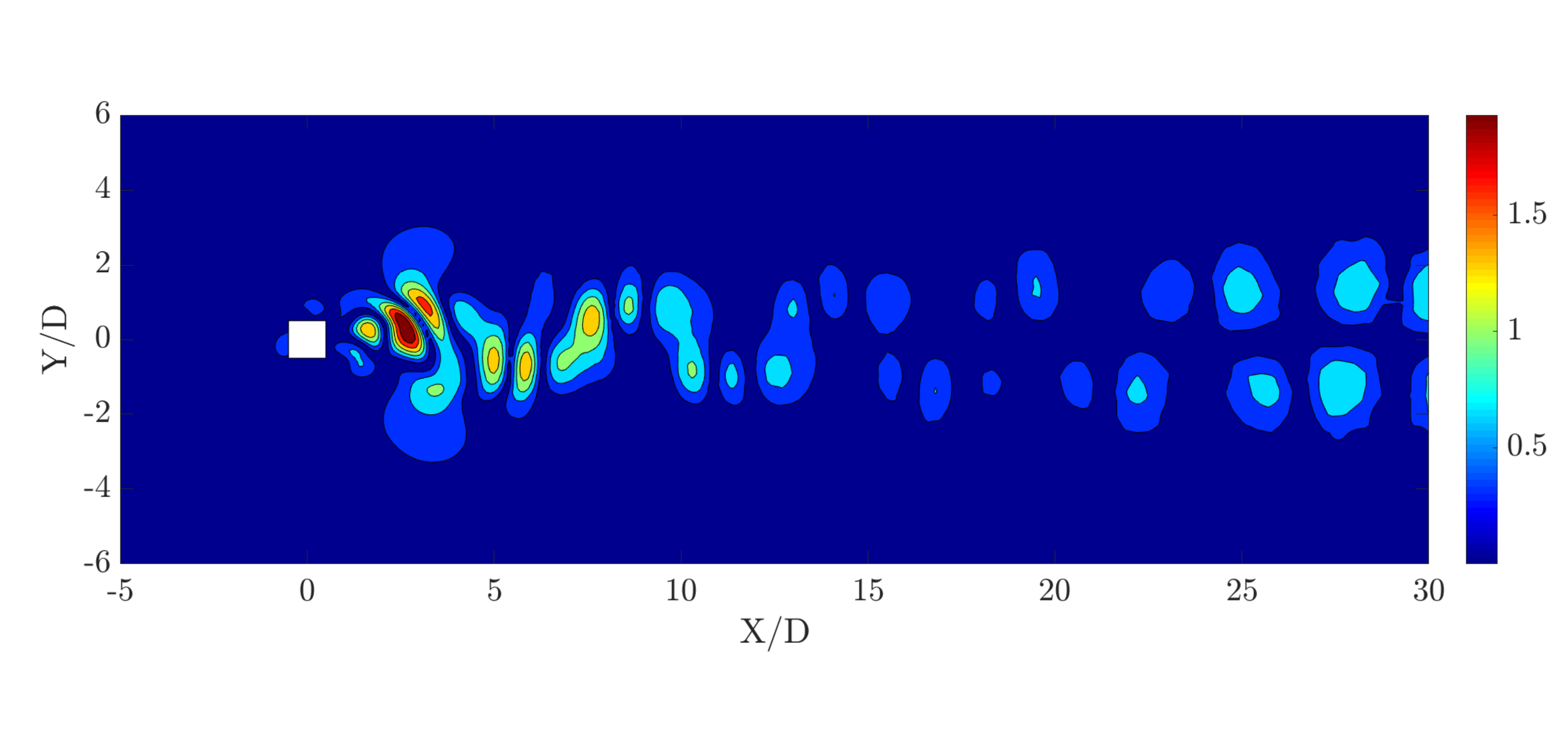}
\caption{}
\end{subfigure}
\caption{Comparison between POD reconstruction and FOM result: pressure distribution obtained by (a) full-order model, (b) linear POD reconstruction, and (c) the relative error (\%). Maximum error percentage is less than 2\%. The highly nonlinear near-wake region and the vortex cores have the highest error. The pressure values are normalized by $\frac{1}{2}\rho^{\mathrm{f}}U_{\infty}^2$. The flow is from left to right.}
\label{PODError}
\end{figure}

Using the POD procedure, we successfully decompose the flow field into physically significant features. We reconstruct the same field combining these modes in the linear POD technique: such that, utilizing Eq. (\ref{eq:linearPOD}) for the pressure field, gives
${P}(t) \approx \overline{{P}} + \sum_{j=1}^{r} \hat{y}_j(t)\b{v}_j$.
Figure \ref{PODError} illustrates the pressure distribution at $tU_{\infty}/D=100$ using the linear POD reconstruction.
The recovered POD mode is compared with the result obtained from the full-order model. A good match with a maximum relative local error $< 2\%$ can be seen  in figure \ref{PODError}. To quantify the accuracy of the entire flow field recovery, the normalized root mean square (rms) error  of the entire distribution is considered. The rms error $\epsilon^{rms}$ is given by:
\begin{equation}
\epsilon^{rms} = \frac{\sqrt{\sum (P_{FOM} - P_{POD})^2/n_c}}{|\overline{P_{100}}|}\times 100,
\end{equation}
where $P_{FOM}$ and $P_{POD}$ are the pressure values of the mesh nodes extracted from the full-order model and the POD reconstruction, respectively and $|\overline{P_{100}}|$ is the mean pressure of the field. When 9 modes are used, this error is $\epsilon^{rms} = 3.78\%$. 
In this linear reconstruction, the highest error is observed at the regions known to exhibit a nonlinear variation, such as the near-wake region, the shear layer and the vortex cores. Next, we analyze the POD-DEIM technique to improve the accuracy in these nonlinear flow features using the snapshot sequence and their respective DEIM points.
\subsection{Nonlinear POD-DEIM reconstruction}
\begin{figure}
\begin{subfigure}[]{\textwidth}
\centering
\includegraphics[trim={0 0 0 0},clip,scale=0.4]{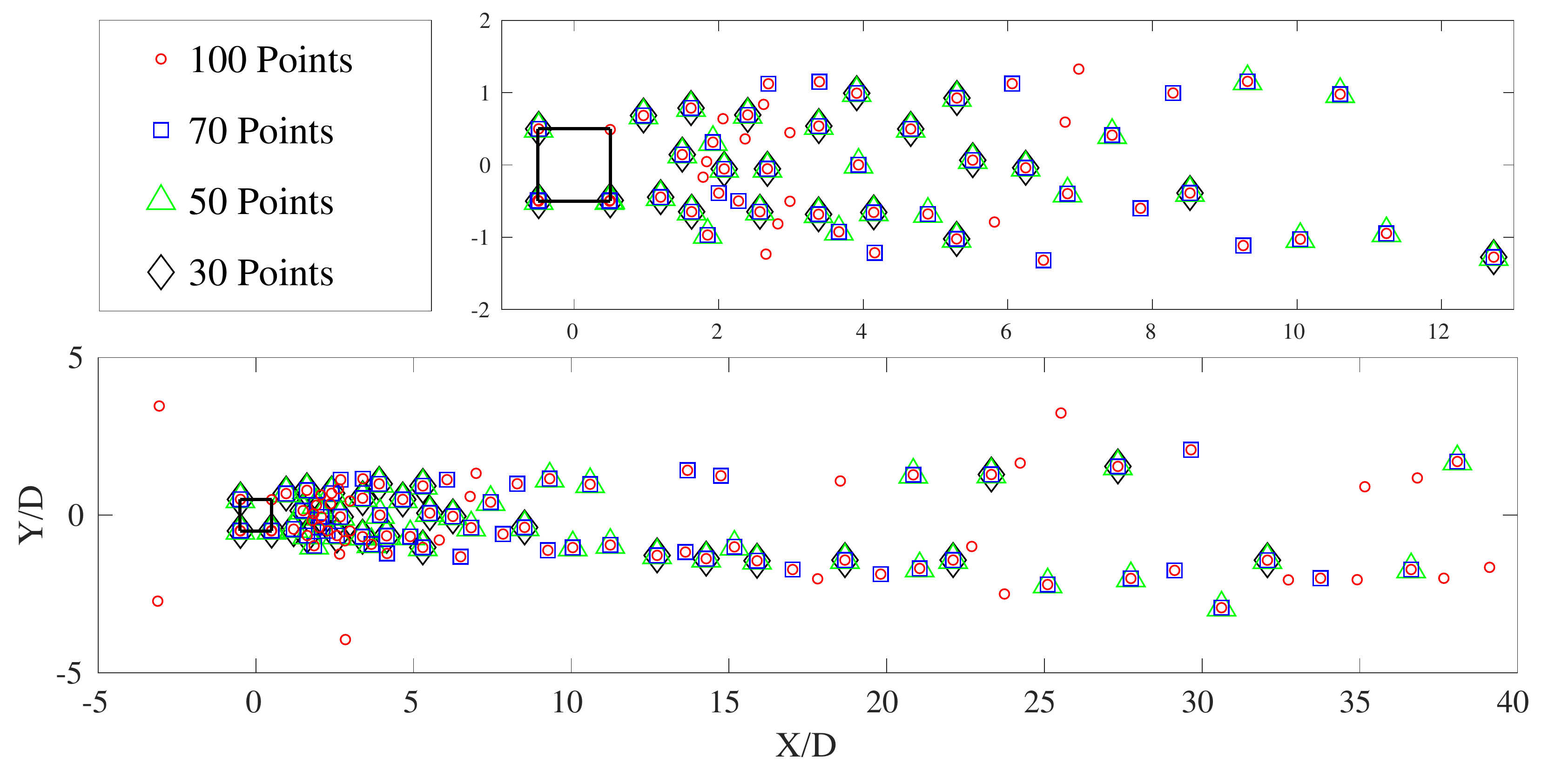}
\caption{}
\end{subfigure}
\begin{subfigure}[]{\textwidth}
\centering
\includegraphics[trim={0 0.5cm 0 0.5cm},clip,scale=0.5]{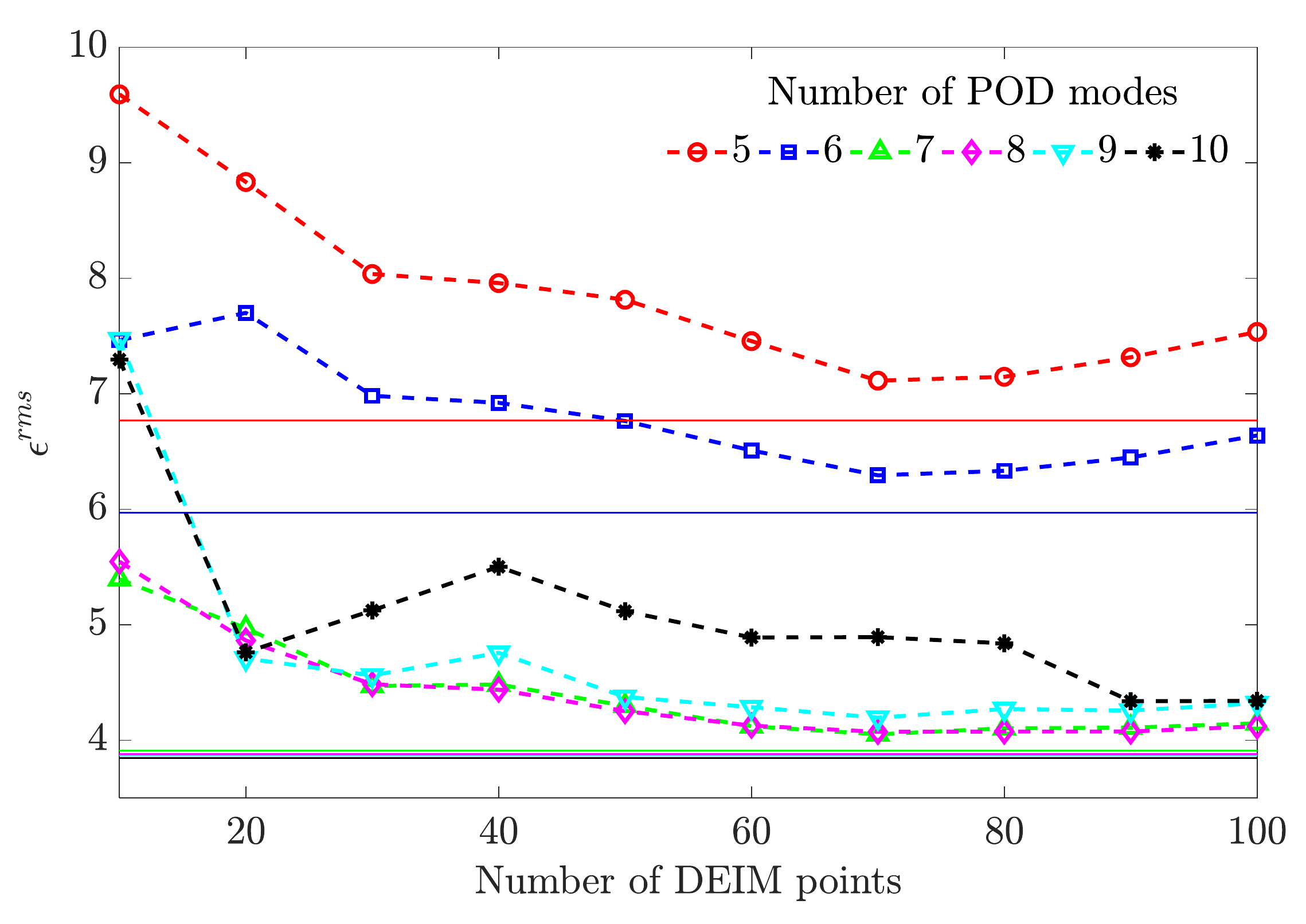}
\caption{}
\end{subfigure}
\caption{(a) DEIM best points, the top right inset illustrates the near-wake DEIM points. (b) Performance of POD-DEIM compared with linear POD, solid lines denote the linear POD error. The least error is observed when 70 DEIM points are used with 7 POD modes. 
}
\label{DEIM_Points}
\end{figure}

\begin{figure}
\begin{subfigure}[]{\textwidth}
\centering
\includegraphics[trim={2cm 2.5cm 1cm 3.5cm},clip,scale=0.45]{Porig100.pdf}
\caption{}
\end{subfigure}
\begin{subfigure}[]{\textwidth}
\centering
\includegraphics[trim={2cm 2.5cm 1cm 3.5cm},clip,scale=0.45]{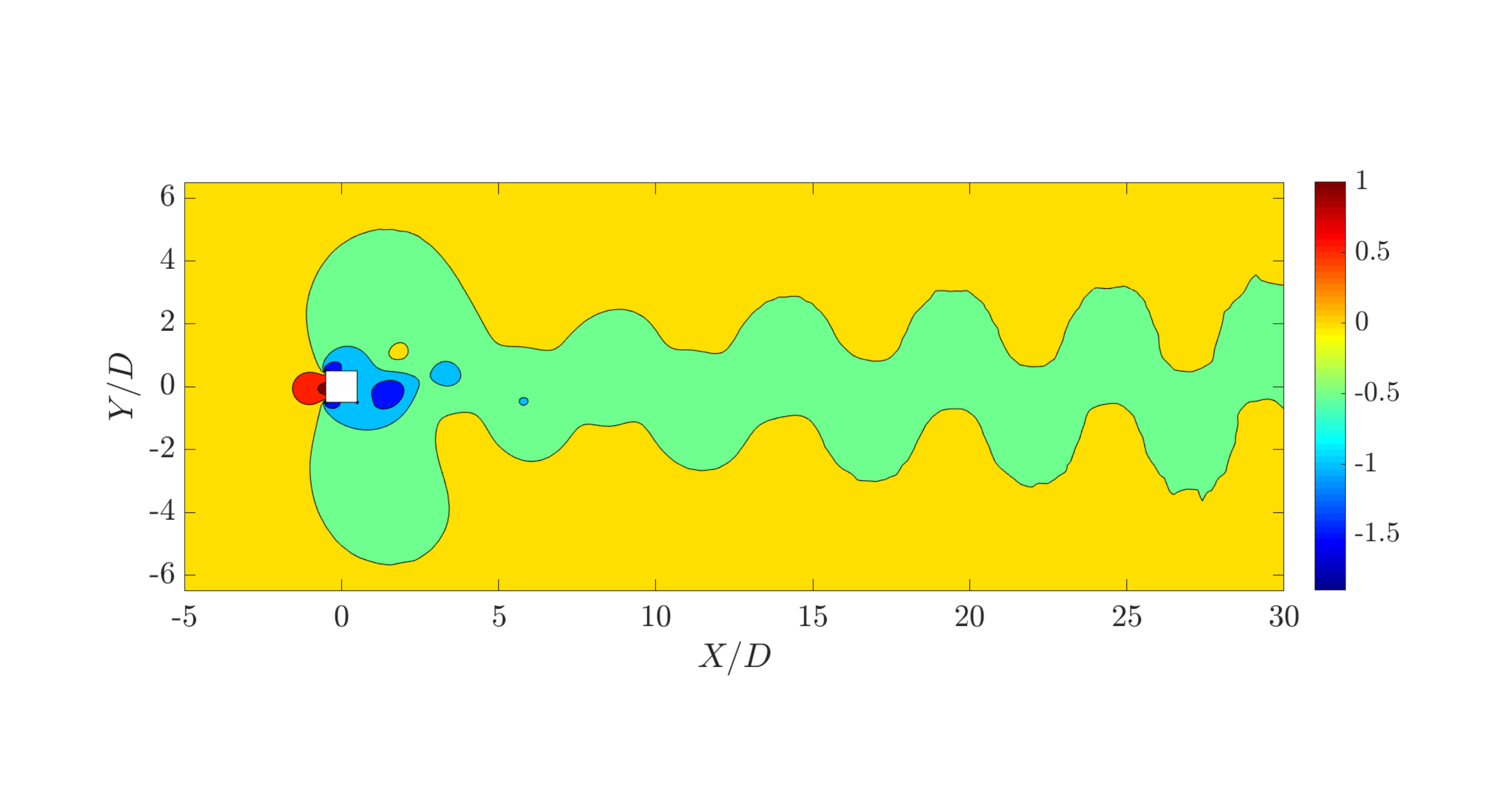}
\caption{}
\end{subfigure}
\begin{subfigure}[]{\textwidth}
\centering
\includegraphics[trim={0.5cm 2cm 0 1.8cm},clip,scale=0.405]{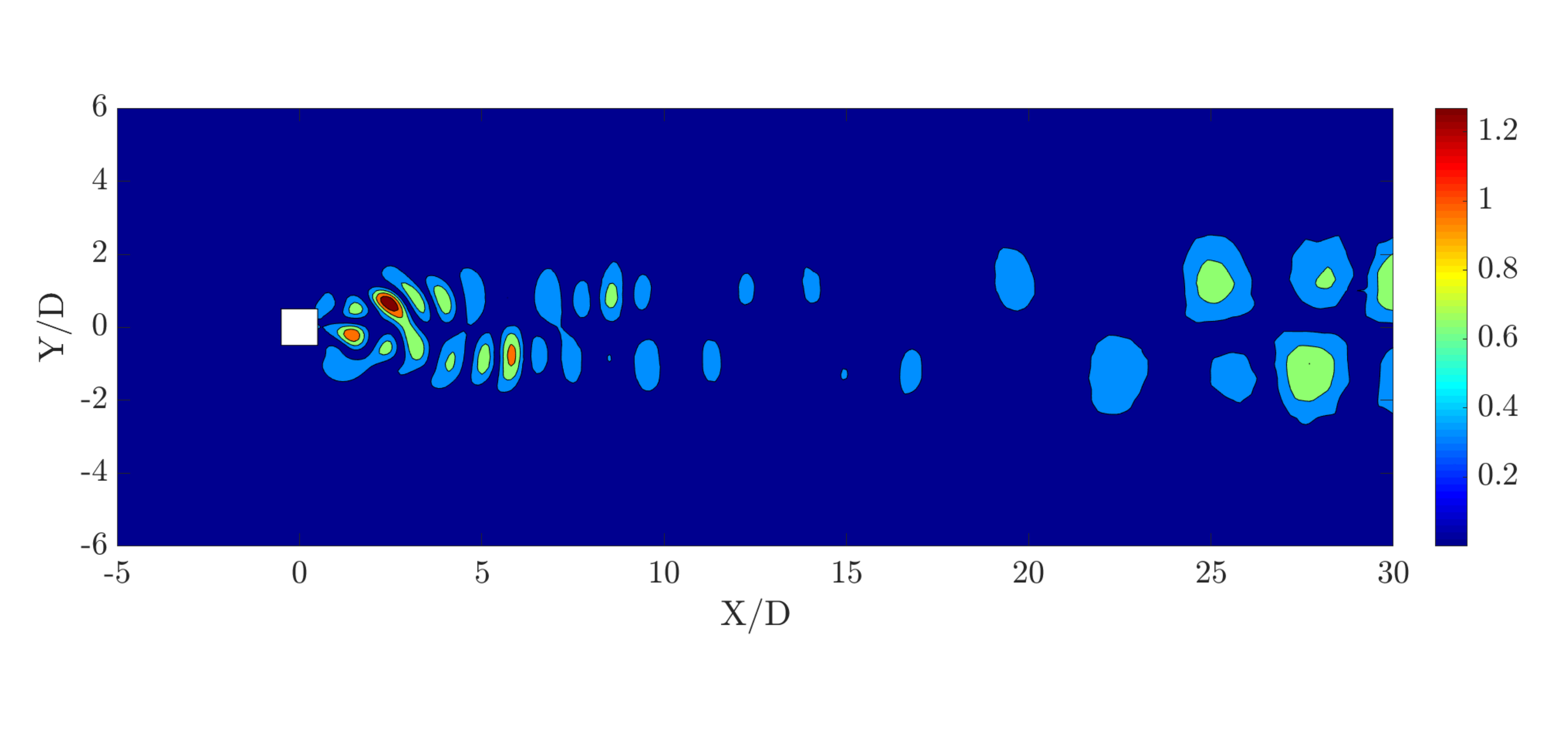}
\caption{}
\end{subfigure}
\caption{Comparison of POD-DEIM and FOM results: pressure distribution obtained by (a) full-order model, (b) POD-DEIM  reconstruction; and (c) the relative error of the reconstruction (\%). Maximum error percentage is less than 1.5\%. The error in the highly nonlinear regions has reduced compared to the linear-POD reconstruction. The pressure values are normalized by $1/2\rho^{\mathrm{f}}U_{\infty}^2$. The flow is from left to right.}
\label{DEIMRecon}
\end{figure}

The linear POD reconstruction has the highest error in the nonlinear regions. To reduce this error, more POD modes should be added to the reconstruction which makes the POD-based reconstruction very expensive.
Instead, when the DEIM technique is used, it reduces the calculation load while properly capturing the nonlinearity of the field variable. 
The DEIM utilizes two POD bases using the snapshot method, namely a first POD basis $\bs{\mathcal{V}}$ from the snapshot sequence, and a second basis $\bs{\mathcal{U}}$ from the nonlinear snapshots via the DEIM points.
However, unlike the linear POD reconstruction, the accuracy does not necessarily improve with the number of DEIM points and the number of POD modes employed. Using many DEIM points result in adding contributions from some non-significant indices. In figure \ref{DEIM_Points}a, it is clearly seen that for 100 DEIM points there are few mesh points which lie away from the significant nonlinear region are taken into the calculation. Further in Table (\ref{DEIMnwPr}), we quantify the number of points in the nonlinear-wake region as we increase the number of DEIM points. It is clear that the percentage of points in the critical region decrease as we include more points for the DEIM calculation. Owing to the nonlinear combination of the POD modes, including additional insignificant modes can increase the total error. As shown in figure \ref{DEIM_Points}b, the lowest $\epsilon^{rms} = 4.05\%$ can be obtained when 70 DEIM points are used with 7 POD modes. It further establishes that the linear POD reconstruction is generally accurate in a global sense, i.e. the entire flow field reconstruction, in contrast to the nonlinear POD-DEIM. However, figure \ref{DEIMRecon} demonstrates the reconstructed pressure distribution at $tU/D = 100$ using the optimum number of points and POD modes. There is a significant reduction in the local error as it allows to capture the nonlinear regions more accurately.

\begin{table}
\centering
\caption{Distribution of DEIM points in the near cylinder wake}
\label{DEIMnwPr}
\begin{tabular}{c|C{3.5cm}|c}
Number of DEIM points & Number of points in near cylinder wake & \%   \\ \hline
30             & 21                & 70.0 \\
50             & 31                & 62.0 \\
70             & 41                & 58.6 \\
100            & 54                & 54.0 \\ 
\end{tabular}
\end{table}

Apart from the global and local accuracy, we further assess the computational time consumed by the linear POD and nonlinear POD-DEIM reconstructions. The detailed analysis is presented in Appendix B. Theoretically, the DEIM reconstruction process should be $\approx 64$ times faster than linear POD reconstruction and the total DEIM process should be $\approx 3.14$ times faster. In the actual computations, when just the reconstructions are considered, the DEIM is $9.28$ times faster than the linear POD. When the total processes are compared, DEIM has a speedup of $3.98$. In terms of accuracy, DEIM is more accurate in a local sense since it captures the nonlinearities better than the linear POD reconstruction. However, when the entire fluid domain is considered, the linear POD method is more accurate than the DEIM. It is likely that the DEIM introduces unnecessary nonlinearities to the potential regions slightly changing the reconstructed field values. When decomposing and reconstructing the laminar flow fields, both linear and nonlinear methods perform to a satisfactory level. Both methods are capable of reaching the required threshold in a reasonable computational time while accurately capturing the flow features of the wake. Here onwards, we employ the POD-DEIM  since it has an improved accuracy when capturing the nonlinearities in the flow field at a lower computational cost.

\subsection{Drag and lift modes}
\begin{figure}
\centering
\begin{subfigure}[]{\textwidth}
\centering
\includegraphics[trim={0 0 0 1cm},clip,scale=0.3]{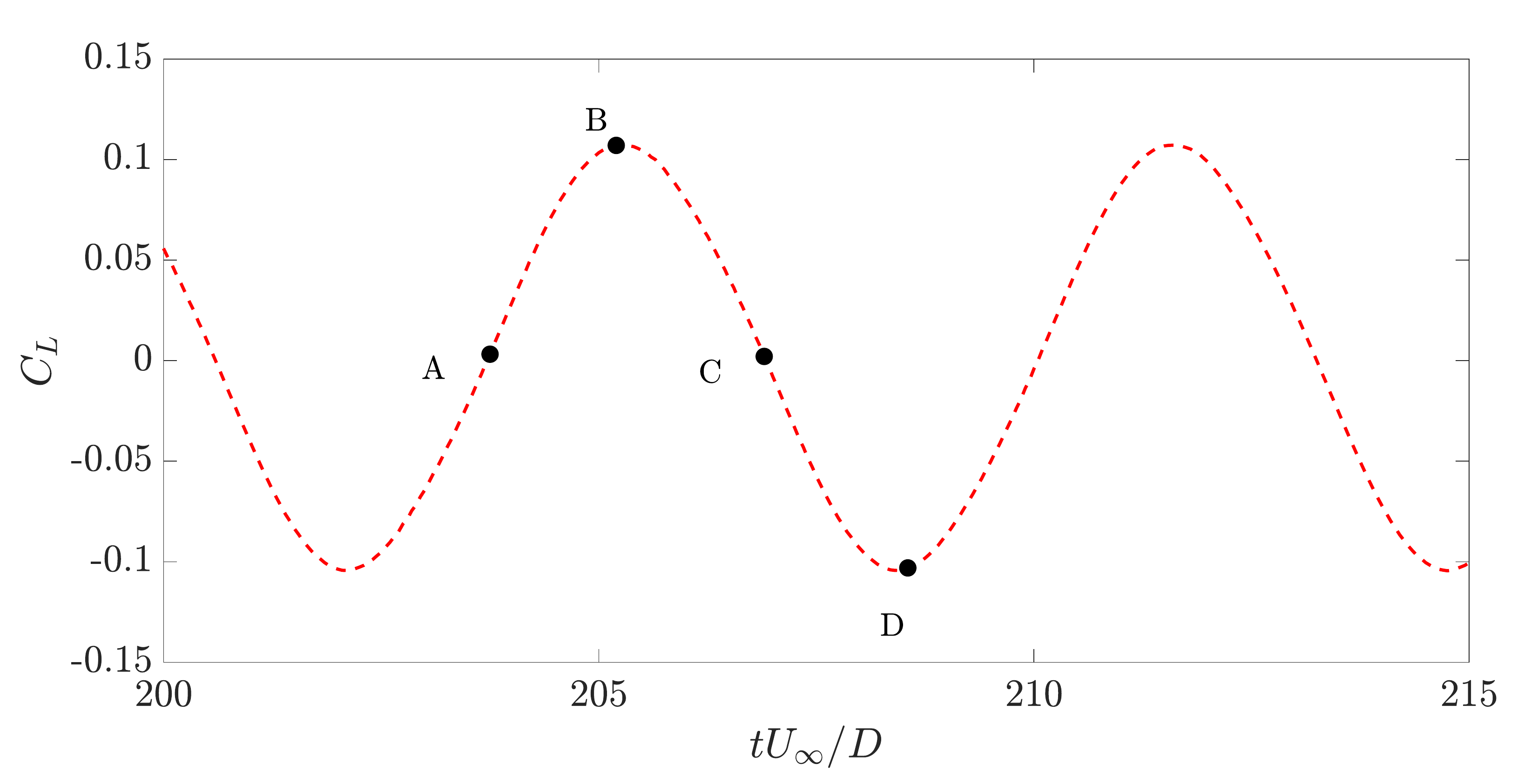}
\caption{}
\end{subfigure}
\begin{subfigure}[]{0.5\textwidth}
\centering
\includegraphics[trim={7cm 0 9.75cm 0},clip,scale=0.28]{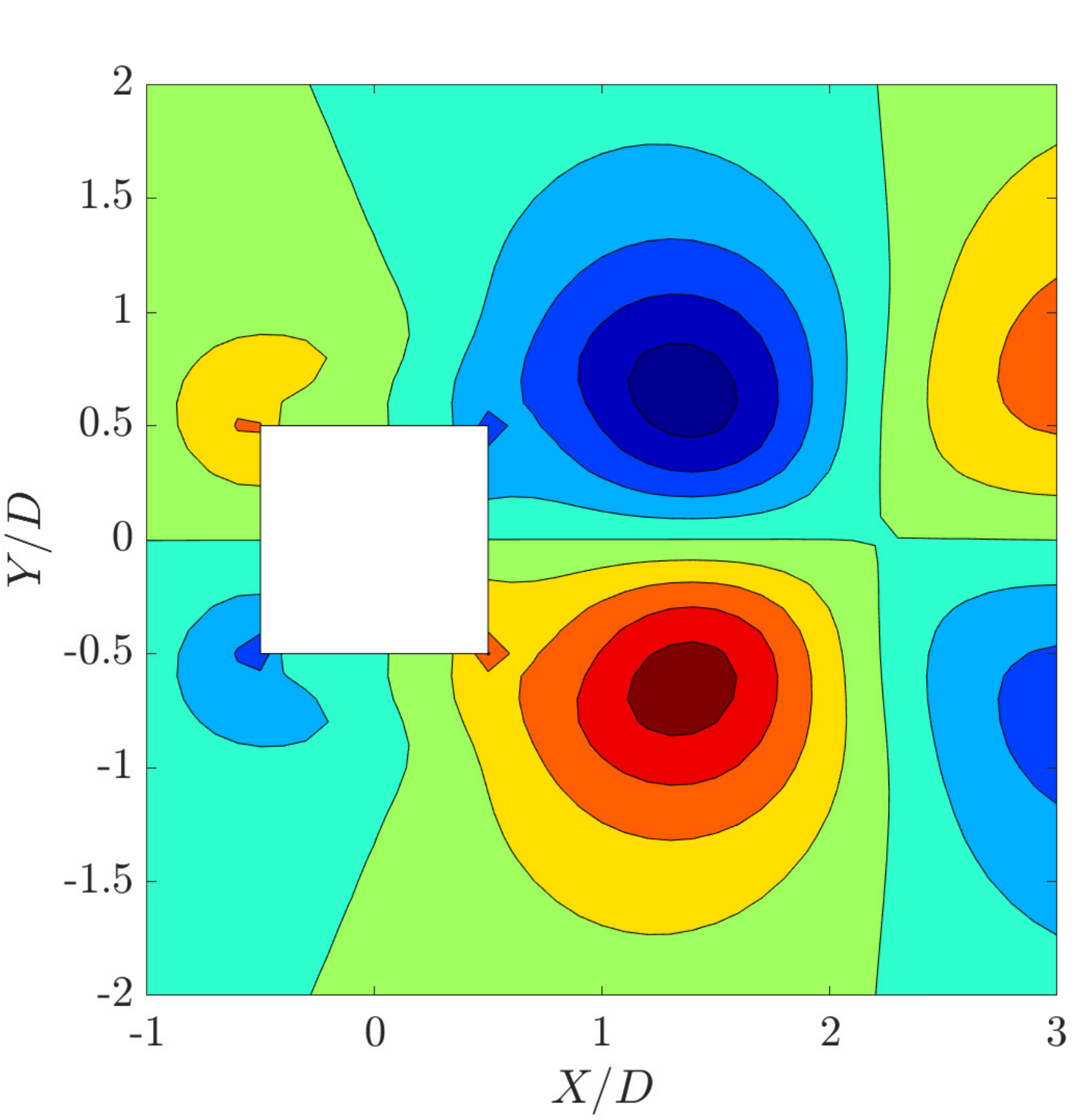}
\caption{Point A: zero lift (increasing)}
\end{subfigure}~
\begin{subfigure}[]{0.5\textwidth}
\centering
\includegraphics[trim={7cm 0 9.75cm 0},clip,scale=0.28]{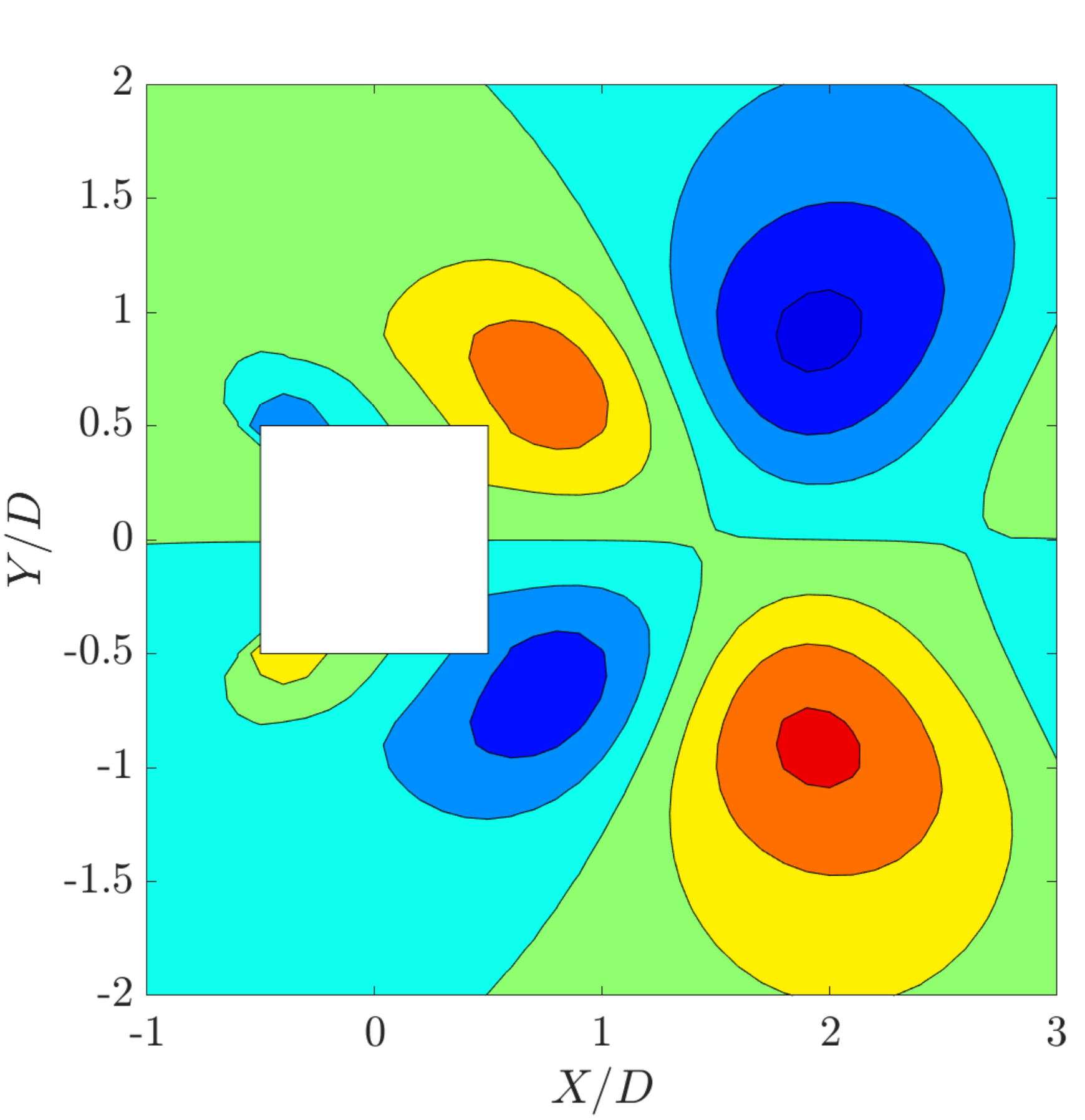}
\caption{Point B: maximum lift}
\end{subfigure}
\begin{subfigure}[]{0.5\textwidth}
\centering
\includegraphics[trim={7cm 0 9.75cm 0},clip,scale=0.28]{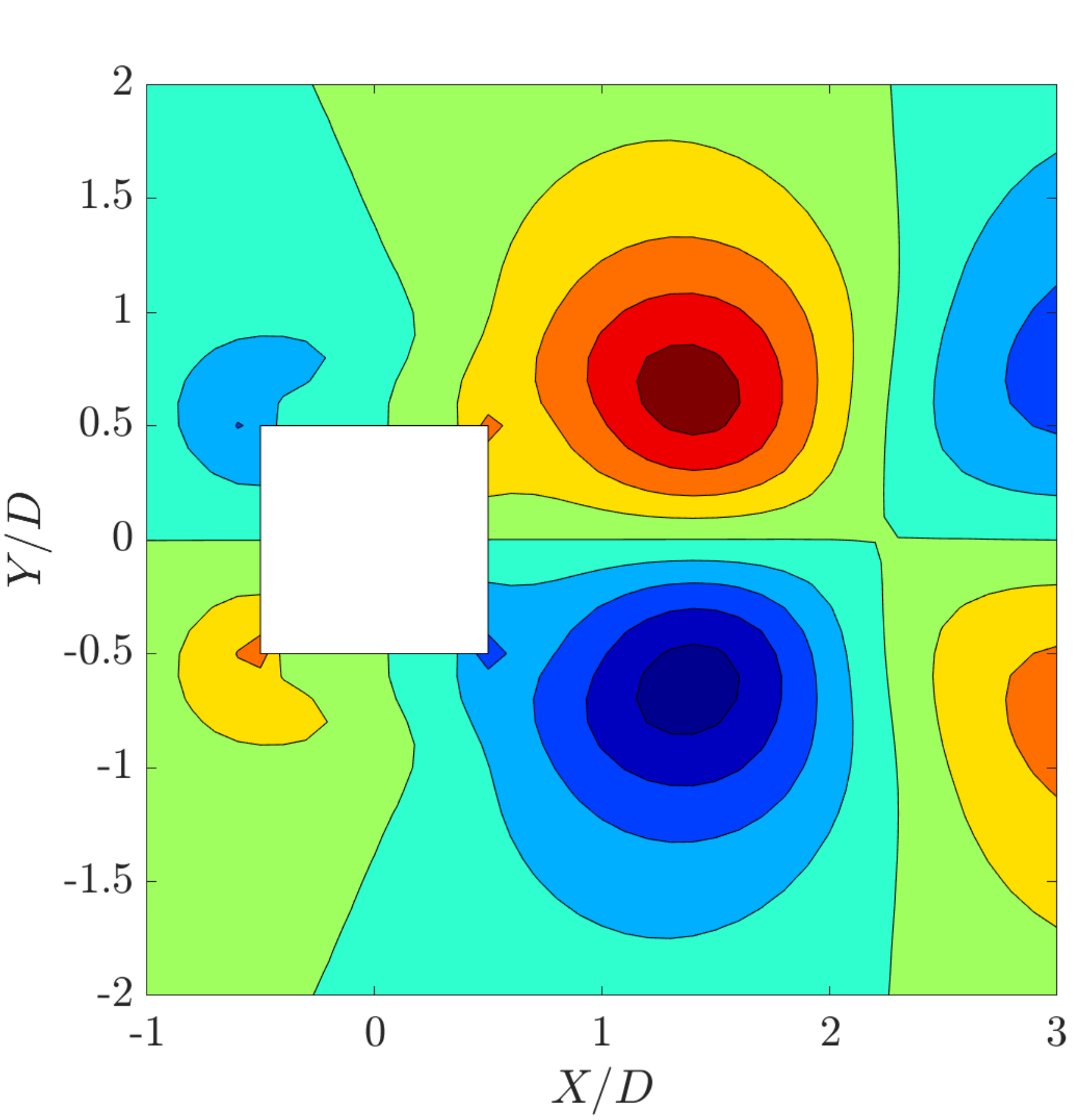}
\caption{Point C: zero lift (decreasing)}
\end{subfigure}~
\begin{subfigure}[]{0.5\textwidth}
\centering
\includegraphics[trim={7cm 0 9.75cm 0},clip,scale=0.28]{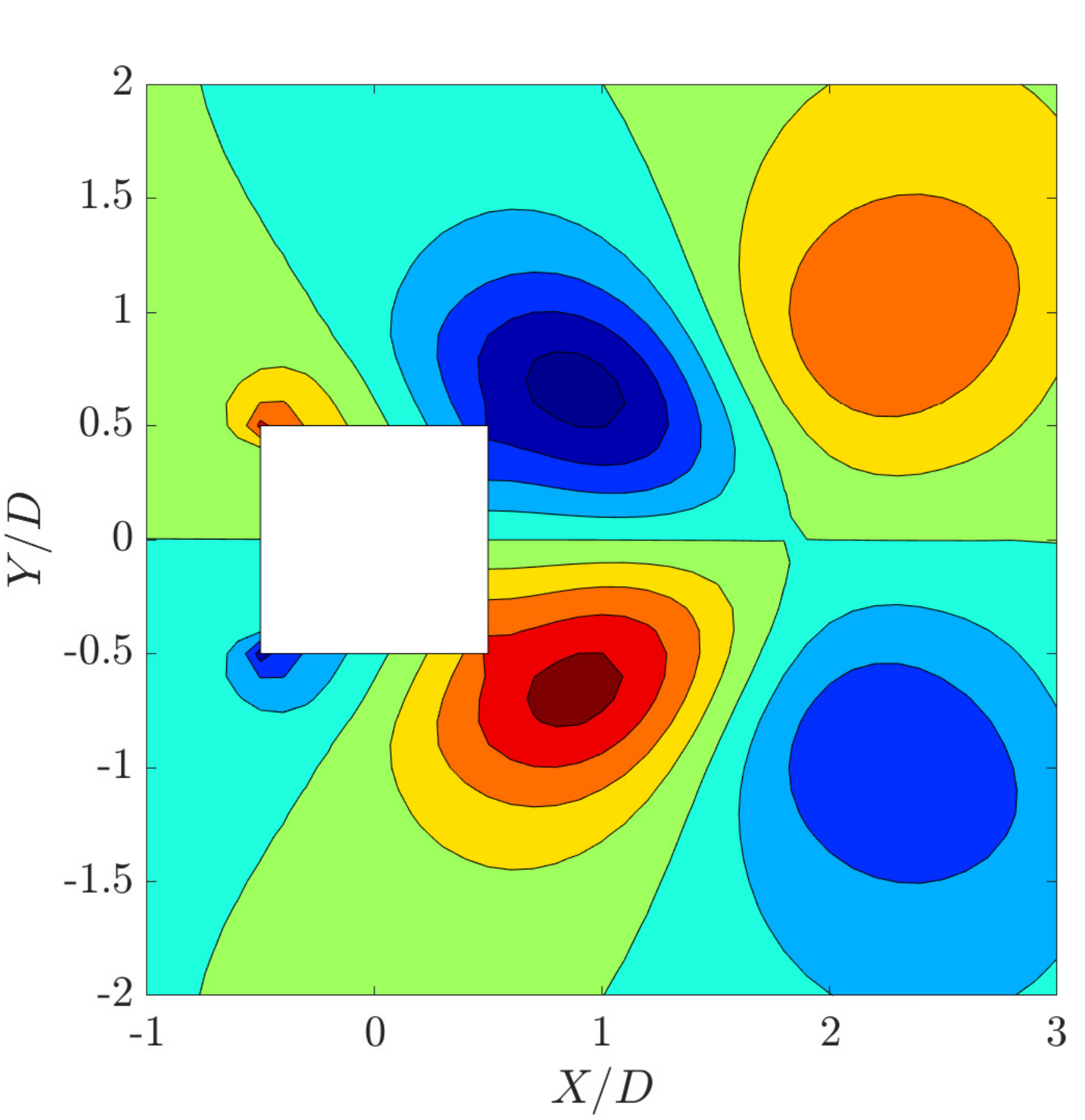}
\caption{Point D: minimum lift}
\end{subfigure}
\caption{Reconstruction of the lift modes in the near cylinder region for $U_r = 6.0$: (a) lift variation, (b) zero lift (increasing) at $tU_{\infty}/D=203.75$, (c) maximum lift at $tU_{\infty}/D=205.25$, (d) zero lift (decreasing) at $tU_{\infty}/D=206.85$, and (e) minimum lift at $tU_{\infty}/D=208.50$. The pressure values are normalized by $1/2\rho^{\mathrm{f}}U_{\infty}^2$. The contours levels are from  $-0.4$ to $0.4$ in increments of $0.1$. The flow is from left to right.}
\label{LiftModes}
\end{figure}

In this section, we analyze the behavior of different modes in the near cylinder region and explain the exclusive nature of their contributions to the drag and lift forces exerting on the oscillating cylinder in a uniform flow. Herein, the vortex shedding modes are referred to as the lift modes, while the shear layer and the near wake represent the drag modes. Figure \ref{LiftModes} displays the combined variation of the lift modes during a single cycle of lift. Note that the motion of the cylinder is not shown for this reconstruction as the POD modes are time invariant. The lift modes vary in an alternating manner in four quadrant. The variation is anti-symmetric about the streamwise centerline. In the maximum lift case (Point B in Figure \ref{LiftModes}a), the positive pressure force difference (i.e. $+Y$ direction) in the downstream quadrants dominates the small negative difference in the upstream quadrants and vice versa for the minimum lift (Point D). In the zero lift cases (Point A and C), the upstream and downstream pressure force differences tend to become equally strong and they cancel each other. 
The lift modes vary in such a way that the force on the top 2 quadrants is equal in magnitude and opposite in direction to the force on the bottom 2 quadrants. Due to this force cancellation, the vortex shedding (lift) modes have no contribution to the drag force. Hence, the FFT of the drag force does not contain the corresponding harmonic of the natural frequency ($f_n$).

\begin{figure}
\centering
\begin{subfigure}[]{\textwidth}
\centering
\includegraphics[trim={0 0 0 1cm},clip,scale=0.3]{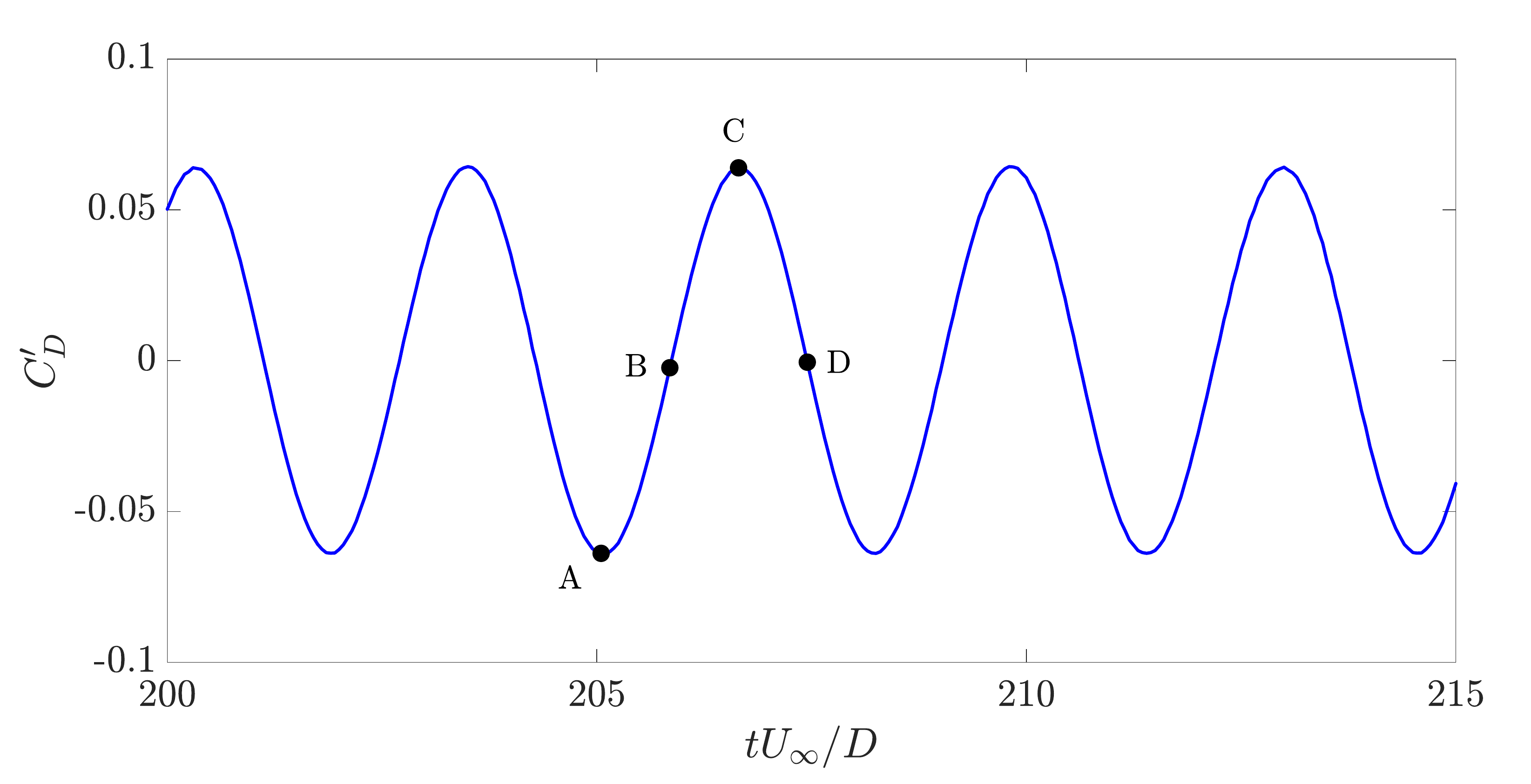}
\caption{}
\end{subfigure}
\begin{subfigure}[]{0.5\textwidth}
\centering
\includegraphics[trim={3cm 0 6.95cm 0},clip,scale=0.275]{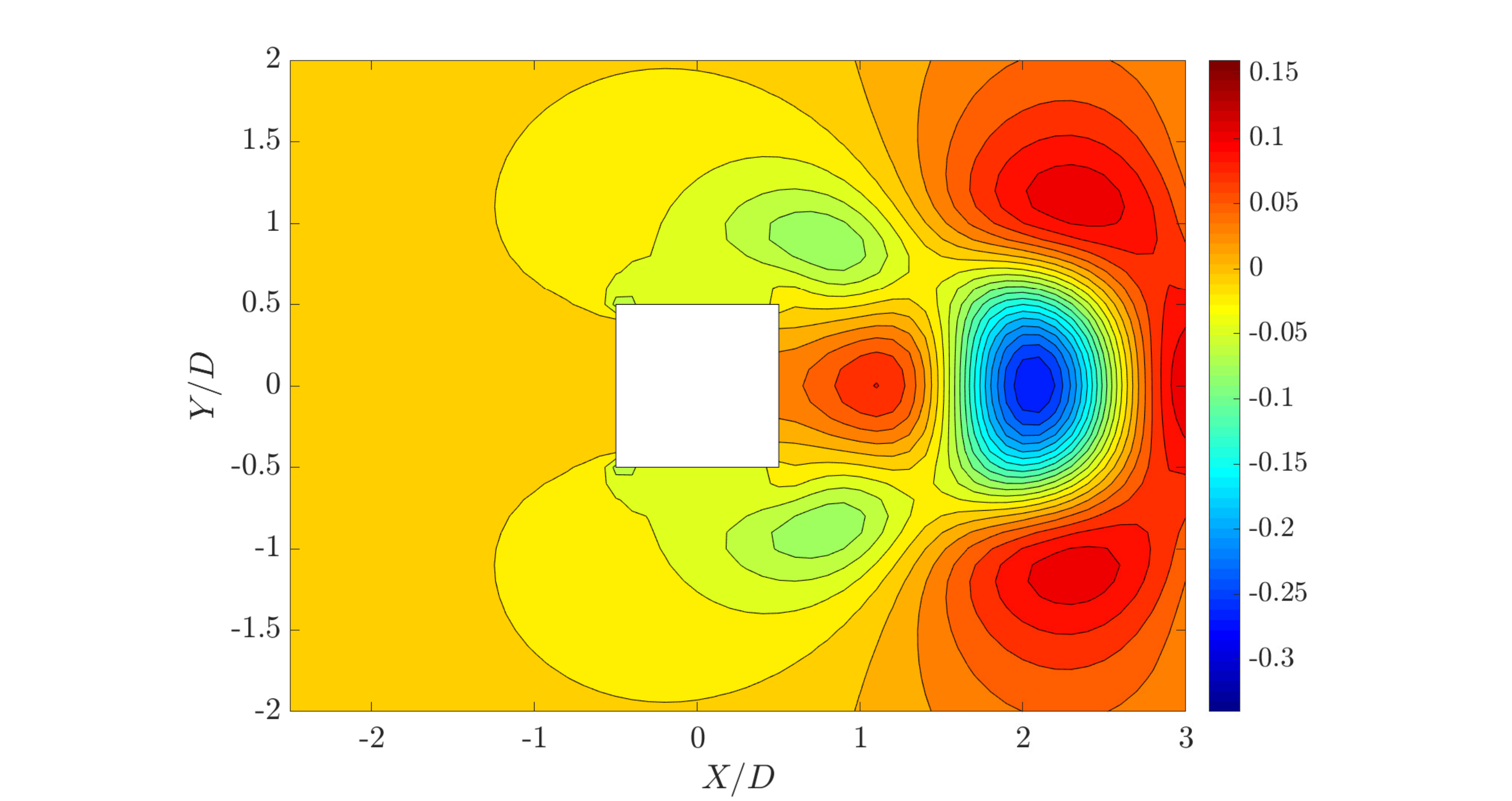}
\caption{Point A: minimum drag fluctuation}
\end{subfigure}~
\begin{subfigure}[]{0.5\textwidth}
\centering
\includegraphics[trim={3cm 0 6.95cm 0},clip,scale=0.275]{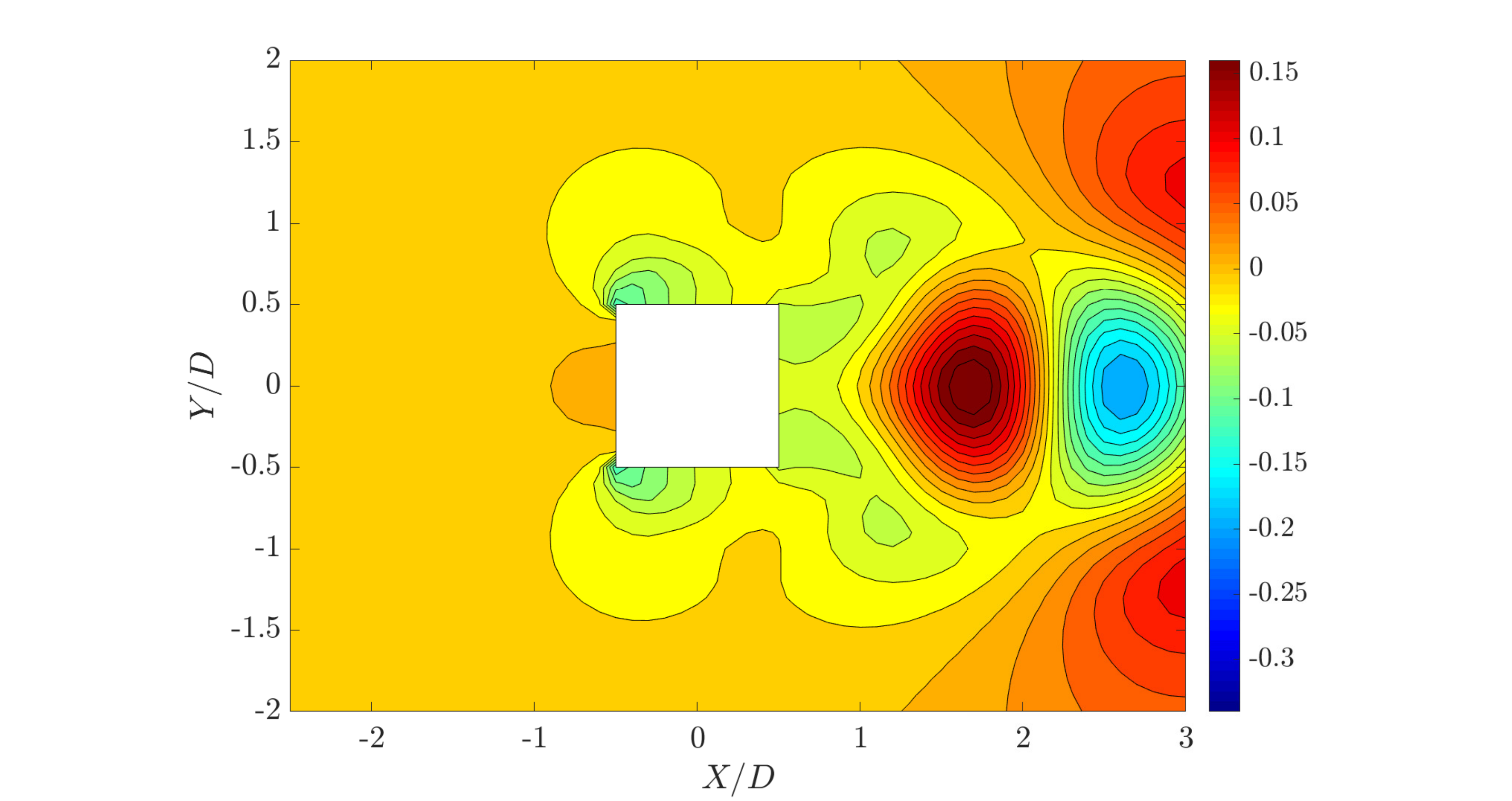}
\caption{Point B: zero drag fluctuation (increasing)}
\end{subfigure}
\begin{subfigure}[]{0.5\textwidth}
\centering
\includegraphics[trim={3cm 0 6.95cm 0},clip,scale=0.275]{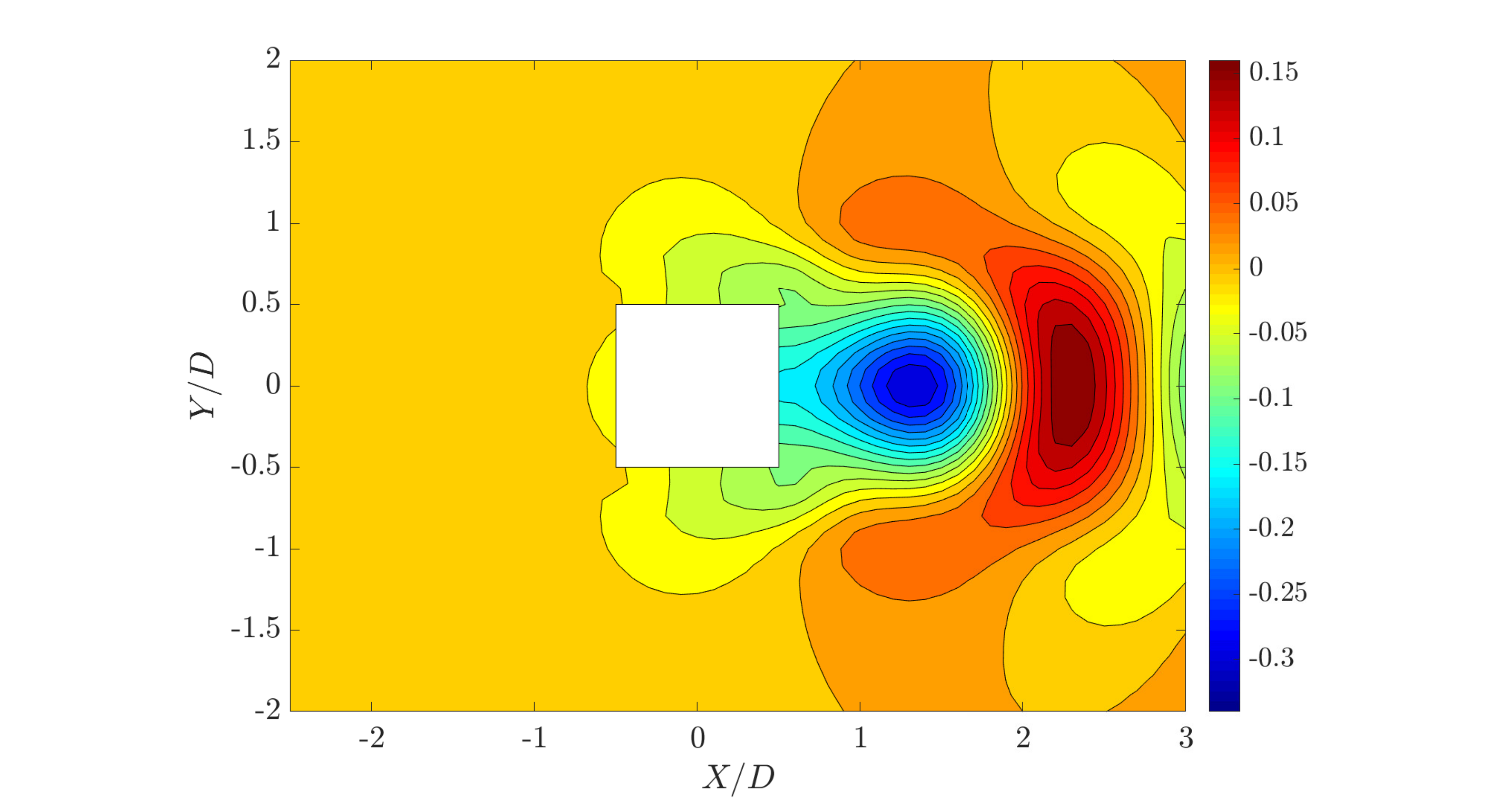}
\caption{Point C: maximum drag fluctuation}
\end{subfigure}~
\begin{subfigure}[]{0.5\textwidth}
\centering
\includegraphics[trim={3cm 0 6.95cm 0},clip,scale=0.275]{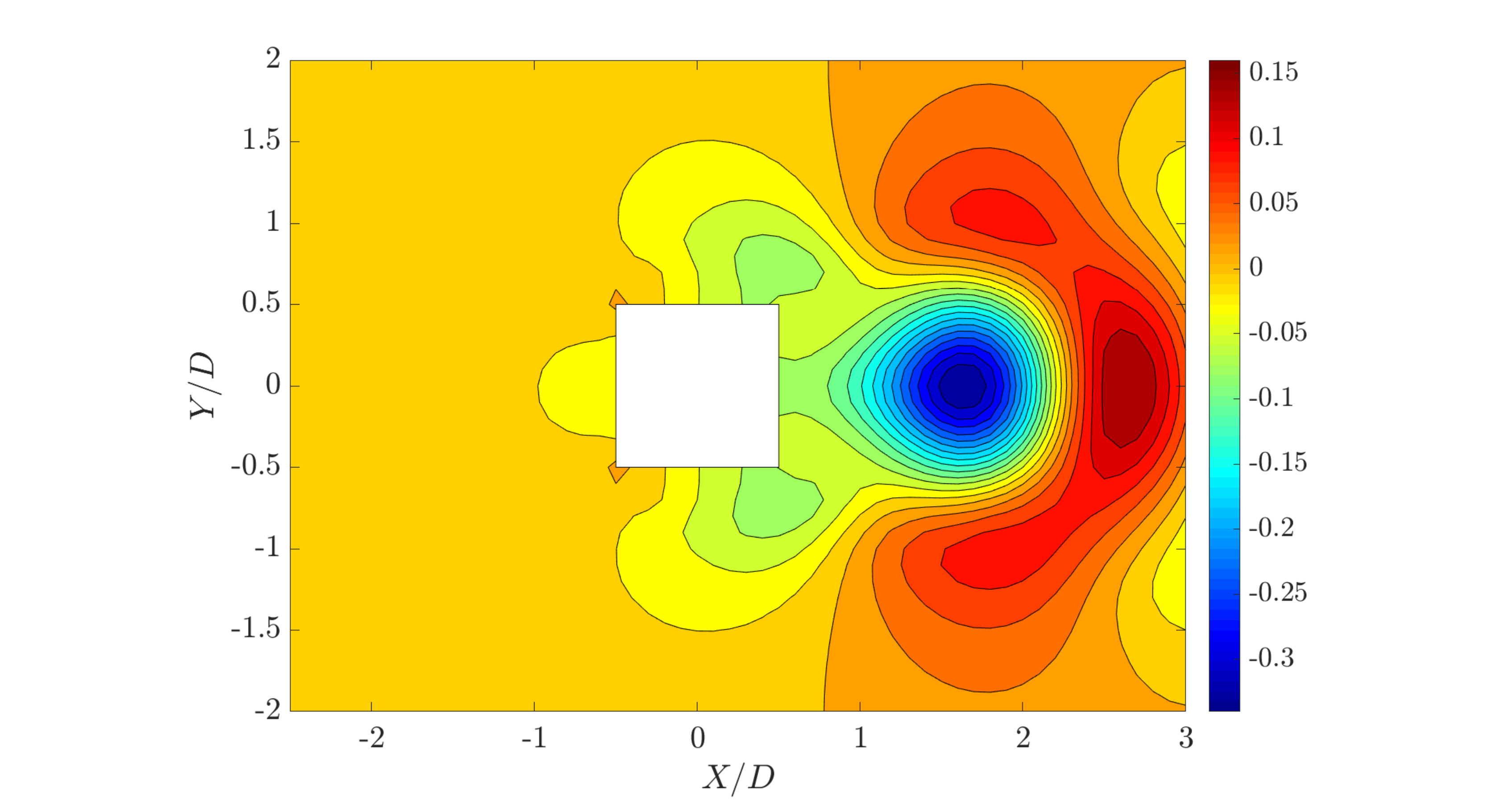}
\caption{Point D: zero drag fluctuation (decreasing)}
\end{subfigure}
\caption{Reconstruction of the drag modes in the near cylinder region for $U_r = 6.0$: (a) fluctuation of drag, (b) minimum drag fluctuation at $tU_{\infty}/D=205.05$, (c) zero drag fluctuation (increasing) at $tU_{\infty}/D=205.85$, (d) maximum drag fluctuation at $tU_{\infty}/D=206.70$, (e) zero drag fluctuation (decreasing) at $tU_{\infty}/D=207.40$. The pressure values are normalized by $1/2\rho^{\mathrm{f}}U_{\infty}^2$. The contours levels are from  $-0.34$ to $0.16$ in increments of $0.025$. The flow is from left to right.}
\label{Fig:DragModes}
\end{figure}
Figure \ref{Fig:DragModes} describes the variation of drag modes with the fluctuation of the drag force. The drag fluctuation is defined as $C_D' = C_D(t) - \overline{C_D}$. The drag modes vary symmetrically around the wake centerline hence offer no contribution to the lift. Similar to the lift modes, this explains the absence of a $2f_n$ harmonic in the lift. The maximum drag fluctuation (Point C) is higher than the minimum drag fluctuation (Point A). Further, at the zero drag fluctuation points (B and D), the magnitude of the drag fluctuation remains a positive value. This further confirms that the drag modes exert a non-zero mean drag on the bluff body apart from the drag force due to the base flow. 

With the aforementioned observations, the decomposition of fluid force on a moving bluff body based on the contributions from different POD modes can be expressed as
\begin{equation}
F_j (t) = {F_j^0} + \sum_{i=1}^{n_{rj}} b_j^i (t) F_j^i,
\end{equation}
where $F_j$ is the force in a particular direction ($j=x$ for in-line and $j=y$ for transverse). $F_j^0$ is the time independent contribution from the mean field and $F_j^i$ is the time independent pressure fluctuation contribution calculated for the $i^{th}$ mode. While $b_j^i (t)$ is the time dependent coefficient of the $i^{th}$ mode for the force in direction $j$, $n_{rj}$ is the number of POD modes with a significant contribution to the particular force. Using the snapshot data, we can determine $b_j^i(t)$ for the streamwise and transverse forces as
\begin{equation}
 b_j^i (t) = \frac{F_j (t) - F_j^0}{F_j^i n_{rj}}.
\end{equation}
The complete description of this force decomposition and its usage is provided in Appendix A. 

\begin{table}
\centering
\caption{Force contributions from the mean field and POD modes. $\overline{b_j^i}$ denotes the time averaged force coefficient.}
\label{PODForce}
\begin{tabular}{m{0.65cm}|m{4.7cm}|m{4.7cm}|>{\centering\arraybackslash}m{1.4cm}}
  & \centering{Near cylinder field}& \centering{Schematic pressure distribution}& \centering{$F_j^i$, $\overline{b_j^i}$} 
\tabularnewline
\hline
\rotatebox{90}{Mean field} & \centering \includegraphics[trim={4.5cm 6.25cm 23.8cm 4.5cm},clip,scale=0.7]{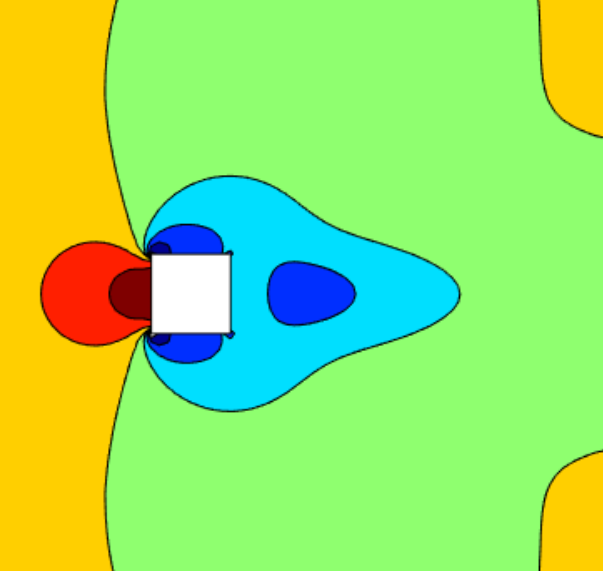}
& \centering \includegraphics[trim={0 0 0 0},clip,scale=0.15]{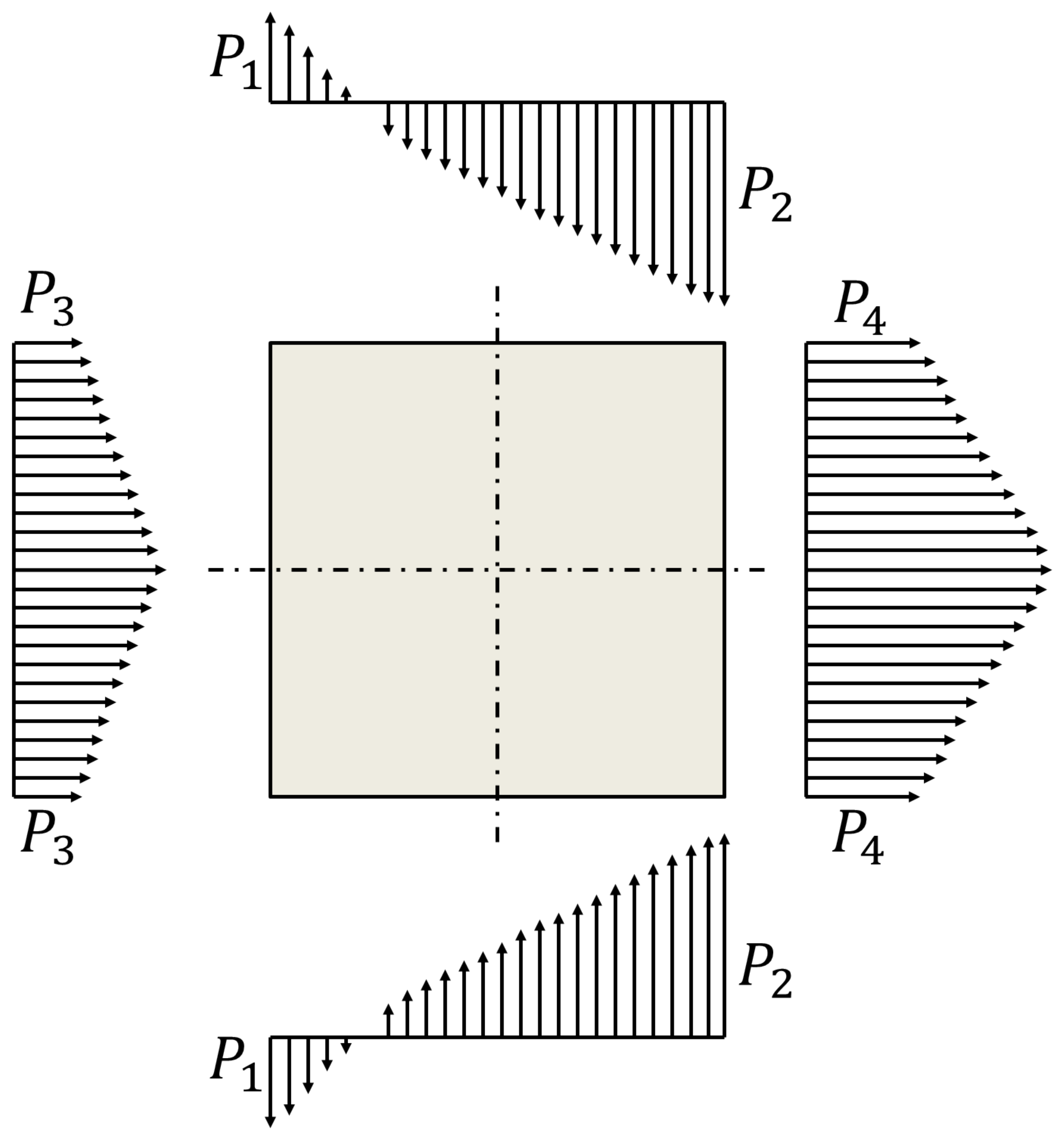} & ${F_x^0} \neq 0$, ${F_y^0} \approx 0 $\\
\hline
\rotatebox{90}{Vortex shedding modes}  &  \centering \includegraphics[trim={7cm 6.75cm 21.5cm 6.15cm},clip,scale=0.75]{Pmode1.pdf} & 
\centering \includegraphics[trim={0.5cm 0 0 0},clip,scale=0.15]{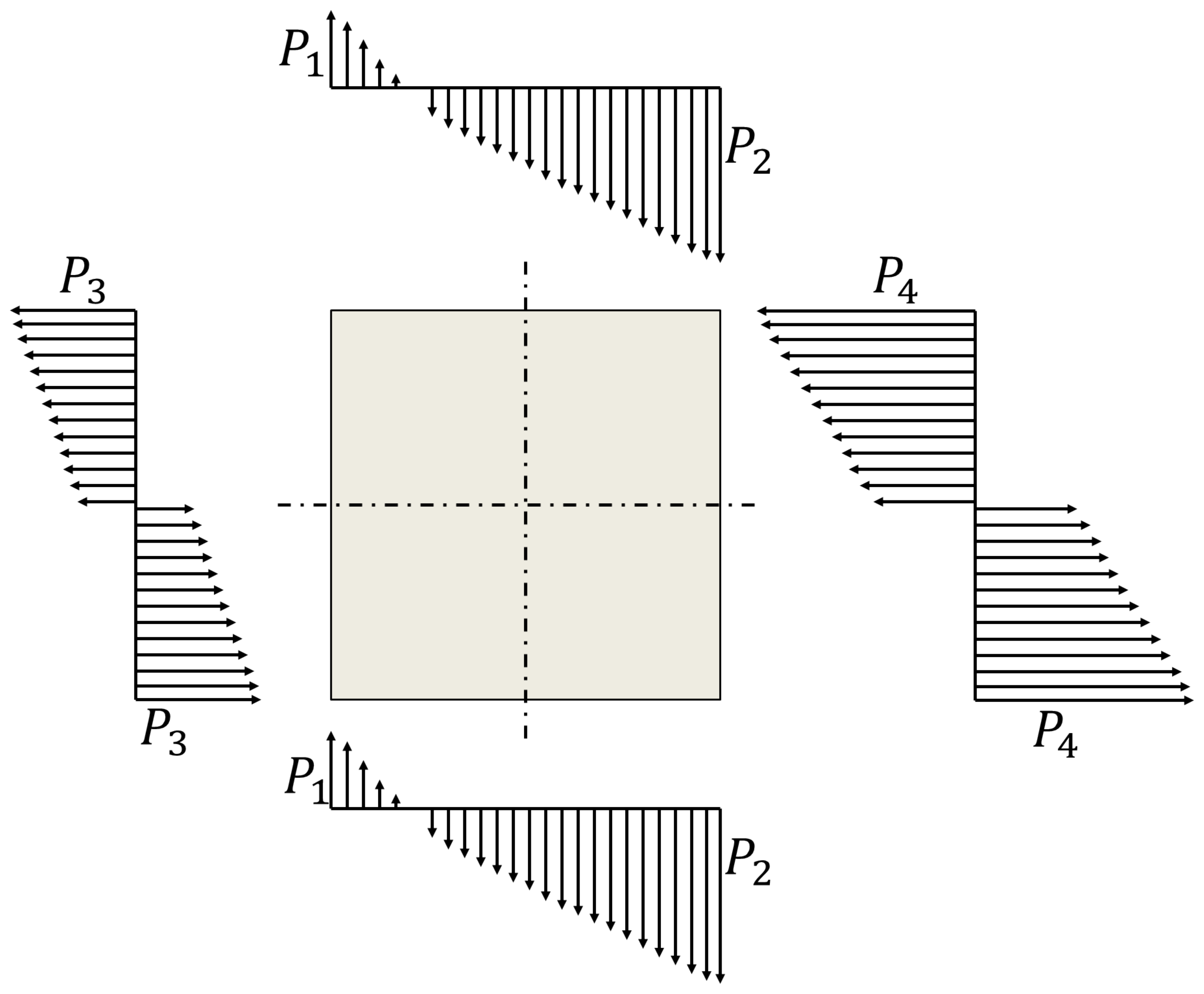} & $F_x^i \approx 0 $, $F_y^i \neq 0$, $\overline{b_y^i} \approx 0$\\
\hline
\rotatebox[origin=c]{90}{\parbox[c]{3cm}{\centering Shear layer and near-wake modes}}
& \centering \includegraphics[trim={5.75cm 7cm 17.75cm 6.4cm},clip,scale=0.4]{Pmode2.pdf} 
\centering \includegraphics[trim={5.75cm 7cm 17.75cm 6.4cm},clip,scale=0.4]{Pmode5.pdf}
& \centering \includegraphics[trim={0 0 0 0},clip,scale=0.15]{DragModes.pdf} & $F_x^i \neq 0$, $\overline{b_x^i} \neq 0$, $F_y^i \approx 0 $ \\
\end{tabular}
\end{table}
Table (\ref{PODForce}) summarizes the qualitative analysis of the contributions from the mean field and the modes to the drag and lift forces. The mean-field has a symmetric pressure distribution about the wake centerline, hence contributes solely to the time-independent component of the drag force. The vortex shedding modes have an anti-symmetric pressure distribution throughout the time history, hence they have no drag force contribution. We observe that these lift force contributions have a near zero mean (similar to the lift variation) as well. The shear layer and near-wake modes have the same qualitative properties of the mean field, however their contributions to the drag force are time-dependent.
The POD modes provide a deep insight into important flow features and their contribution to the wake dynamics. It is important to investigate their variation with different flow conditions, i.e. the parameters mentioned in Eq. (\ref{eq:DampingRatio}). The variation of POD modes and their contribution to wake dynamics with the reduced velocity is examined in the next section, with the goal to explain the role of the wake features in sustaining the synchronized wake-body motion. 

\section{Wake feature interaction and sustenance of VIV lock-in}
\begin{figure}
\centering
\pgfplotsset{every tick label/.append style={font=\Large}}
\pgfplotsset{every tick label/.append style={scale=1}}
\centering
\begin{subfigure}{0.48\textwidth}
\centering
\begin{tikzpicture}[trim axis left, 
trim axis right, 
scale=0.55, 
baseline]
\begin{axis}[
    xlabel={\Large $U_r$},
    ylabel={\Large $A_y^{rms}/D$},
    xmin=3.7, xmax=12.5,
    ymin=0, ymax=0.21,
    xtick={2,4,6,8,10,12},
    ytick={0,0.05,0.1,0.15,0.2},
    yticklabel style={/pgf/number format/fixed, /pgf/number format/precision=2},
    width = 10cm,
    height = 7cm,
]
 
\addplot[
    color=black,
    solid,
    mark=square,
    very thick,
    mark size = 4,
    ]
    coordinates {
(4,0.0256)(4.5,0.0441)(5,0.1947)(5.5,0.1480)(6,0.1121)(7,0.0675)(8,0.0501)(9,0.0422)(10,0.0378)(11,0.0350)(12,0.0332)
    };
\end{axis}
\end{tikzpicture}
\caption{}
\label{fig:AyvsUrm}
\end {subfigure}
~
\begin{subfigure}[h]{0.48\textwidth}
\centering
\begin{tikzpicture}[trim axis left, trim axis right, scale=0.55]
\begin{axis}[
    ylabel={\Large $C_L^{rms}$},
    xmin=3.7, xmax=12.5,
    ymin=-0.02, ymax=0.55,
    xtick={2,4,6,8,10,12},
    xticklabels={,,}
    legend pos=north east,
    legend style={draw=none},
    ylabel near ticks, 
    yticklabel pos=right,
    ytick pos=right,
    width = 10cm,
    height = 7cm,
]
 
\addplot[
    color=red,
    dashed,
    mark=o,
     very thick,
     mark size = 4,
     mark options=solid,
    ]
    coordinates {
    (4,0.2578)(4.5,0.3044)(5,0.5259)(5.5,0.1878)(6,0.0738)(7,0.0303)(8,0.0700)(9,0.0887)(10,0.0991)(11,0.1055)(12,0.1099)
    };
    \addlegendentry{\large $C_L^{rms}$}; \label{LiftF}
    \end{axis}
\begin{axis}[
    xlabel={\Large $U_r$},
    ylabel={\Large $C_D^{rms}$},
    xmin=3.7, xmax=12.5,
    ymin=-0.02, ymax=0.18,
    xtick={2,4,6,8,10,12},
    ytick={0,0.05,0.10,0.15},
    yticklabel style={/pgf/number format/fixed, /pgf/number format/precision=2},
    legend pos=north east,
    legend style={draw=none},
    ytick pos=left,
       width = 10cm,
    height = 7cm,
]
 
\addplot[
    color=blue,
    solid,
    mark=square,
     very thick,
    mark size = 4,
    ]
    coordinates {
    (4,0.0025)(4.5,0.0067)(5,0.1610)(5.5,0.0890)(6,0.0453)(7,0.0160)(8,0.0103)(9,0.0085)(10,0.0077)(11,0.0072)(12,0.0070)
    };
    \addlegendentry{\Large $C_D^{rms}$}
    \addlegendimage{/pgfplots/refstyle=LiftF}\addlegendentry{\Large $C_L^{rms}$}
\end{axis}
\end{tikzpicture}
\caption{}
\label{fig:CDm}
\end{subfigure}
 
\begin{subfigure}[h]{0.50\textwidth}
\centering
\includegraphics[trim={0.7cm 0 0cm 0},clip,scale=0.365]{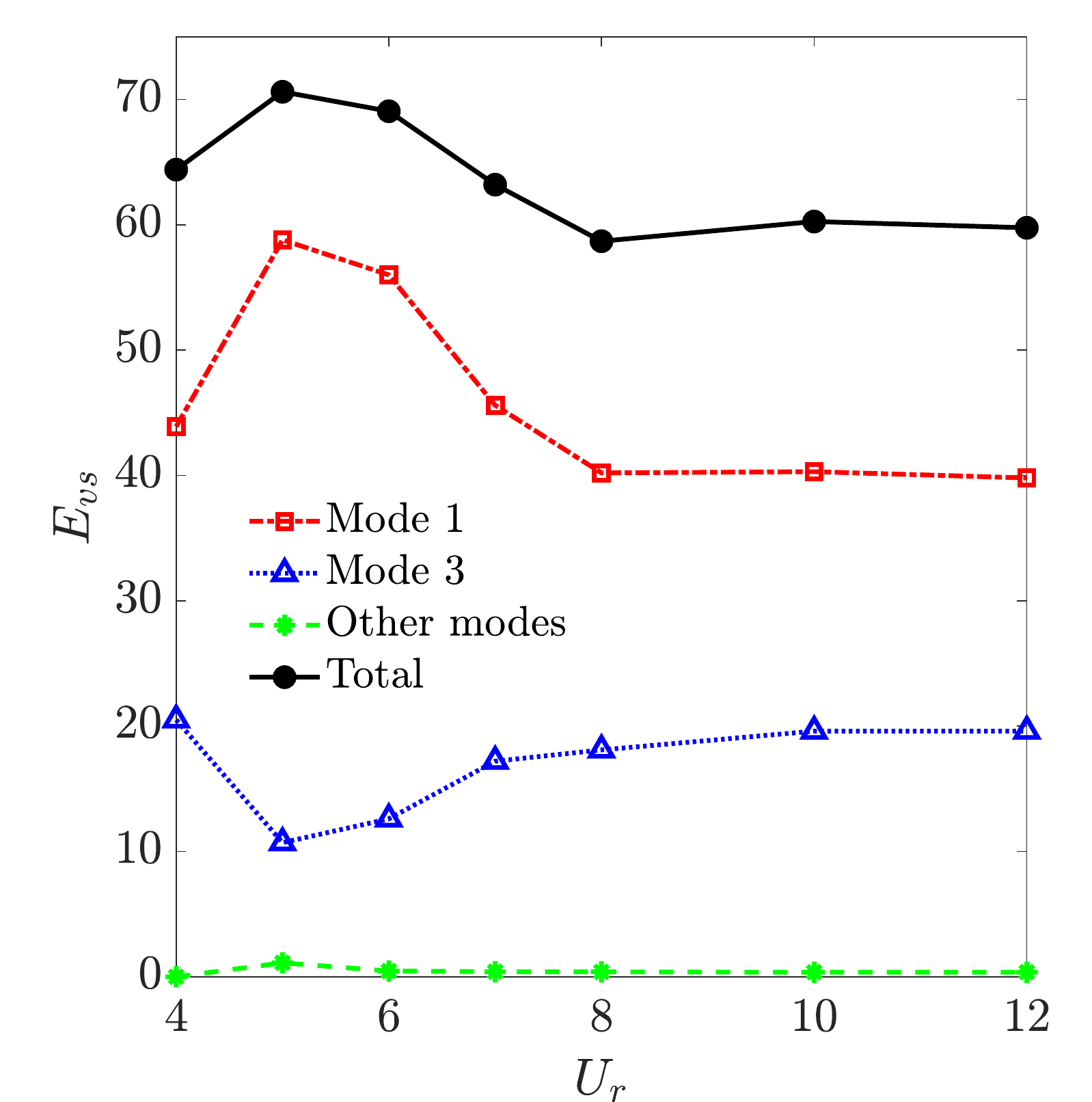}
\caption{}
\end{subfigure}~
\begin{subfigure}[h]{0.50\textwidth}
\centering
\includegraphics[trim={0.7cm 0 0cm 0},clip,scale=0.365]{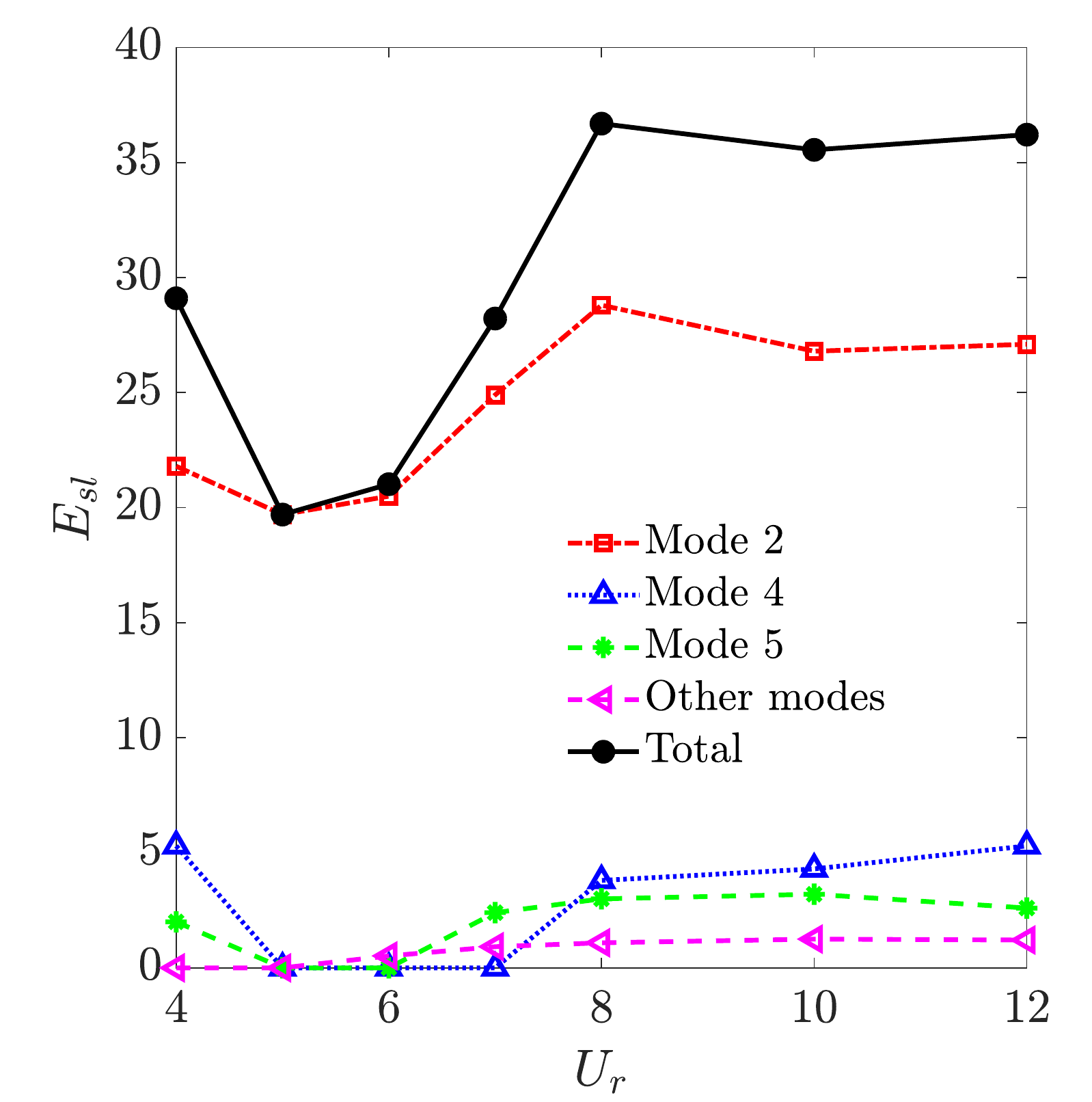}
\caption{}
\end{subfigure}
\begin{subfigure}[h]{0.50\textwidth}
\centering
\includegraphics[trim={0.7cm 0 0cm 0},clip,scale=0.365]{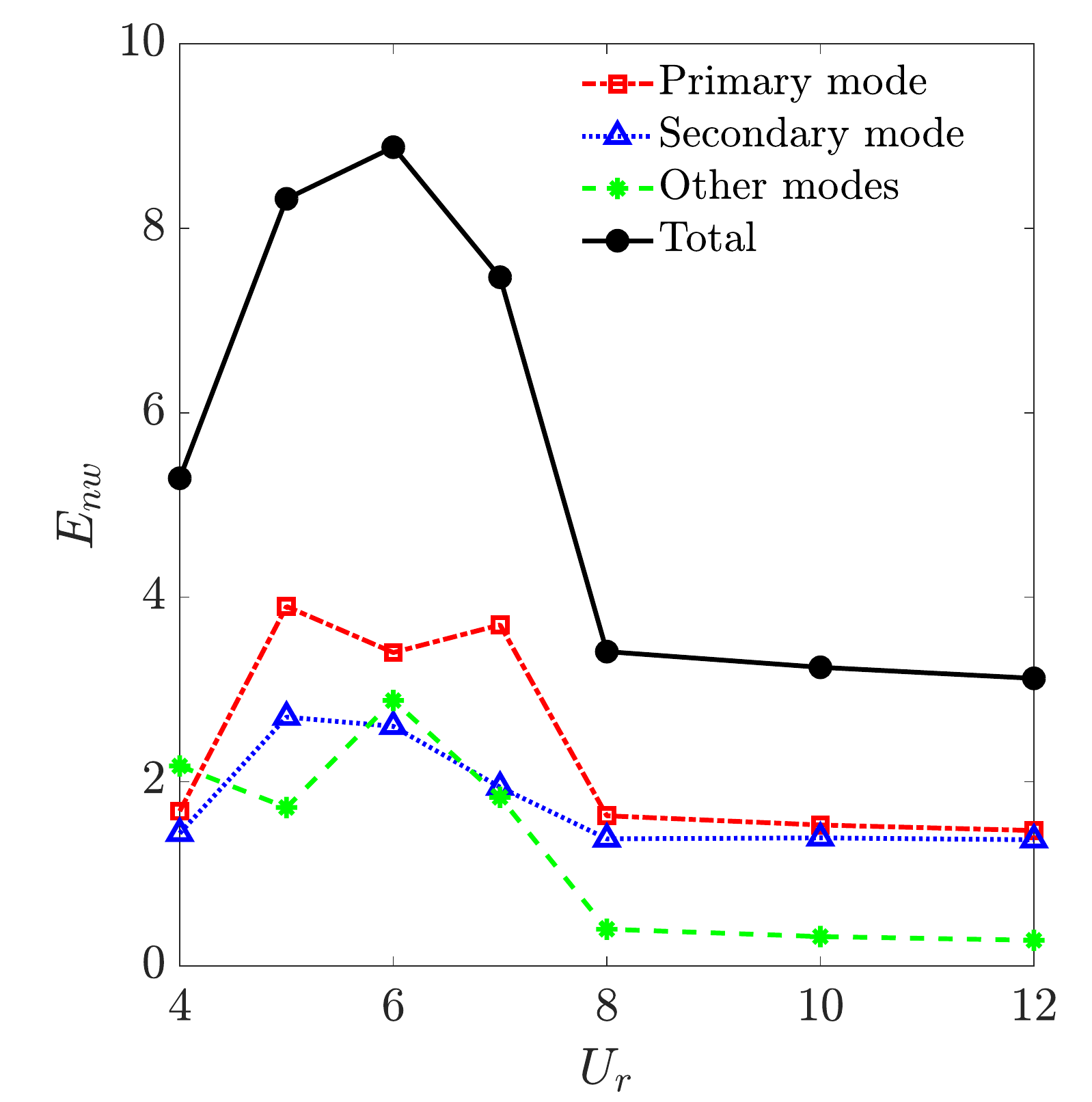}
\caption{}
\end{subfigure}~
\begin{subfigure}[h]{0.50\textwidth}
\centering
\includegraphics[trim={0.4cm 0 0cm 0},clip,scale=0.365]{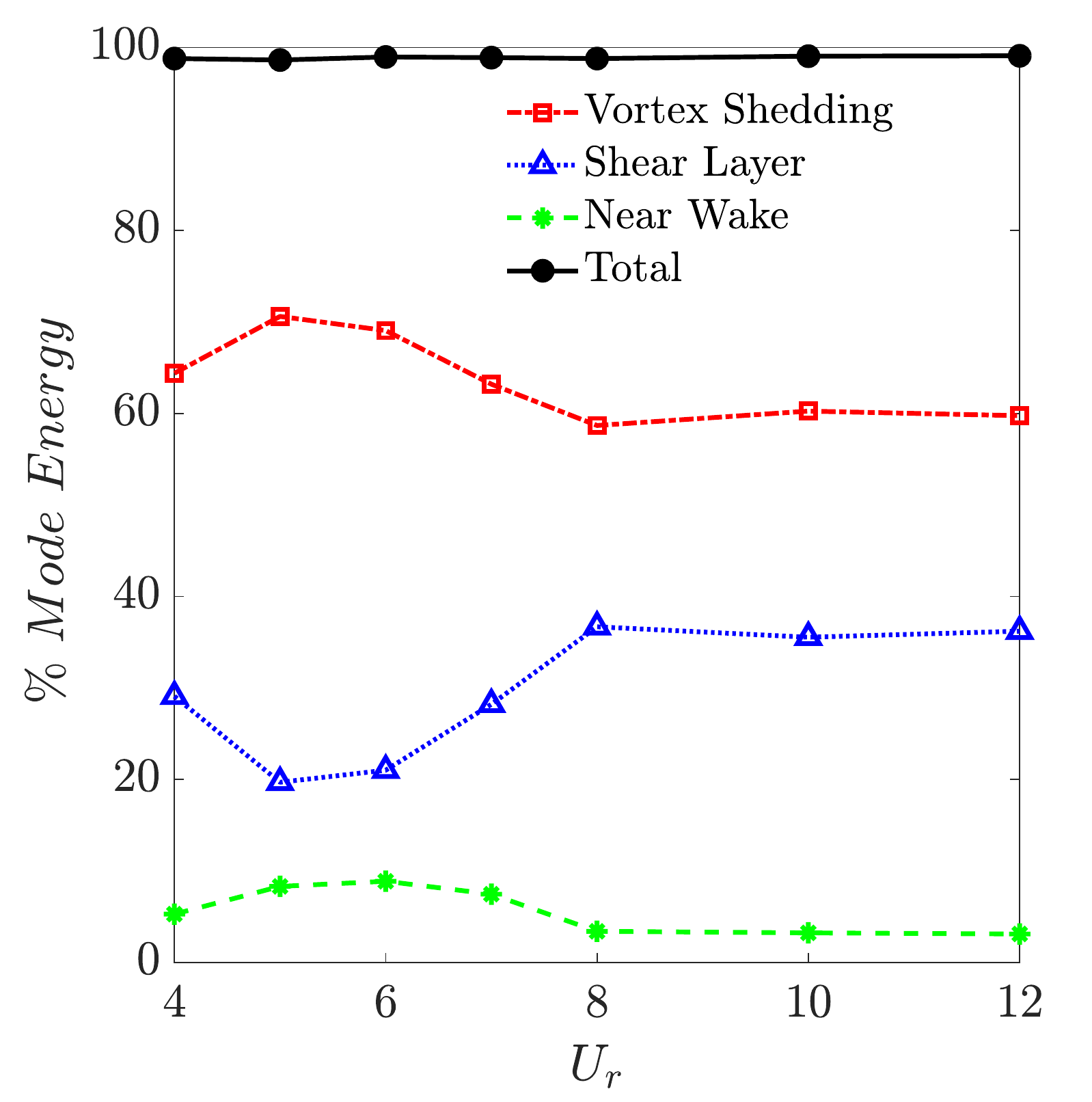}
\caption{}
\end{subfigure}
\caption{Response characteristics and decomposition of wake dynamics for a freely vibrating square cylinder at $m^*=3.0$, $Re=100$ and $\zeta=0$: (a) transverse displacement,   (b) drag and lift force variations,  mode energy contributions from different flow features (c) vortex shedding, (d) shear layer, (e) near-wake, and (f) total mode energy contributions. $E_{vs}, E_{sl}, E_{nw}$ denote the relative mode energy of the wake features as a percentage of the total mode energy. The first 9 modes which contain $99\%$ of the total mode energy are considered.}
\label{PODEVIV}
\end{figure}
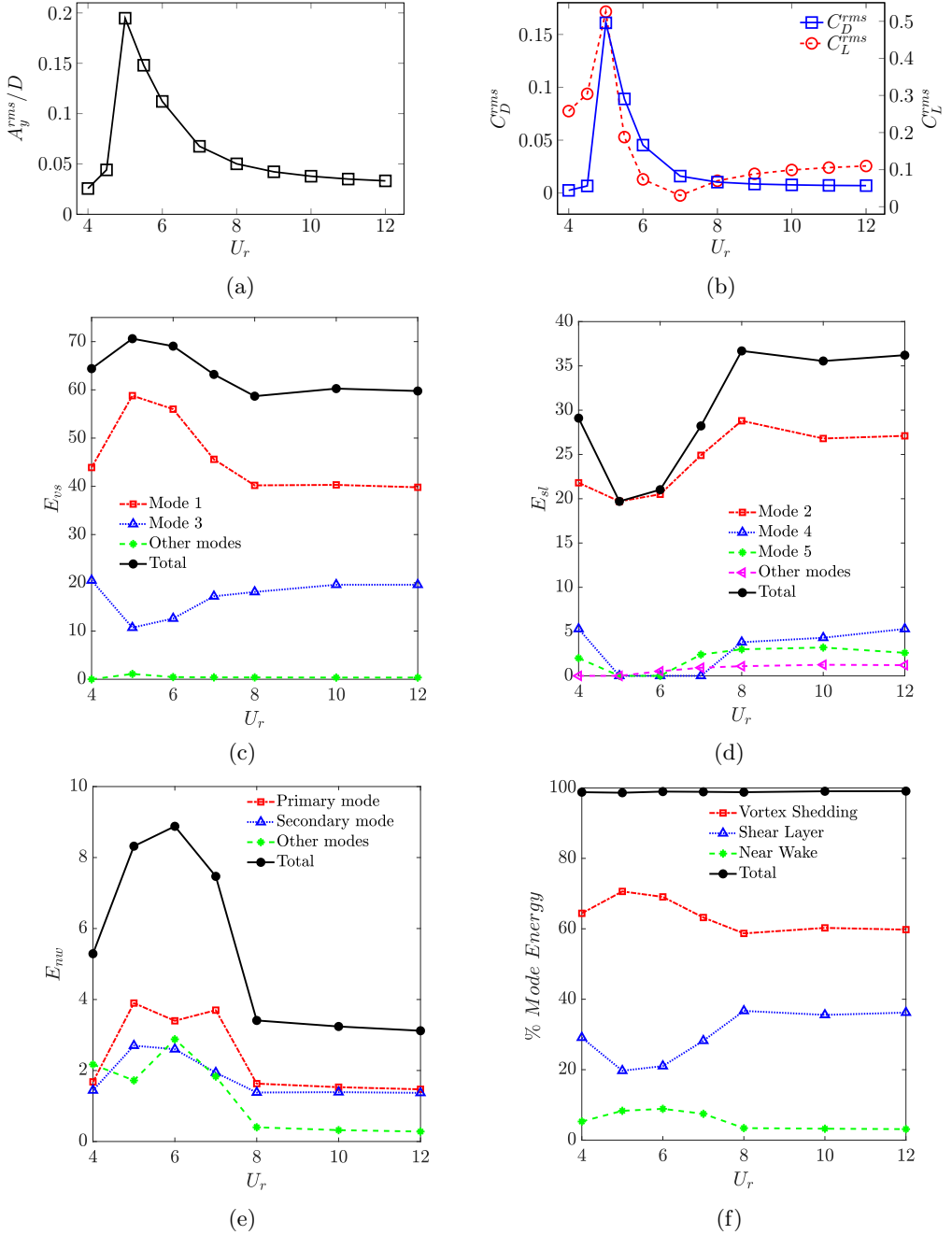

In this section, we investigate the relative contributions from different features to the pressure fluctuations and eventually the fluid forces on the freely oscillating bluff body. When a bluff body is free to oscillate in a current flow it undergoes the lock-in phenomenon: the oscillation amplitude significantly increase when the natural frequency of the bluff body approaches the vortex shedding frequency. In  \cite{miyanawala2016flow}, the lock-in phenomenon for a square cylinder immersed in a laminar flow at $Re=100$ is systematically studied. 

Figures \ref{PODEVIV}a and \ref{PODEVIV}b summarize the bluff body dynamics of a freely vibrating square cylinder. The cylinder undergoes wake-body synchronized lock-in in the range $U_r\in[4.5,7]$ and the peak oscillation occurs at $U_r=5.0$.
Figures \ref{PODEVIV}c-f elucidate the variation of relative contributions from different modes as a function of the reduced velocity ($U_r$). It is interesting to note that the three most energetic modes correspond to the same flow features throughout the $U_r$ range namely the first and third modes (vortex shedding) and the second mode (shear layer). However, the modes 4-10 vary in this regard, where most of these modes correspond to the near-wake phenomena. We quantify the relative energy contribution from each wake feature by summing the mode energy of the corresponding modes, i.e.
\begin{equation}
E_j = \frac{\sum_{i=1}^{n_j}\sigma_i^2}{\sum_{i=1}^k\sigma_i^2}, 
\end{equation}
where $E_j$ is the relative energy contribution from the wake feature ($j=vs$ for the vortex shedding, $j=sl$ for the shear layer and $j=nw$ for the near-wake), $n_j$ is the number of significant modes corresponding to a particular flow feature and $k$ is the total number of modes.
As shown in figure \ref{PODEVIV}c, the total contribution from the vortex shedding increases in the lock-in region. However, the first mode becomes more energetic while the third mode is relatively less energetic in this region. Figure \ref{PODEVIV}d exhibits that the contribution from the shear layer modes reduces significantly in the lock-in region. All the individual shear layer modes also follow a similar trend. The near-wake modes depicted in figure \ref{PODEVIV}e become more energetic in the lock-in region relative to the pre- and post-lock-in regions. Unlike the vortex shedding and shear layer, the primary and secondary near-wake modes have remarkably similar contributions. A summary of the contributions from the 10 most energetic POD modes corresponding to different physical phenomena is illustrated in figure \ref{PODEVIV}f. Note that these 10 modes capture $\approx 99\%$ of the total mode energy. It is clear that the shear layer contributions decrease while the vortex shedding and the near-wake contributions increase in the lock-in region. In the post-lock-in region, the relative contributions from the flow features remain almost constant. However, our intuition poses a question: it is evident that the extreme transverse motion in the lock-in is sustained by the resonance of the bluff body spring-mass system and the vortex shedding since the transverse motion frequency matches the vortex shedding frequency. Why does not the energy transfer from the fluid to the bluff body make the vortex shedding modes less energetic instead of what we observe here? In that relation, we propose a cycle to explain this counter-intuitive behavior of the decomposed wake features.

\begin{figure}
 	\centering
    \begin{subfigure}{\textwidth}
    \centering
   	\begin{tikzpicture}[decoration={markings,mark=at position 0.5 with {\arrow[scale=2]{>}}},scale=1.1]
    \tikzstyle{every node}=[thick, ellipse, minimum width=60pt, minimum height=10pt, align=center]
   	\node[draw, text width = 3.25cm, align = center, text height=2.5ex, minimum height=2.5em] (VS) at (8,-1.25) {Strengthening of near wake and vortices};
   	\node[draw, text width = 2.5cm, align = center, text height=2.5ex, minimum height=2.5em] (BBV) at (0,-1.25) {Lock-in phenomenon};
   	\node[draw, text width = 3.5cm, align = center, text height=2.5ex, minimum height=2.5em] (SLW) at (4,0) {Wake and shear layer widening};
   	\draw [>=stealth,-, postaction={decorate}, thick, midway, above left] (VS.south) 
    node[xshift = -2cm, yshift = -1.1cm] {Synchronized vortex shedding and \\ bluff body motion, $f_n\approx f_{vs}$} 
    to [in=330,out=210]
    (BBV.south) ;
   	\draw [>=stealth,-, postaction={decorate}, align = center, text width=4.5cm, thick]  (SLW.east)
    node[xshift = 2.25cm, yshift = 0.35cm] {Vorticity transfer to near \\ wake bubble and vortices} to [in=145,out=350](VS.north);
   	\draw [>=stealth,-, postaction={decorate}, thick, midway, below left] (BBV.north) 
    node[xshift = 1.5cm, yshift = 1.3cm] {High amplitude \\ oscillation} 
    to [in=190,out=35]
    ([xshift=0em]SLW.west);
   	\end{tikzpicture}
     \caption{}
     \end{subfigure}
     \begin{subfigure}{0.49\textwidth}
     \centering
     \includegraphics[trim={3.25cm 0cm 2cm 0.3cm},clip,scale=0.2]{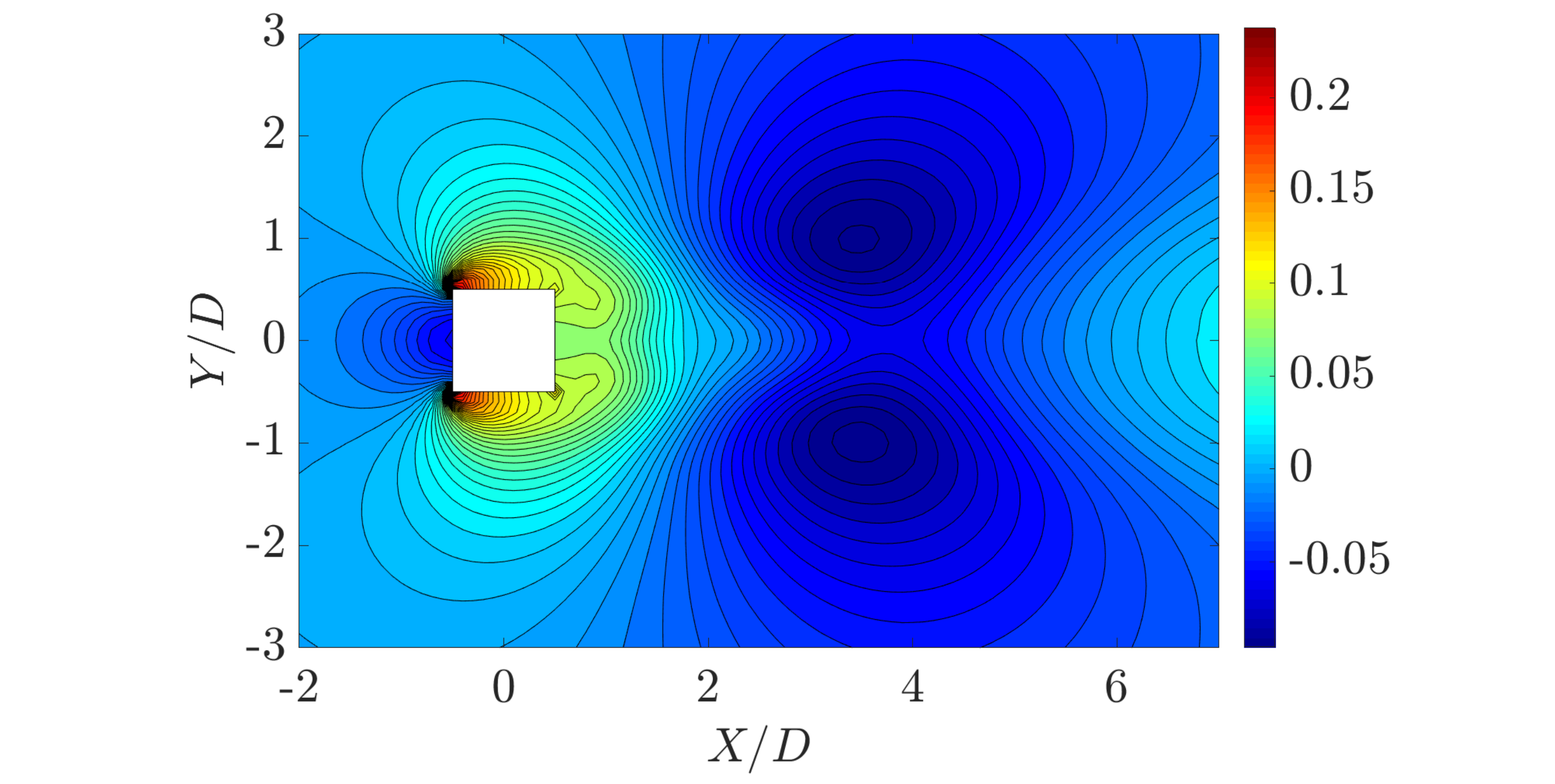}
	\caption{Pre lock-in, $U_r=4.0$}
     \end{subfigure}
     \begin{subfigure}{0.49\textwidth}
     \centering
     \includegraphics[trim={3.25cm 0cm 2cm 0.3cm},clip,scale=0.2]{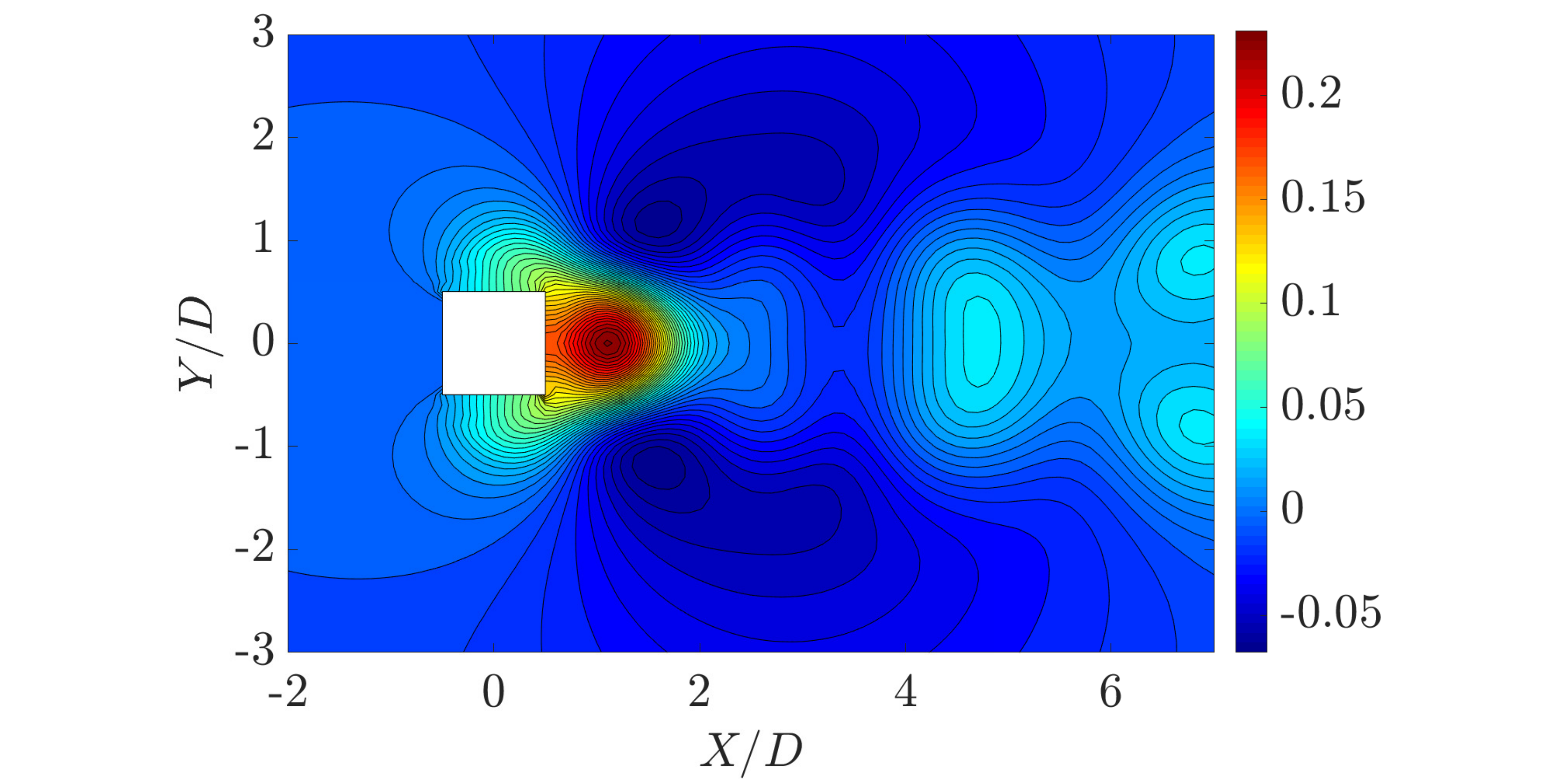}
	\caption{Lock-in, $U_r=5.0$}
     \end{subfigure}
     \begin{subfigure}{0.49\textwidth}
     \centering
     \includegraphics[trim={3.25cm 0cm 2cm 0.3cm},clip,scale=0.2]{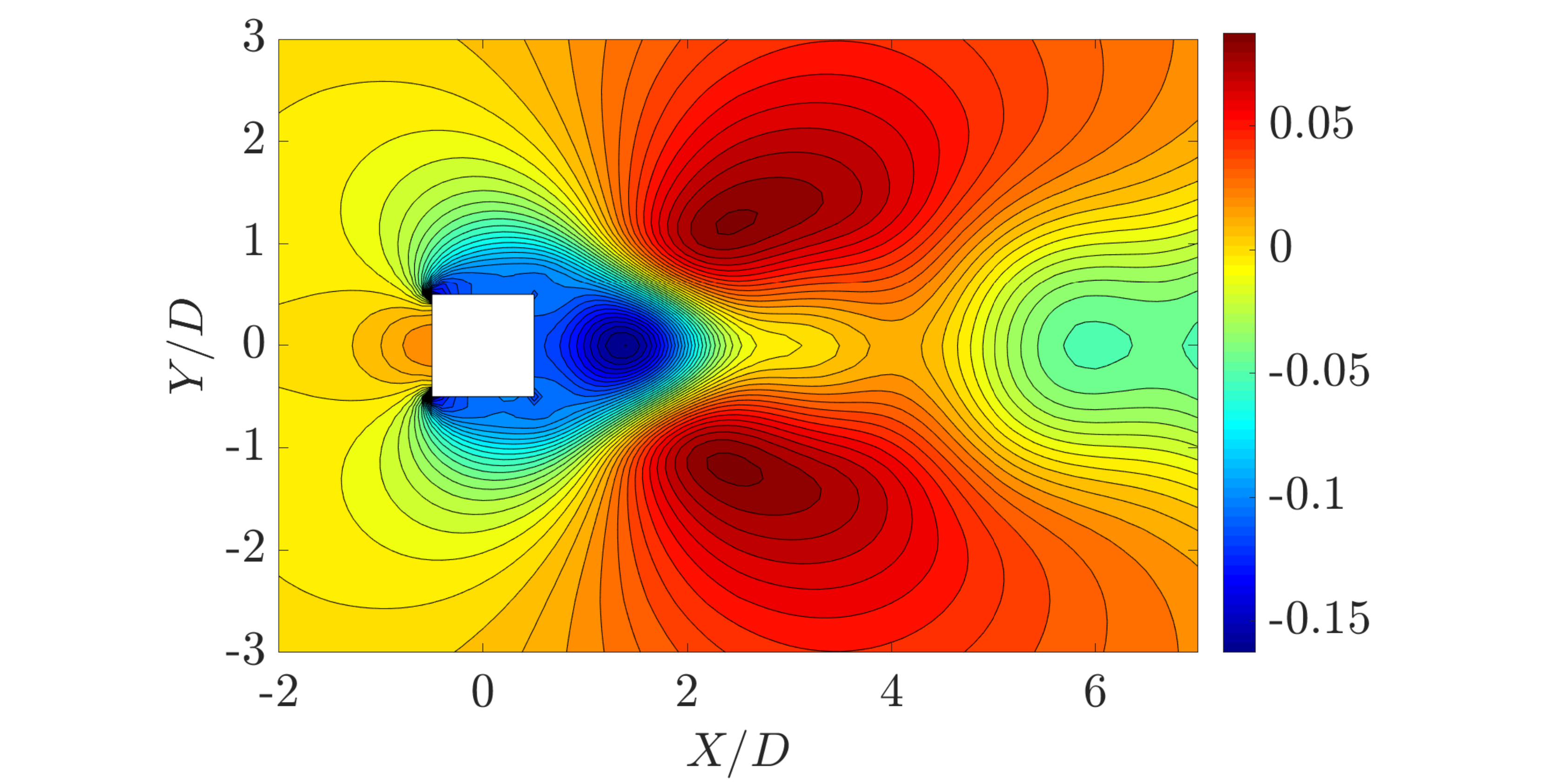}
	\caption{Lock-in, $U_r=6.0$}
     \end{subfigure}
     \begin{subfigure}{0.49\textwidth}
     \centering
     \includegraphics[trim={3.25cm 0cm 2cm 0.3cm},clip,scale=0.2]{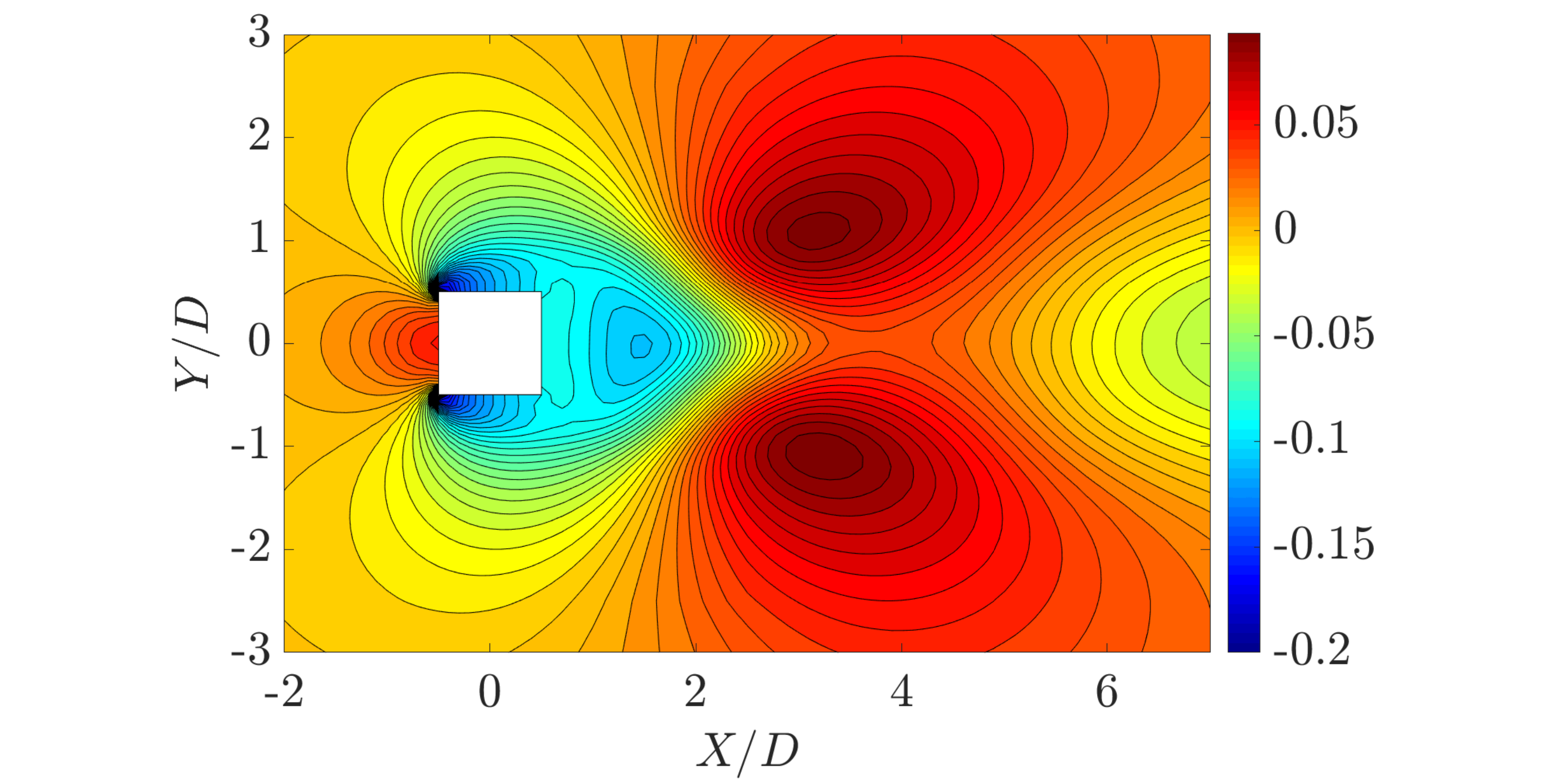}
	\caption{Post lock-in, $U_r=10.0$}
     \end{subfigure}
     \begin{subfigure}{0.49\textwidth}
     \centering
     \includegraphics[trim={0cm 0cm 0cm 0.3cm},clip,scale=0.24]{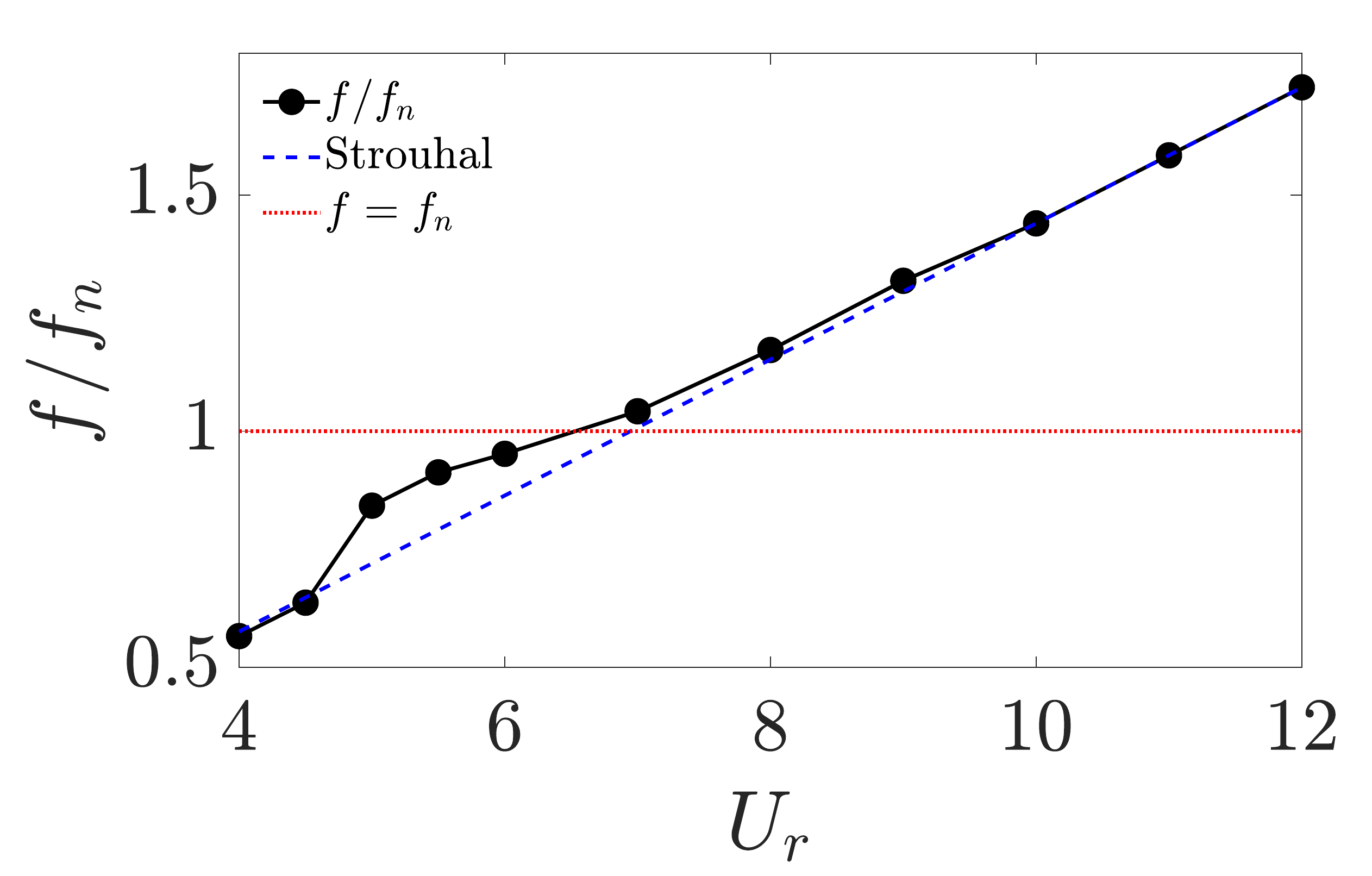}
	\caption{}
     \end{subfigure}~
     \pgfplotsset{every tick label/.append style={font=\Large}}
\pgfplotsset{every tick label/.append style={scale=1}}
\centering
\begin{subfigure}{0.49\textwidth}
\centering
\begin{tikzpicture}[trim axis left, 
trim axis right, 
scale=0.55, 
baseline]
\begin{axis}[
    xlabel={\Large $U_r$},
    ylabel={\Large $\phi$},
    xmin=3.7, xmax=12.5,
    ymin=-5, ymax=185,
    xtick={2,4,6,8,10,12},
    ytick={0,90,180},
    yticklabel style={/pgf/number format/fixed, /pgf/number format/precision=2},
    width = 10cm,
    height = 7cm,
]
 
\addplot[
    color=black,
    solid,
    mark=square,
    very thick,
    mark size = 4,
    ]
    coordinates {
(4,0.0026)(4.5,0.0027)(5,0.0213)(5.5,0.0030)(6,0.6192)(7,179.6543)(8,179.9349)(9,179.9505)(10,179.9742)(11,179.9657)(12,179.9919)
    };
\end{axis}
\end{tikzpicture}
\caption{}
\label{fig:AyvsUrm}
\end {subfigure}
     \caption{(a) Schematic of the interaction between bluff body motion, vortex shedding and shear layer instability in the lock-in region. (b-e) Primary shear layer mode variation for different $U_r$ regimes. The flow is from left to right. In the lock-in regime, the high gradient region shrinks in the in-line ($X$) and expands in the transverse ($Y$) direction. (f) Cylinder vibration frequency variation with $U_r$. (e) Phase difference ($\phi$) between the fluid force and bluff-body motion. The onset of phase jump from ${0}\degree$ to ${180}\degree$ coincide with the sign change of the primary shear layer mode (c-d).}
     \label{Fig:WakeInst}
\end{figure}
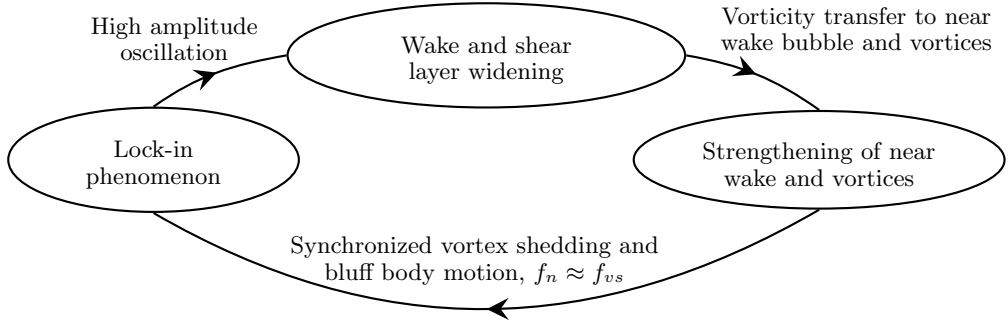
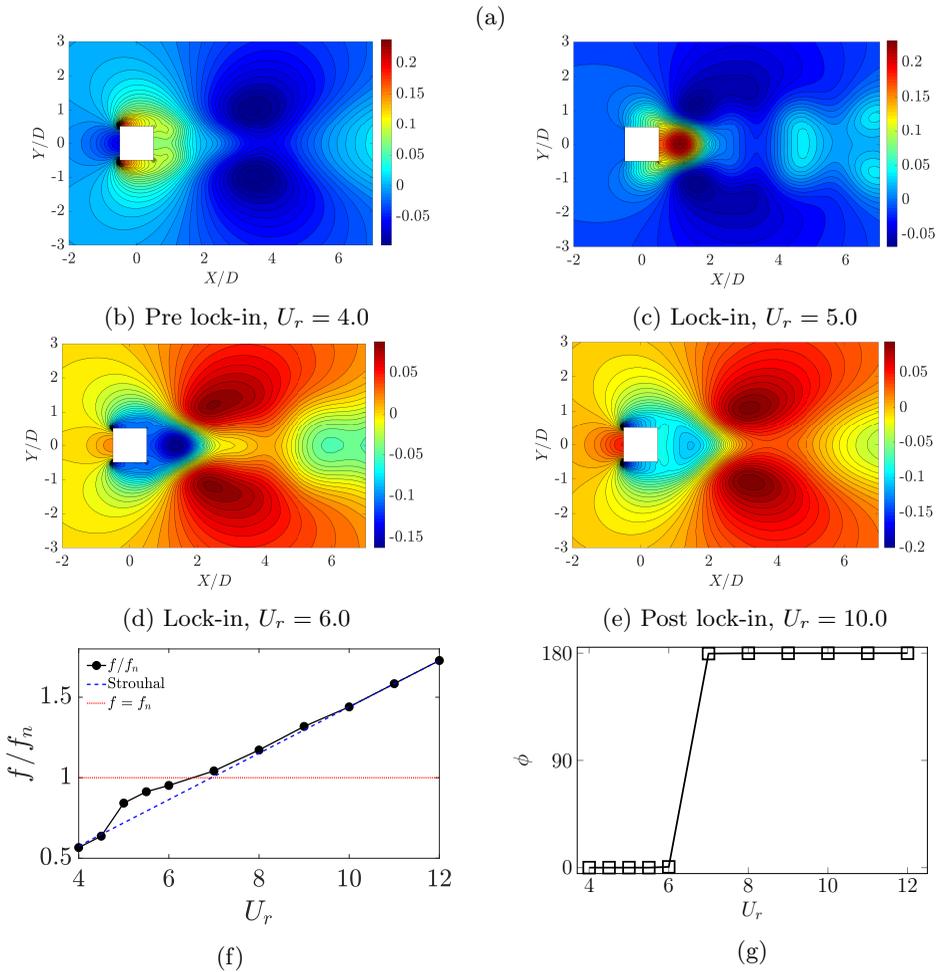
Figure \ref{Fig:WakeInst}a elaborates the interaction between the wake features and the bluff body motion. When the vortex shedding synchronizes with the bluff body motion, it causes the bluff body to undergo a relatively high-amplitude motion. This widens the wake and eventually the shear layer, decreasing the velocity gradients. This causes the shear layer to give away vorticity flux to the vortex shedding process, intensifying the vortices and the near-wake bubble. The strengthening of vortices increases the in-phase forces with the motion, i.e., the surrounding fluid flow tends to supply higher energy to the structure. As illustrated in \cite{jauvtis2004effect}, the force $F_v$ and the energy transfer rate $\dot{e_v}$ due to the principal vortices can be analyzed using the following simple analytical relations:
\begin{gather}
F_v = \rho^{\mathrm{f}} \Gamma U_v, \\
\dot{e_v} = \rho^{\mathrm{f}} \Gamma U_v \dot{y},
\end{gather}
where $\Gamma$ is the vortex strength, $U_v$ is the streamwise velocity of the predominant vortex relative to the bluff body and $\dot{y}$ is the transverse velocity of the bluff body. It is clear that the increase in the vortex strength will increase the forces and energy transfer to the bluff body. The widening of the high gradient shear layer region in the lock-in regime can be seen in figures \ref{Fig:WakeInst}b-e, which demonstrate the primary shear layer mode for the different $U_r$ cases. In the pre-lock-in regime, the near-wake region is positive compared to the shear layer region. When $U_r=5.0$, the maximum amplitude case, it is clear that the high gradient region has shrunk in the streamwise direction and expanded in the transverse direction. Consequently, the near-wake region and the shear layer region interchanges the distribution when $U_r=6.0$, i.e. the near-wake region is negative compared to the shear region. This sign change continues to the post-lock-in regime, where the high gradient region extends to the streamwise direction and becomes narrower in the transverse direction. 

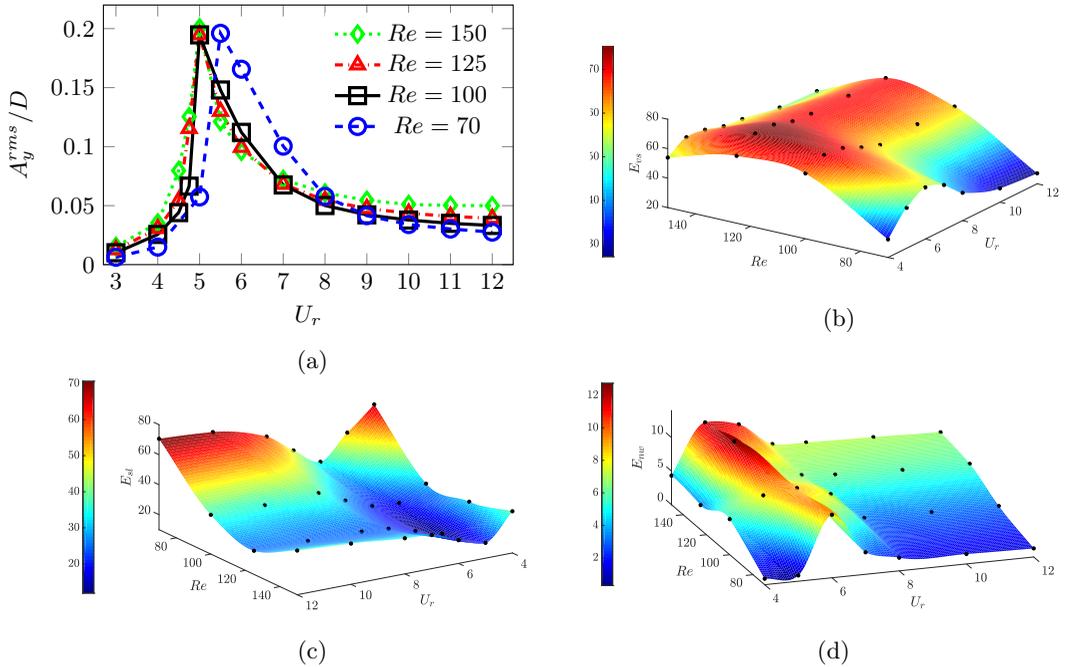
\begin{figure}
\centering
\pgfplotsset{every tick label/.append style={scale=1}}
\begin{subfigure}{0.5\textwidth}
\centering
\begin{tikzpicture}[trim axis left, 
trim axis right, 
baseline]
\begin{axis}[
    xlabel={$U_r$},
    ylabel={$A_y^{rms}/D$},
    xmin=2.7, xmax=12.5,
    ymin=0, ymax=0.22,
    xtick={2,3,4,5,6,7,8,9,10,11,12},
    ytick={0,0.05,0.1,0.15,0.2},
    legend pos=north east,
    legend style={draw=none},
    yticklabel style={/pgf/number format/fixed, /pgf/number format/precision=2},
    width = 7cm,
    height = 5cm,
]

\addplot[
    color=green,
    dotted,
    mark=diamond,
    very thick,
    mark size = 3,
    mark options=solid,
    ]
    coordinates {
(3,0.0154)(4,0.0353)(4.5,0.0797)(4.75,0.1253)(5,0.2007)(5.5,0.1212)(6,0.0960)(7,0.0722)(8,0.0607)(9,0.0545)(10,0.0510)(11,0.0502)(12,0.0499)
};
\addlegendentry{{$Re=150$}}

\addplot[
    color=red,
    dash dot,
    mark=triangle,
    very thick,
    mark size = 3,
    mark options=solid,
    ]
    coordinates {
(3,0.0126)(4,0.0306)(4.5,0.0542)(4.75,0.1154)(5,0.1948)(5.5,0.1299)(6,0.0994)(7,0.0681)(8,0.0545)(9,0.0476)(10,0.0436)(11,0.0409)(12,0.0392)
};
\addlegendentry{{$Re=125$}}

\addplot[
    color=black,
    solid,
    mark=square,
    very thick,
    mark size = 3,
    mark options=solid,
    ]
    coordinates {
(3,0.0101)(4,0.0256)(4.5,0.0441)(4.75,0.0665)(5,0.1947)(5.5,0.1480)(6,0.1121)(7,0.0675)(8,0.0501)(9,0.0422)(10,0.0378)(11,0.0350)(12,0.0332)
};
\addlegendentry{{$Re=100$}}
\addplot[
    color=blue,
    dashed,
    mark=o,
    very thick,
    mark size = 3,
    mark options=solid,
    ]
    coordinates {
(3,0.0063)(4,0.0149)(5,0.0574)(5.5,0.1960)(6,0.1656)(7,0.1004)(8,0.0578)(9,0.0415)(10,0.0340)(11,0.0302)(12,0.0276)
    };
    \addlegendentry{{$Re=70$}}

\end{axis}
\end{tikzpicture}
\label{fig:AyRe}
\caption{}
\end{subfigure}~
\begin{subfigure}[h]{0.5\textwidth}
\centering
\includegraphics[trim={1cm 0 0 1cm},clip,scale=0.2]{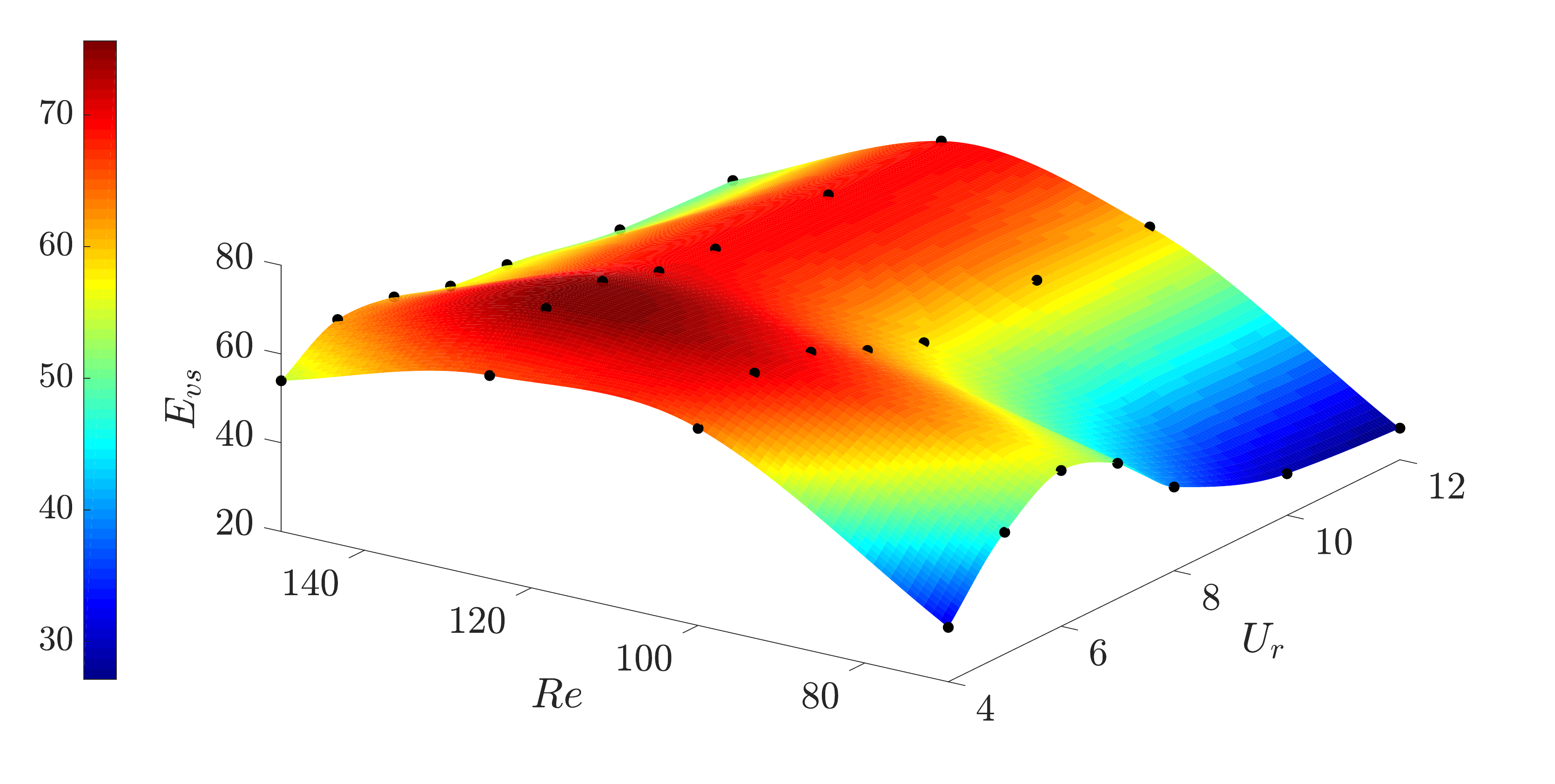}
\caption{}
\end{subfigure}
\begin{subfigure}[h]{0.5\textwidth}
\centering
\includegraphics[trim={1cm 0 0 1cm},clip,scale=0.2]{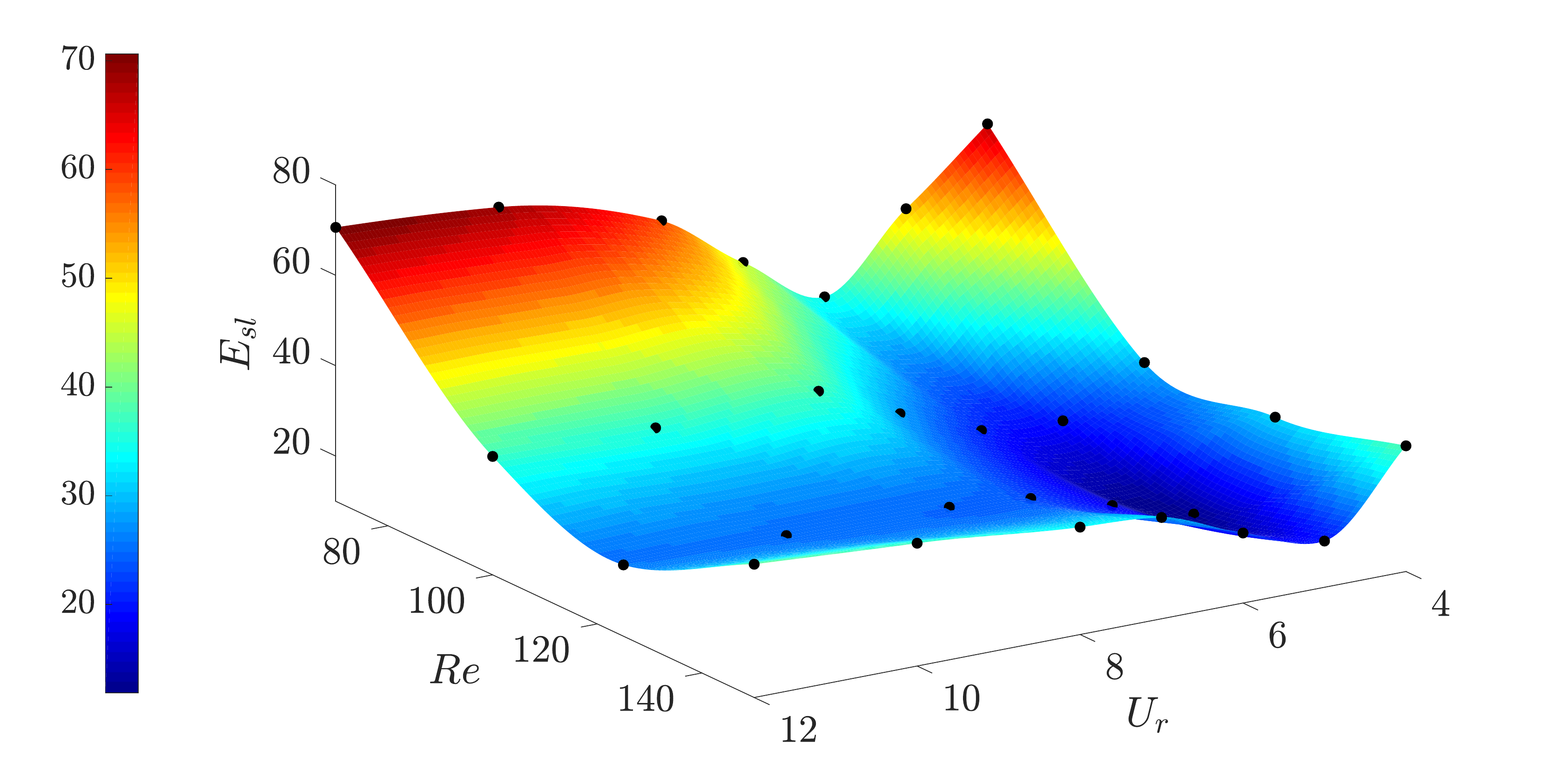}
\caption{}
\end{subfigure}~
\begin{subfigure}[h]{0.5\textwidth}
\centering
\includegraphics[trim={1cm 0 0 1cm},clip,scale=0.2]{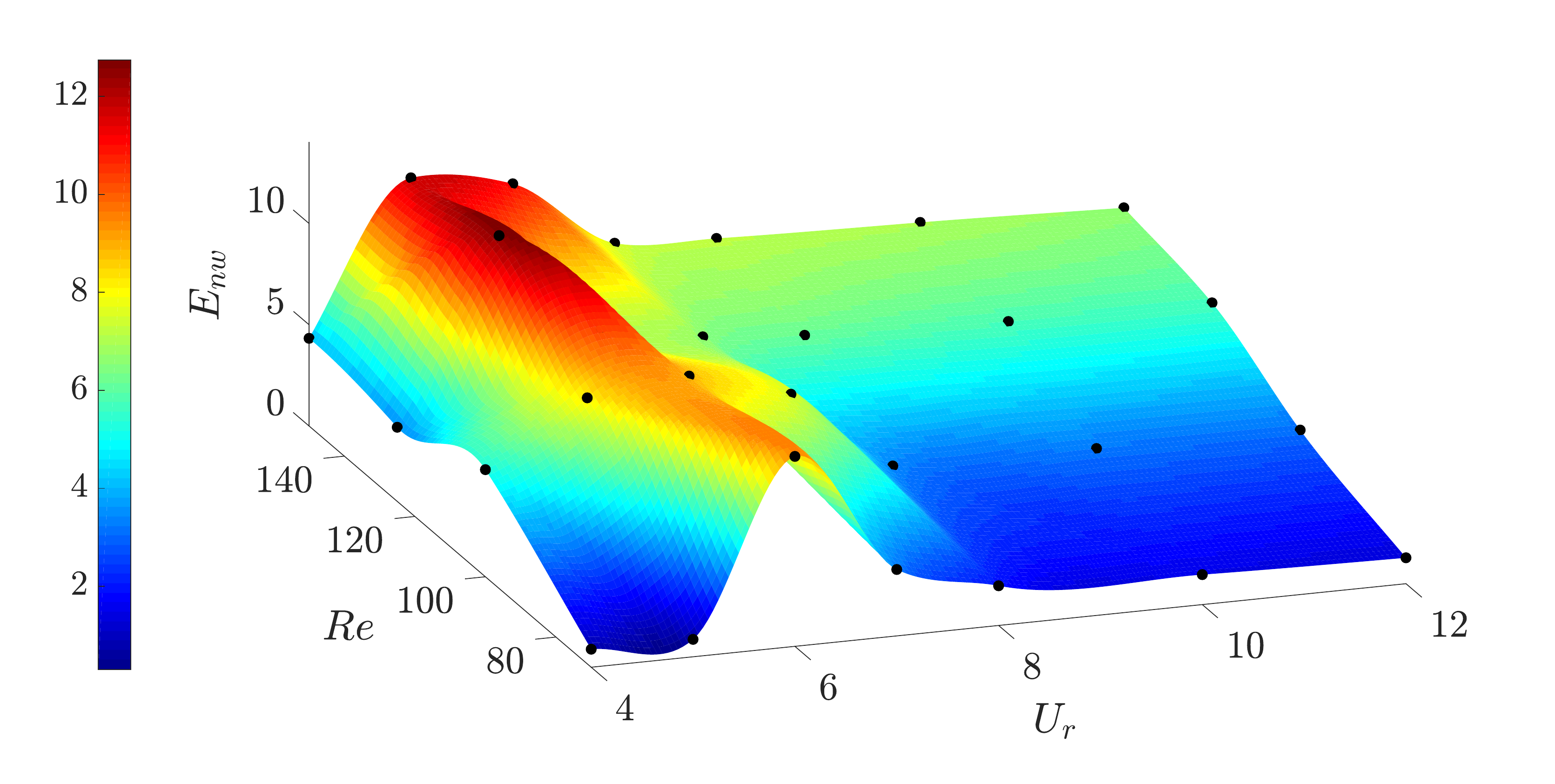}
\caption{}
\end{subfigure}
\caption{(a) Transverse amplitude $A_y^{rms}$ of square cylinder as a function of $U_r$ for $Re \in [70,150]$ at $m^*=3$ and $\zeta =0.0$ and the mode energy contributions from wake features: (b) vortex shedding, (c) shear layer and (d) near-wake bubble at different Reynolds numbers. The vortex shedding and near-wake is energized at the lock-in region while the shear layer mode energy is reduced.}
\label{fig:AyUnsw}
\end{figure}

We further generalize this variation of the wake feature contribution for $Re>Re_{cr} (=44.7$ \citep{yao2017model}). Figure \ref{fig:AyUnsw} demonstrates the bluff body motion response and the modal energy contribution from the large-scale features of the wake. The cylinder motion follows a similar trend for $Re=100, 125$ and $150$ where the lock-in region is detected as $U_r \in [4.5,7]$. This region is slightly shifted to $U_r \in [5.5,8]$ for $Re=70$. In all cases, we observe a maximum of $A_y^{rms} \approx 0.2D$. Regardless of $Re$, the wake features exhibit a similar trend in terms of modal energy. As displayed in figures \ref{fig:AyUnsw}(b) and (d), the vortex shedding and the near-wake modes become more energetic during the lock-in and the shear layer modes become less energetic. This further confirms the proposed interaction cycle for the coupling of the wake features and the bluff body motion.

Using the modal decomposition, we have quantitatively explained the interaction dynamics of the flow features which have been conjectured by many previous studies. For example, many successful VIV suppression techniques are proposed by passive \citep{law2017wake} and active \citep{guan2017control, narendran2018control} methods with the experience based presumption that preventing the interaction between the shear layer, the vortex street and  the near-wake will suppress the synchronized wake-body lock-in phenomena. The cycle proposed above provides a proper physical mechanism for the success of those methods: they prevent the vorticity transfer between the shear layer and the vortex shedding and/or near-wake bubble, which breaks the self-sustenance of the wake interaction cycle. This understanding of the wake features and their interactions will be vital for the development of effective suppression methods and devices for flow-induced vibrations.
In this analysis, we observe that the synchronization of the wake and bluff body weakens the shear layer and intensifies the vortices and the near-wake bubble. In the aforementioned analysis, there exists a periodic vortex shedding for the stationary and pre-/post-lock-in regimes. In what follows, we investigate whether the perturbation of the near-wake bubble via flexibility can sustain the synchronized wake-body interaction at below critical $Re$ flow wherein the well-defined periodic vortex shedding does not exist for the stationary counterpart.

\section{Synchronized wake-body interaction at below critical Re}
At very low $Re$, the flow past a bluff body is two-dimensional, steady and symmetric with respect to the wake centerline. The near-wake bubble attached to its surface is the essential feature below the critical Reynolds number $Re_{cr}$, which is formed by the steady separation from the sharp corners of a square cylinder. Two symmetric and counter-rotating recirculation zones are present in the wake bubble. As $Re$ increases above the critical value, a Hopf bifurcation sets in and the flow becomes periodic via vortex shedding process. For circular and square cylinders \cite{park2016flow} demonstrated these values to be $Re_{cr} = 46.8$ and $44.7$ respectively, which were further confirmed by \cite{yao2017model}. Interestingly, when the bluff body is free to vibrate, \cite{meliga2011asymptotic} predicted for a circular cylinder that this unstable boundary will hold when $m^*>1000$ and the Hopf bifurcation will occur at much lower $Re$ for low $m^*$ values. The authors further conjectured that for a circular cylinder there will be a limiting $Re  \sim 32$, below which the wake flow will be 2D and steady regardless of the mass ratio $m^*$.

\begin{figure}
\input{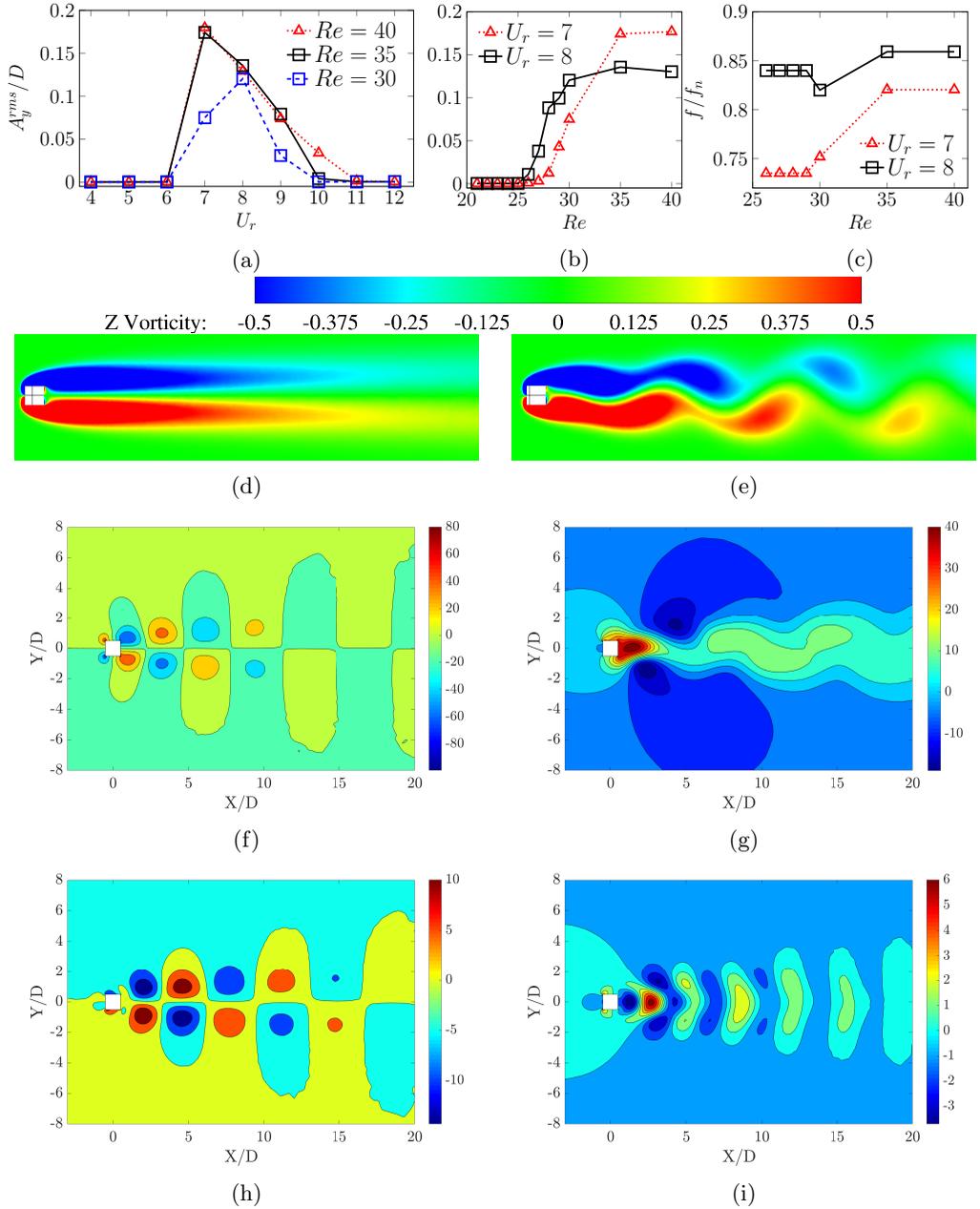}
\caption{Response characteristics of wake-body interaction for $Re<Re_{cr}$:  dependence of transverse amplitude on $U_r$ and $Re$ (a) reduced velocity range $U_r \in [4,12]$ for  $Re\in[30,40]$, (b) $Re\in[20,40]$ for $U_r=7,8$ (c) vibration frequency of the bluff body, (d) the spanwise $Z$-vorticity for a non-synchronized case ($U_r=4.0, Re=40$), (e) unsteady wake in a synchronized case ($U_r=7.0, Re=40$) at $tU_{\infty}/D=120$. (f-i) POD modes containing $\sim99\%$ of the mode energy of the synchronized wake-body case: $U_r=7.0, Re=40$. The flow is from left to right.}
\label{fig:AyReLow}
\end {figure}
Herein, we observe that for some $Re<Re_{cr}$ the spring-mounted square cylinder undergoes significant synchronized wake-body motion for a specific range of $U_r$. Figure \ref{fig:AyReLow}a illustrates a variation of high amplitude motion  for $Re = 30, 35 $ and $40$. When $Re$ becomes closer to $Re_{cr}$, the synchronization regime widens and the highest amplitude $U_r$ shifts from $U_r=8$ to $U_r=7$. In contrast to $Re>Re_{cr}$ cases, we observe no motion of the cylinder in pre- and post-synchronization regimes. We further examine the conjecture of \cite{meliga2011asymptotic} and demonstrate that for a square cylinder this synchronized motion is present when $Re \geq 26$ (figure \ref{fig:AyReLow}b). Additional analysis on these synchronized motion cases revealed that the wake-body system synchronizes to a frequency slightly less than the natural frequency $f_n$ of the bluff body similar to the $Re>Re_{cr}$ lock-in regime (figure \ref{fig:AyReLow}c). We observe that, for all synchronized motion cases, the wake is unsteady with some vortex shedding patterns. 
For example, we can compare the representative $Z$-vorticity contours for the zero motion and the synchronized motion cases displayed in figure \ref{fig:AyReLow}d and (\ref{fig:AyReLow}e), respectively.  The zero motion case is almost identical to the stationary cylinder counterpart, while the synchronized motion case is similar to the lock-in scenario for $Re>Re_{cr}$. Notwithstanding, the vortex formation length is considerably high for this below $Re_{cr}$ configuration. We further decompose the unsteady wake of the synchronized motion case and examine similar features as $Re>Re_{cr}$ cases, i.e. the vortex shedding (figure \ref{fig:AyReLow}f and h), the shear layer (figure \ref{fig:AyReLow}g) and the near-wake bubble (figure \ref{fig:AyReLow}i). 

\pgfplotsset{every tick label/.append style={font=\normalsize}}
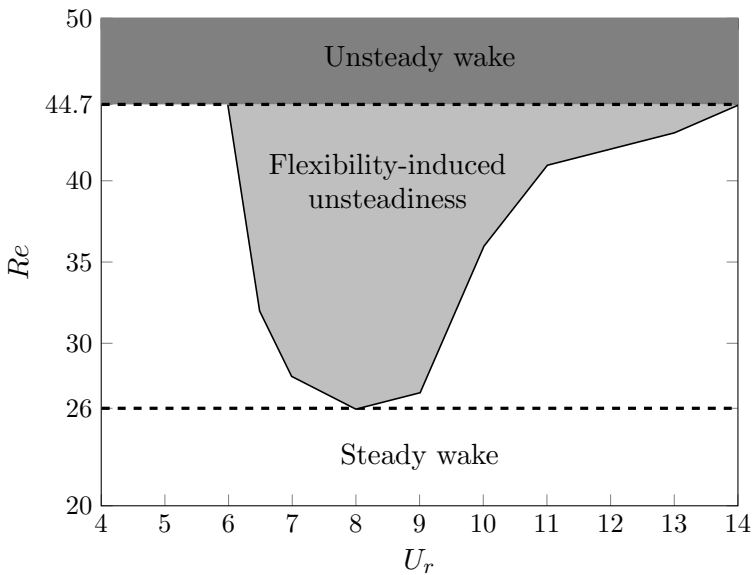
\begin{figure}
\centering
\begin{tikzpicture}[trim axis left, 
trim axis right, 
scale=1, 
baseline]
\begin{axis}[
    xlabel={\large{$U_r$}},
    ylabel={\large{$Re$}},
    xmin=4, xmax=14,
    ymin=20, ymax=50,
    xtick={4,5,6,7,8,9,10,11,12,13,14},
    ytick={20,26,30,35,40,44.7,50},
    legend pos=north east,
    legend style={draw=none},
    yticklabel style={/pgf/number format/fixed, /pgf/number format/precision=2},
    width = 10cm,
    height = 8cm,
]
   \addplot[
    color=black,
    solid,
    very thick,
    ]
    coordinates {
(6,44.7)(6.5,32)(7,28)(8,26)(9,27)(10,36)(11,41)(12,42)(13,43)(14,44.7)
    };
\node[draw=none,align=center,font=\large] at (axis cs:9,23) {Steady wake};
\addplot[
    color=black,
    dashed,
    very thick,
    ]
    coordinates {
(0,26)(16,26)
    };
\addplot[
fill=lightgray,draw=none] coordinates {(6,44.7)(6.5,32)(7,28)(8,26)(9,27)(10,36)(11,41)(12,42)(13,43)(14,44.7)(6,44.7)}\closedcycle;
\node[draw=none,align=center,font=\large] at (axis cs:8.5,40) {Flexibility-induced \\ unsteadiness};
\addplot[fill=gray,draw=none] coordinates {(0,44.7)(15,44.7)(15,51)(0,51)(0,44.7)}\closedcycle;
\node[draw=none,align=center,font=\large] at (axis cs:9,47.5) {Unsteady wake};
\addplot[
    color=black,
    dashed,
    very thick,
    ]
    coordinates {
(0,44.7)(15,44.7)
    };
\end{axis}
\end{tikzpicture}
\caption{Demarcation of wake unsteadiness for a freely vibrating square cylinder at $m^*=3$. For the stationary square cylinder, the wake is steady for $Re<26$ and for $Re>44.7$ the wake becomes unsteady, as shown by dashed lines. }
\label{fig:UnsBound}
\end{figure}
These observations constitute the basic requirement for the wake-bluff body synchronized motion: the bluff body should have an optimal amount of flexibility (i.e., not too rigid nor too flexible) and the flow needs to have sufficiently large inertia (i.e., higher $Re$) to trigger the unsteadiness in the near-wake bubble. This particular $Re$ is lower than the $Re_{cr}$ for a fixed bluff body. This means that the flexibility of the solid body provides an avenue for the wake and the spring-mounted body to synchronize eventually causing the wake to be unsteady. From this numerical experiment, we can deduce that the flexibility of the bluff body is the primary factor driving the synchronized wake-body motion, neither the vortex shedding nor the shear layer roll-up. Hence the most critical wake feature for the onset of wake-body synchronization is the near-wake bubble. When the $Re$ is very low ($<26$) this bubble remains steady and the counter-rotating recirculation zones  behind the bluff body are stable. For $26 \leq Re \leq 44.7$, it remains same if the bluff body is either too rigid or flexible. However, in this $Re$ regime, when the bluff body is appropriately flexible, slight perturbations cause distortions in the steady wake. These distortions become periodic and begin to synchronize with the bluff body. This synchronization leads to a relatively higher amplitude oscillations at a frequency slightly less than the natural frequency of the solid body in a vacuum. At the same time, due to the vorticity generation by the unsteadiness developed in the wake, the vortices are shed from the downstream end of the wake bubble. 

With the aforementioned findings, we summarize the bluff body wake behavior of $20 \leq Re \leq 50$ regime in figure \ref{fig:UnsBound}. We deduce that the root cause of wake-body synchronization is the frequency lock-in between the natural frequency of the bluff body and the near-wake bubble. This further demonstrates that the unsteadiness of the wake can be induced by the flexibility of the bluff body. Moreover, the unsteady wake alone cannot induce high amplitude bluff body oscillations (e.g., pre- and post-lock-in in $Re>Re_{cr}$). Hence, we can further infer that the wake-body synchronization is induced by the synchronization of the bluff body motion with the near-wake bubble, not with the vortex street. In the next section, we generalize our findings to three-dimensional turbulent flows at moderately high $Re$.

\section{Effect of turbulence}
In this section, we investigate the dynamic decomposition of the wake behind a three-dimensional oscillating square cylinder at $Re =22,000$, wherein the wake
is fully turbulent. Our aim is to understand the role of turbulence when we extend the wake feature interaction cycle to turbulent flow. To retrieve the high-fidelity data at this $Re$, we employ a well-established dynamic subgrid-scale turbulence model in our  finite-element formulation. The filtered Navier-Stokes formulation
and the determination of the subgrid stress term via the dynamic subgrid-scale model is provided in \cite{jaiman_caf2016}. We incorporate this full-order model to generate 3D snapshots of the flow fields and the POD-Galerkin projection is applied on this high-fidelity data set.
At high-$Re$ turbulent wake flow, the aforementioned large-scale organized flow features are fragmented into smaller scales until the scales are fine enough to dissipate by the fluid viscosity. Therefore, small-scale and high-frequency modes can have a significant impact on the overall dynamics for the high-$Re$ turbulent condition, 
in contrast to the low-$Re$ study.
\begin{figure}
\centering
\begin{subfigure}[h]{0.5\textwidth}
\includegraphics[trim={0 0cm 0 0.3cm},clip,scale=0.4]{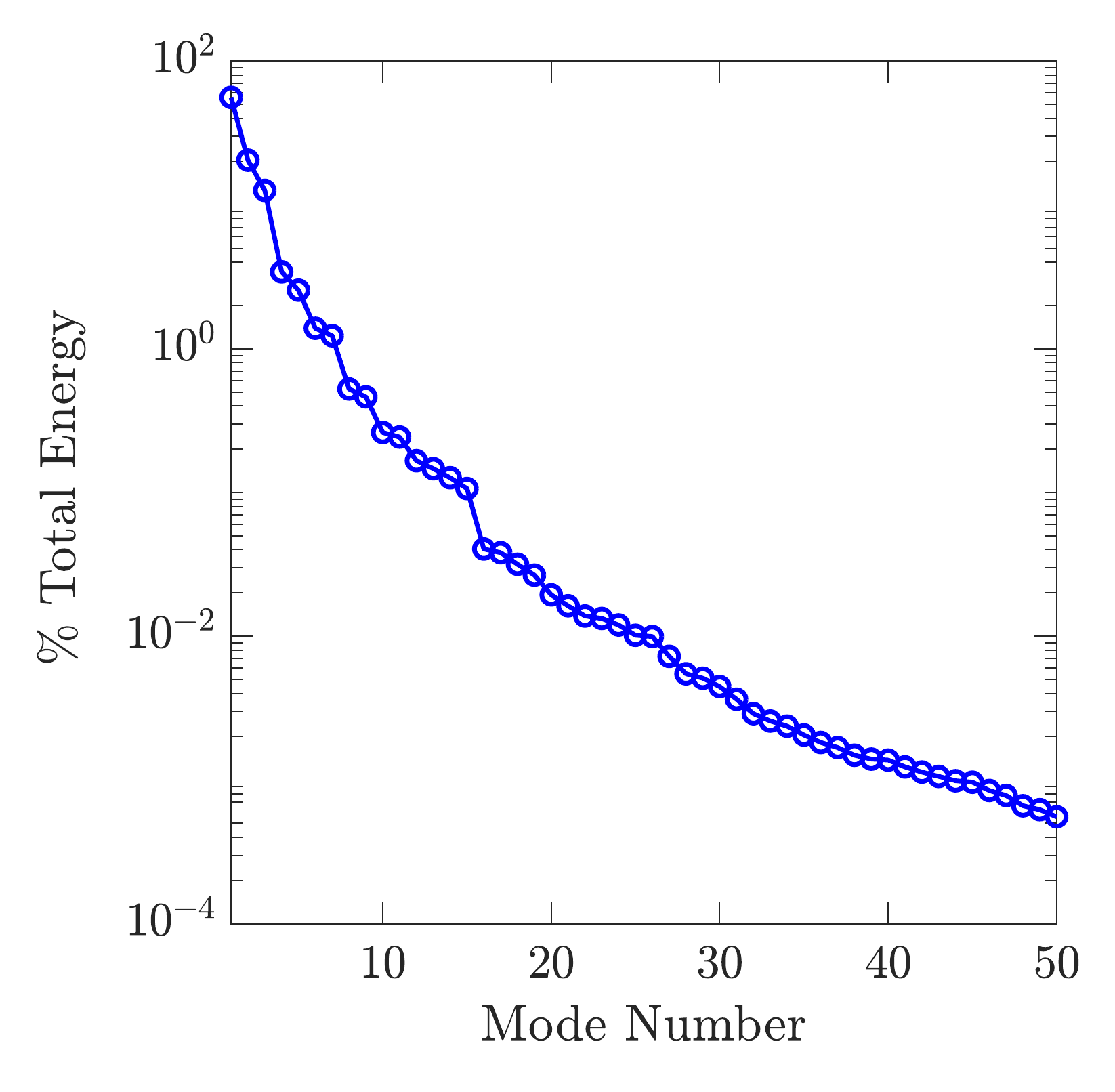}
\caption{}
\end{subfigure}~
\begin{subfigure}[h]{0.5\textwidth}
\includegraphics[trim={0 0cm 0 0.3cm},clip,scale=0.4]{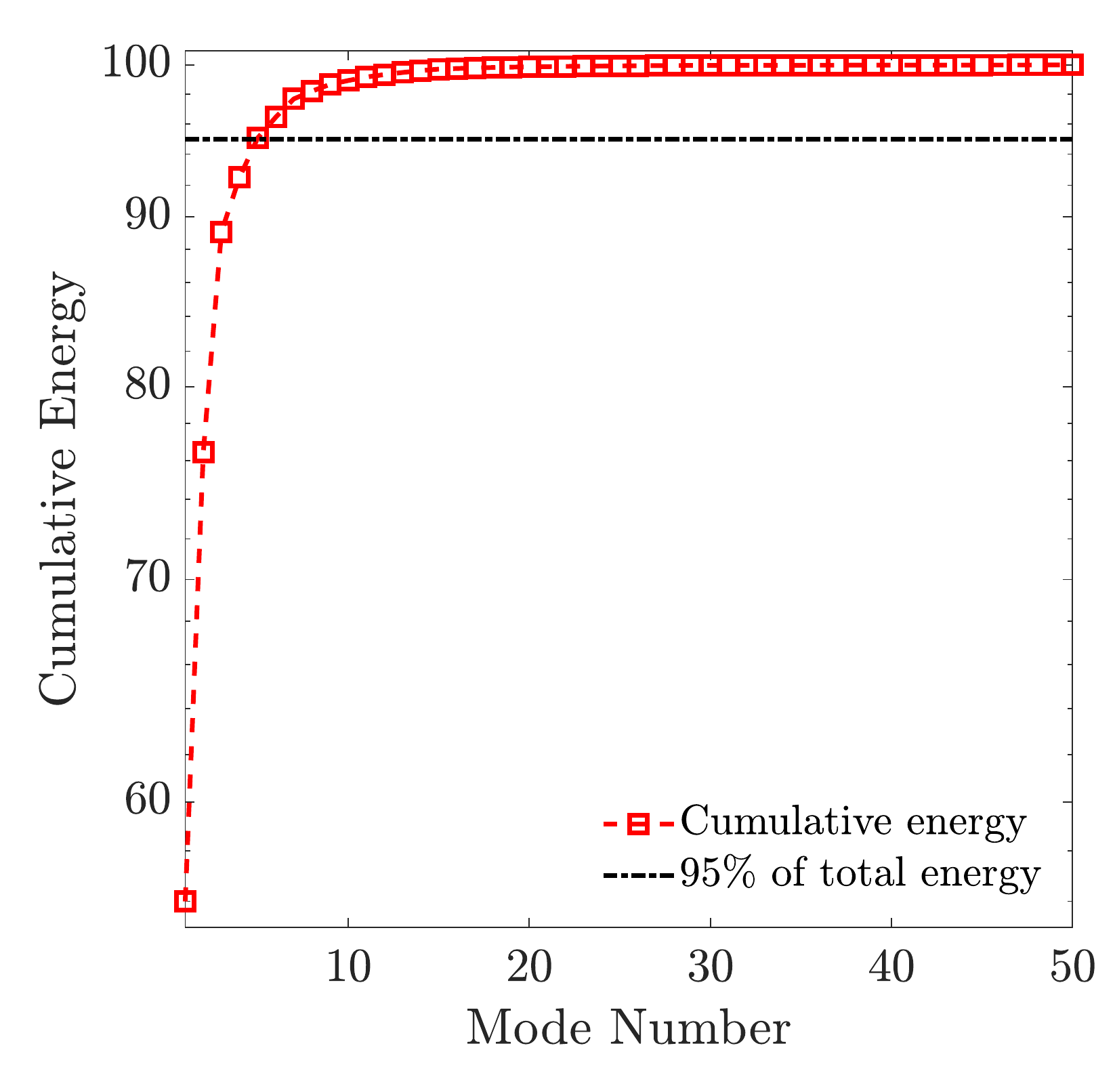}
\caption{}
\end{subfigure}
\caption{The energy distribution of the POD modes for $Re=22,000, m^*=3.0, U_r=6.0$: (a) energy decay of POD modes, and (b) cumulative energy of POD modes. The dashed line represents $95\%$ of the total mode energy.}
\label{PODEnergyHRe}
\end{figure}
Figure \ref{PODEnergyHRe} demonstrates the mode energy distribution for $Re =22,000$. Compared to the low cases, the modal energy is much more distributed among the modes. For instance, the most energetic mode of the $Re=100$ case contains $56\%$ of the total energy while it is $32.88\%$ for the high-$Re$ case at $Re=22,000$. Due to this broadening of the mode energy, the energy decay is less steep. For the low $Re$ cases, the first $5$ modes contain $95\%$ of the total mode energy and the first $9$ modes contain $99\%$. For the high-$Re$ case, a total of $123$ modes is required to capture $95\%$ of the energy and $211$ modes are required for $99\%$. We further investigate this distribution of the mode energy with the presumption that the presence of broadband turbulence is the key factor.

\begin{figure}
\centering
\begin{subfigure}[]{\textwidth}
\centering
\includegraphics[trim={0 0.25cm 0 0},clip,scale=0.2]{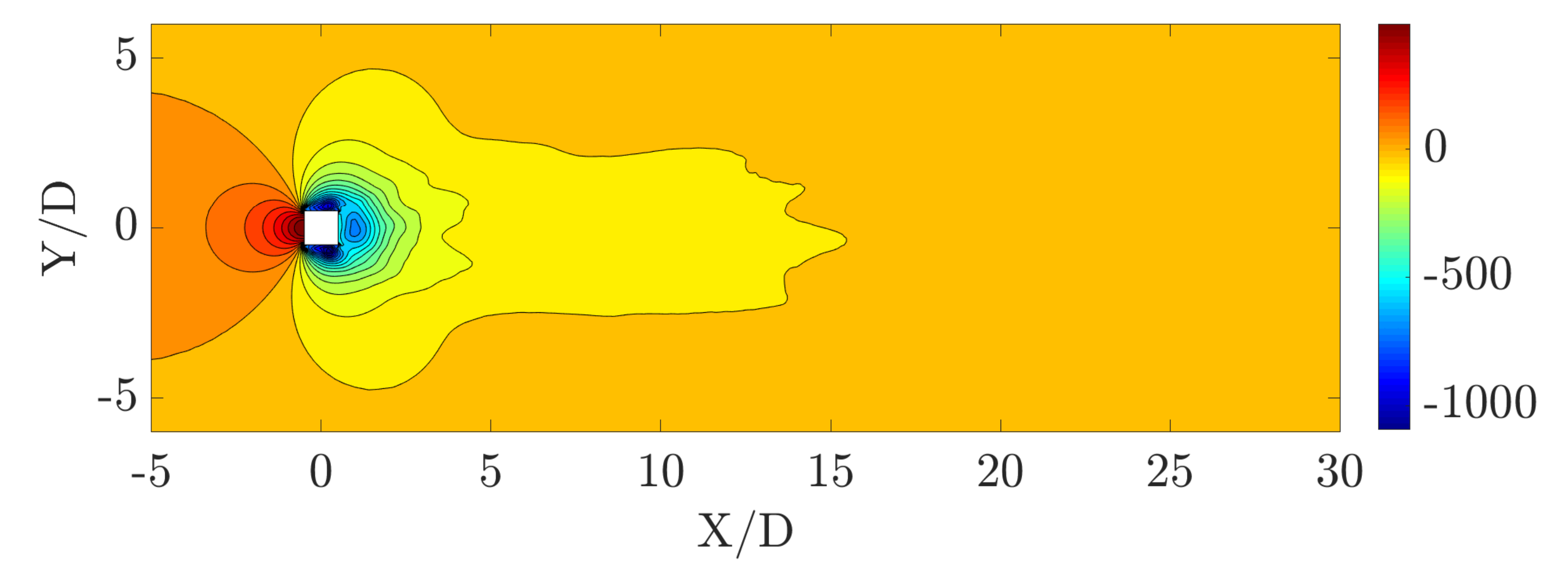}
\caption{Mean Pressure}
\end{subfigure}
\begin{subfigure}[]{0.5\textwidth}
\centering
\includegraphics[trim={0 0.25cm 0 0},clip,scale=0.2]{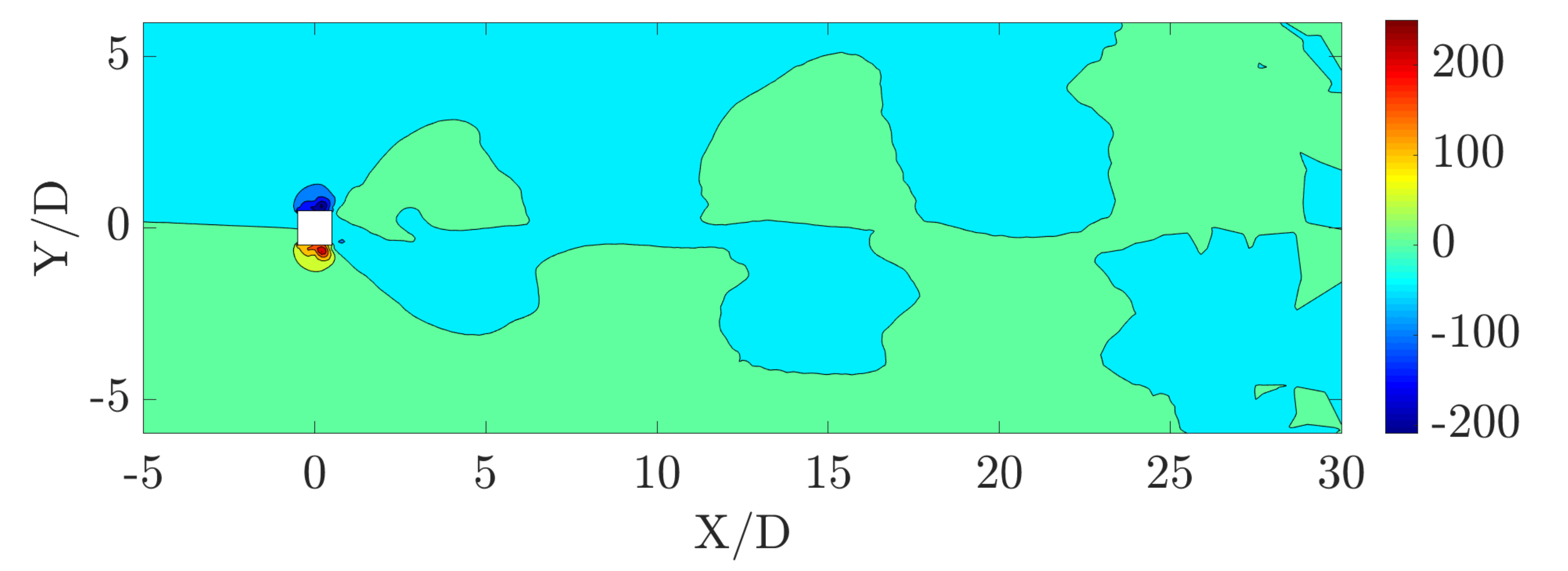}
\caption{Mode 1 (32.88\%)}
\end{subfigure}~
\begin{subfigure}[]{0.5\textwidth}
\centering
\includegraphics[trim={0 0.25cm 0 0},clip,scale=0.2]{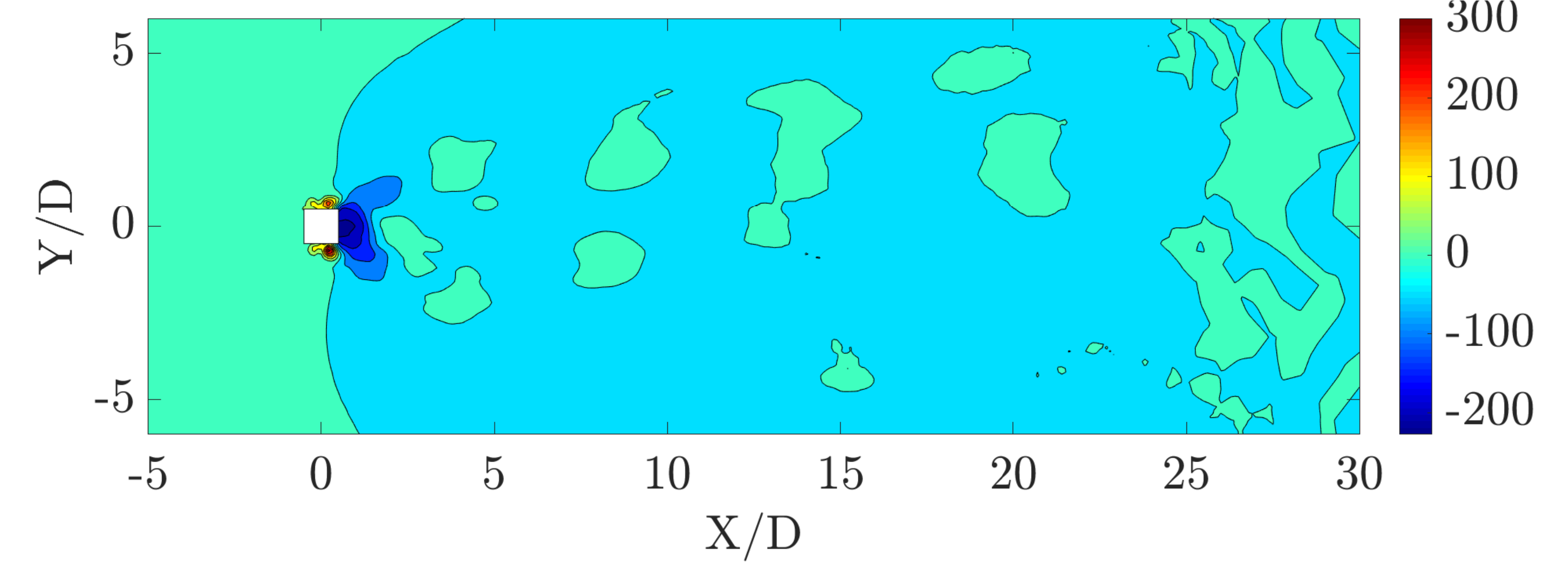}
\caption{Mode 2 (10.58\%)}
\end{subfigure}
\begin{subfigure}[]{0.5\textwidth}
\centering
\includegraphics[trim={0 0.25cm 0 0},clip,scale=0.2]{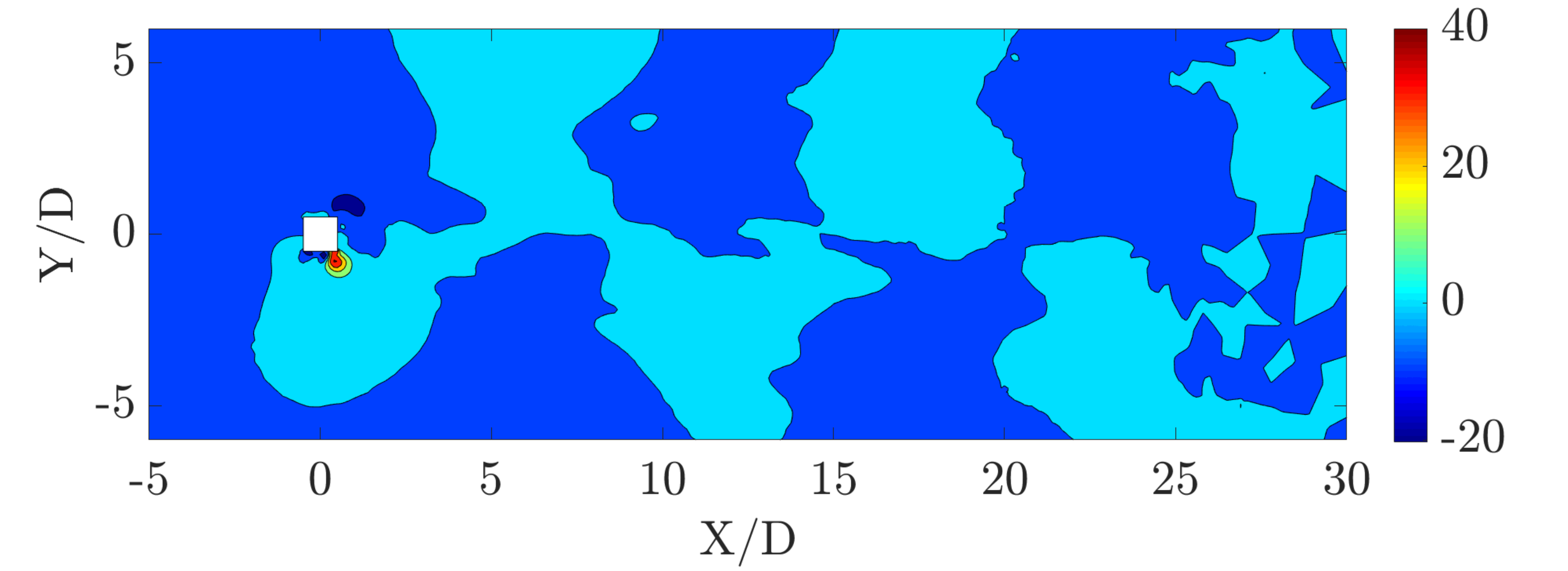}
\caption{Mode 3 (6.94\%)}
\end{subfigure}~
\begin{subfigure}[]{0.5\textwidth}
\centering
\includegraphics[trim={0 0.25cm 0 0},clip,scale=0.2]{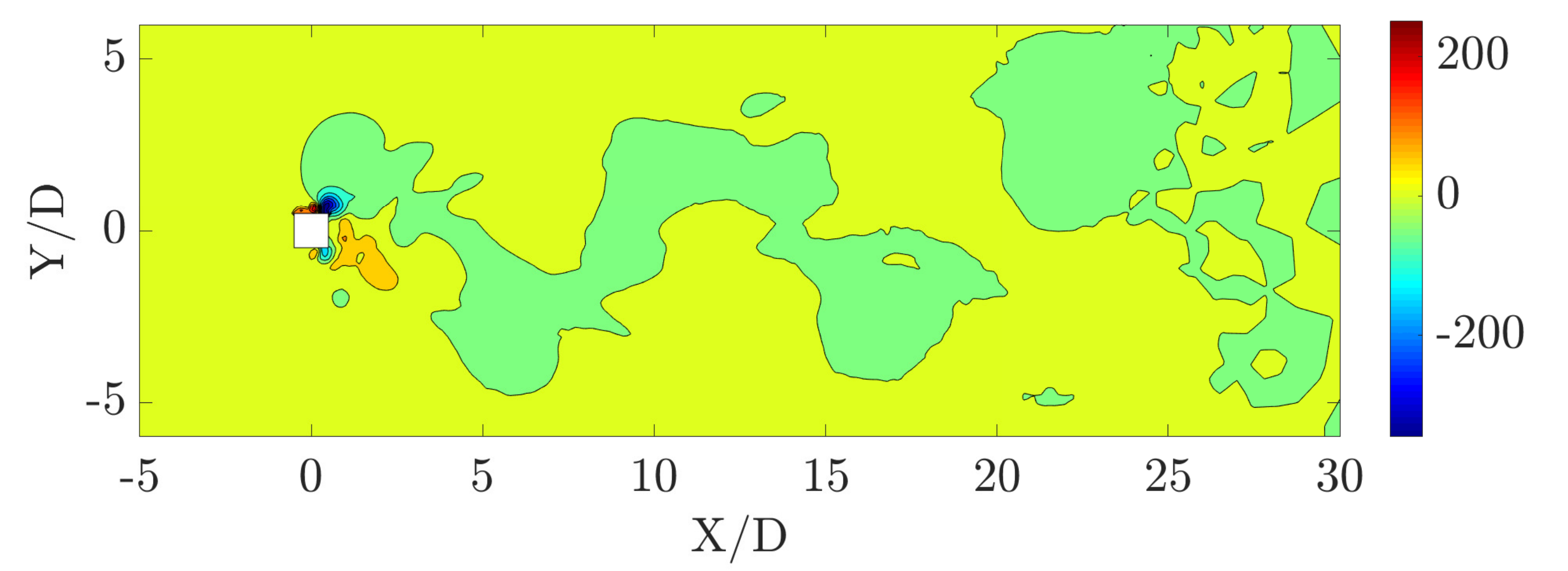}
\caption{Mode 4 (5.37\%)}
\end{subfigure}
\begin{subfigure}[]{0.5\textwidth}
\centering
\includegraphics[trim={0 0.25cm 0 0},clip,scale=0.2]{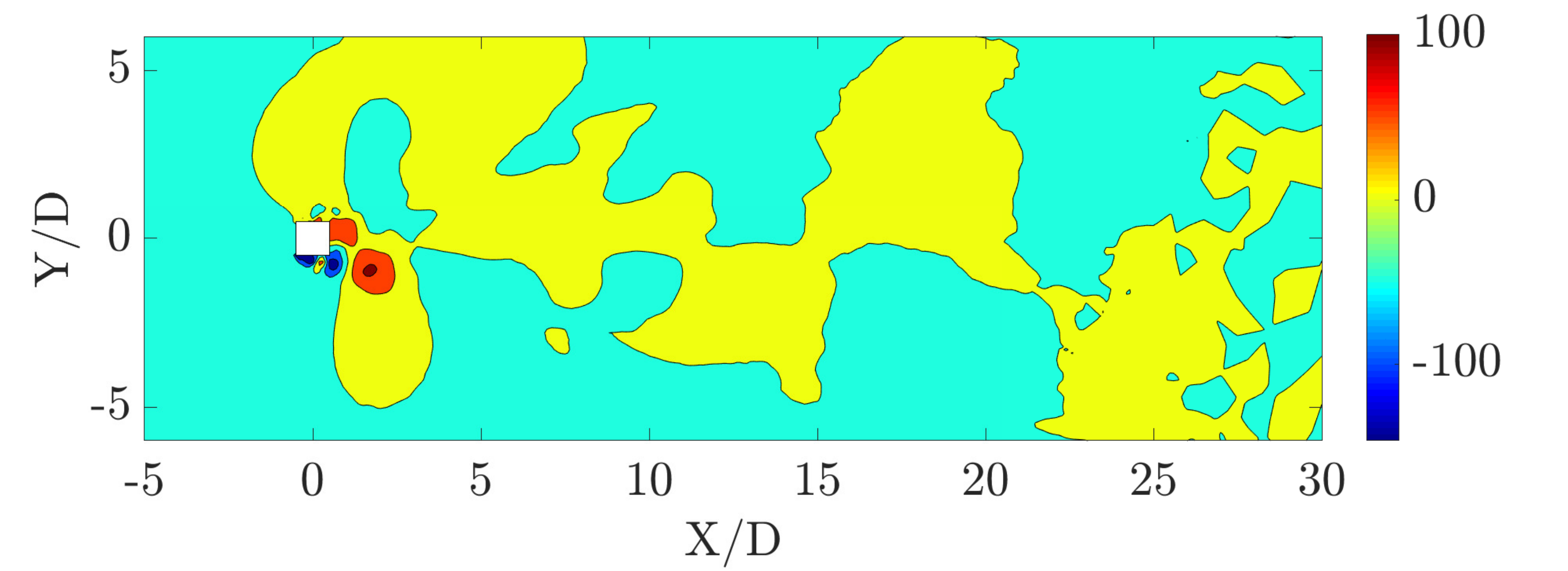}
\caption{Mode 5 (2.73\%)}
\end{subfigure}~
\begin{subfigure}[]{0.5\textwidth}
\centering
\includegraphics[trim={0 0.25cm 0 0},clip,scale=0.2]{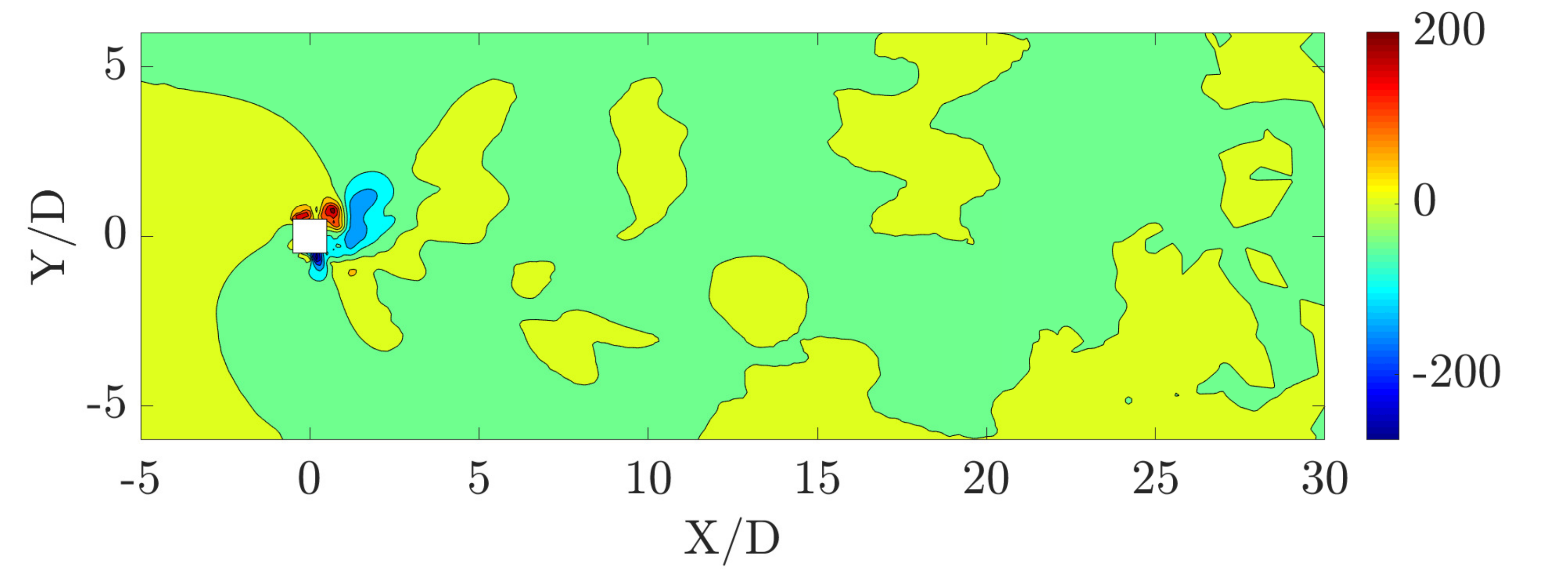}
\caption{Mode 6 (2.46\%)}
\end{subfigure}
\begin{subfigure}[]{0.5\textwidth}
\centering
\includegraphics[trim={0 0.25cm 0 0},clip,scale=0.2]{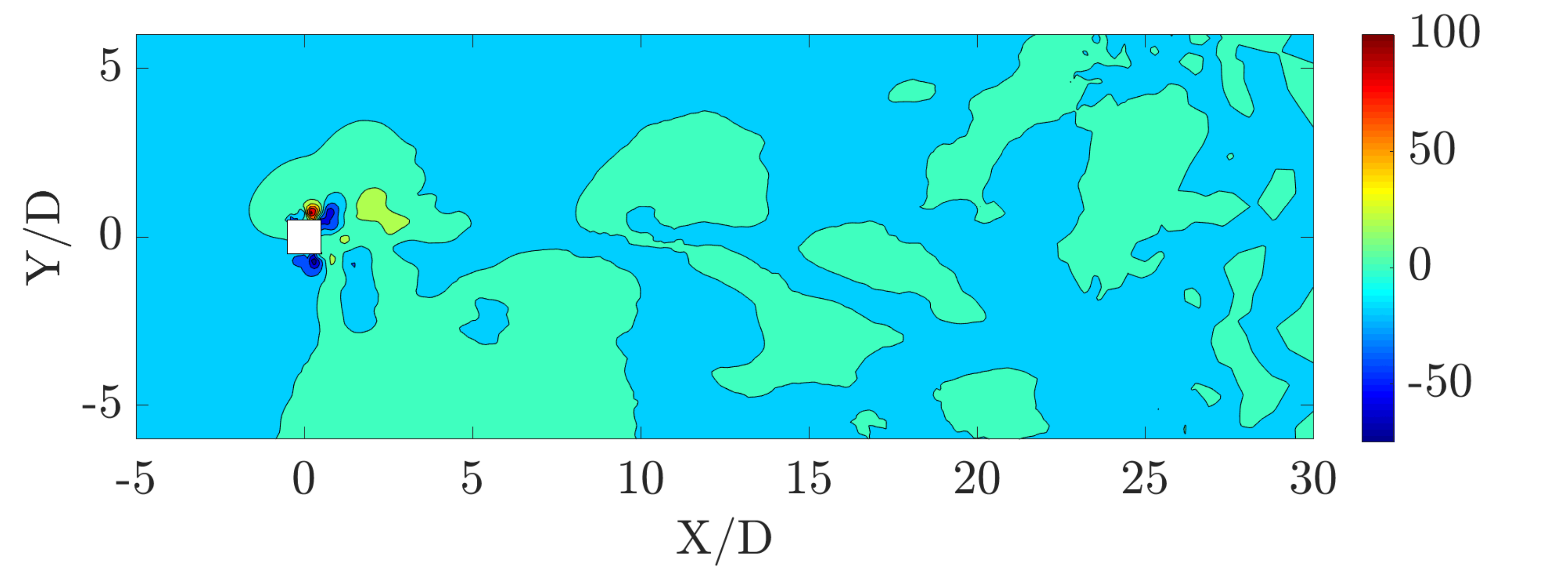}
\caption{Mode 7 (2.16\%)}
\end{subfigure}~
\begin{subfigure}[]{0.5\textwidth}
\centering
\includegraphics[trim={0 0.25cm 0 0},clip,scale=0.2]{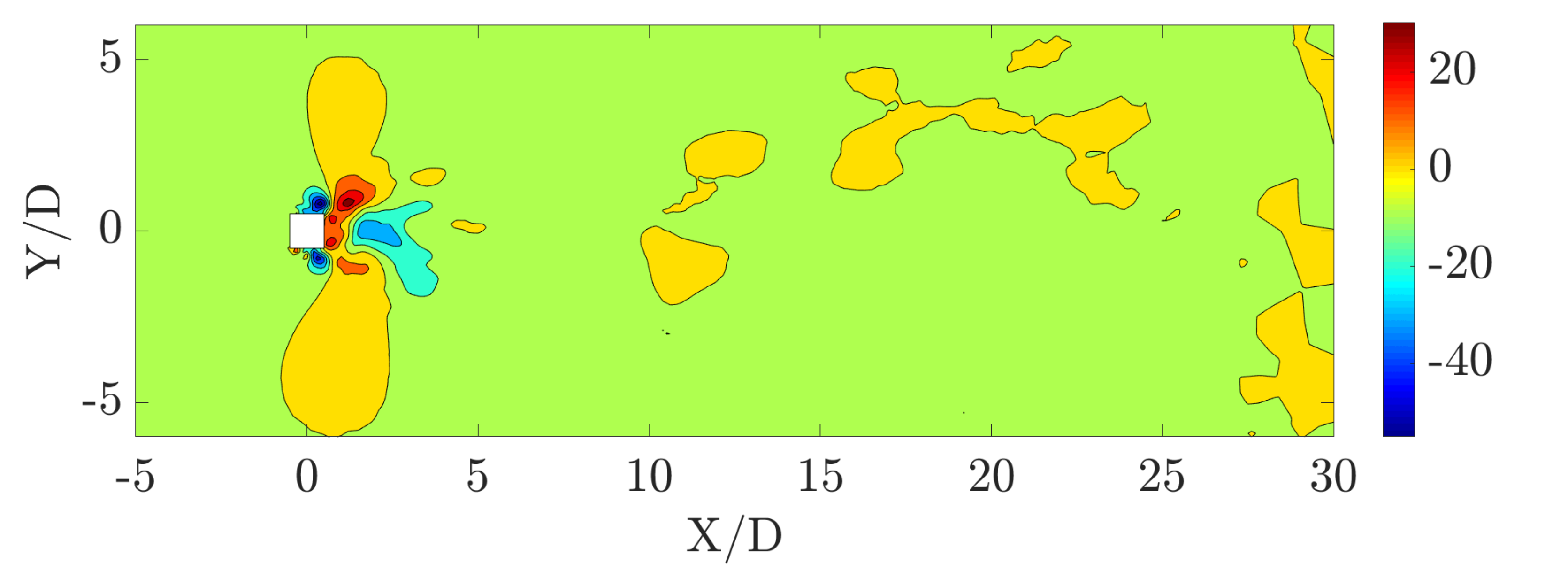}
\caption{Mode 8 (1.83\%)}
\end{subfigure}
\begin{subfigure}[]{0.5\textwidth}
\centering
\includegraphics[trim={0 0.25cm 0 0},clip,scale=0.2]{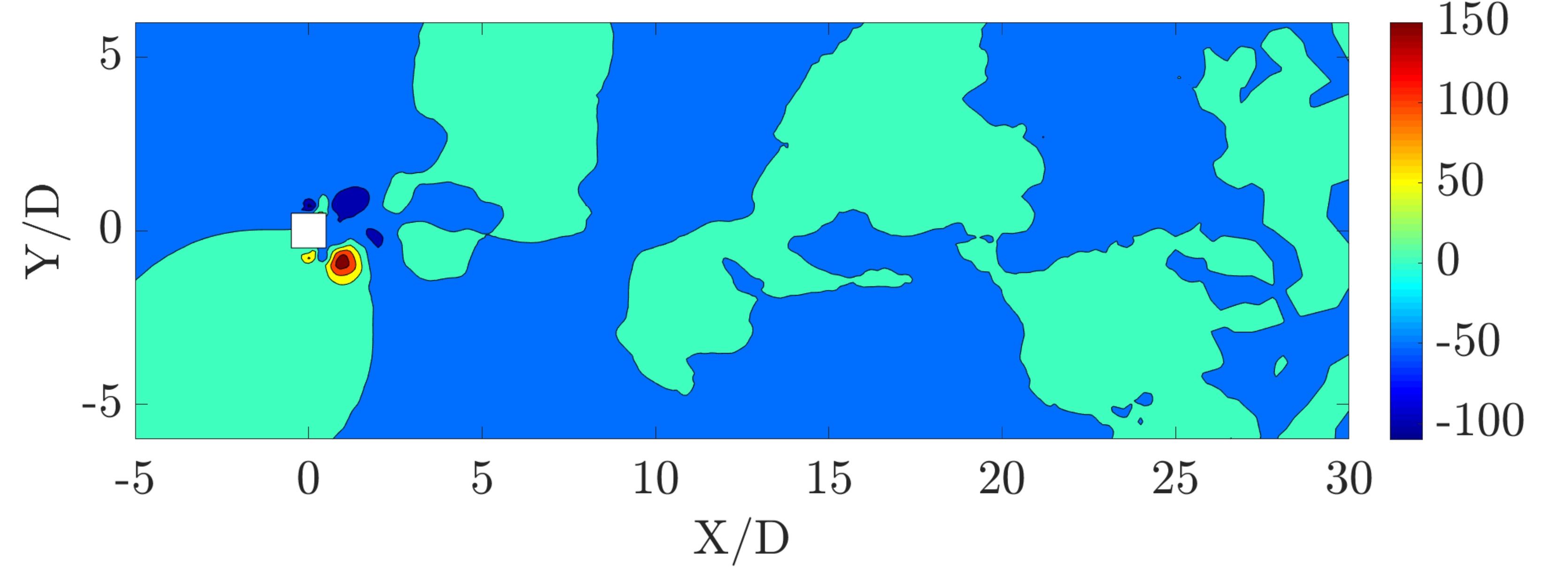}
\caption{Mode 9 (1.65\%)}
\end{subfigure}~
\begin{subfigure}[]{0.5\textwidth}
\centering
\includegraphics[trim={0 0.25cm 0 0},clip,scale=0.2]{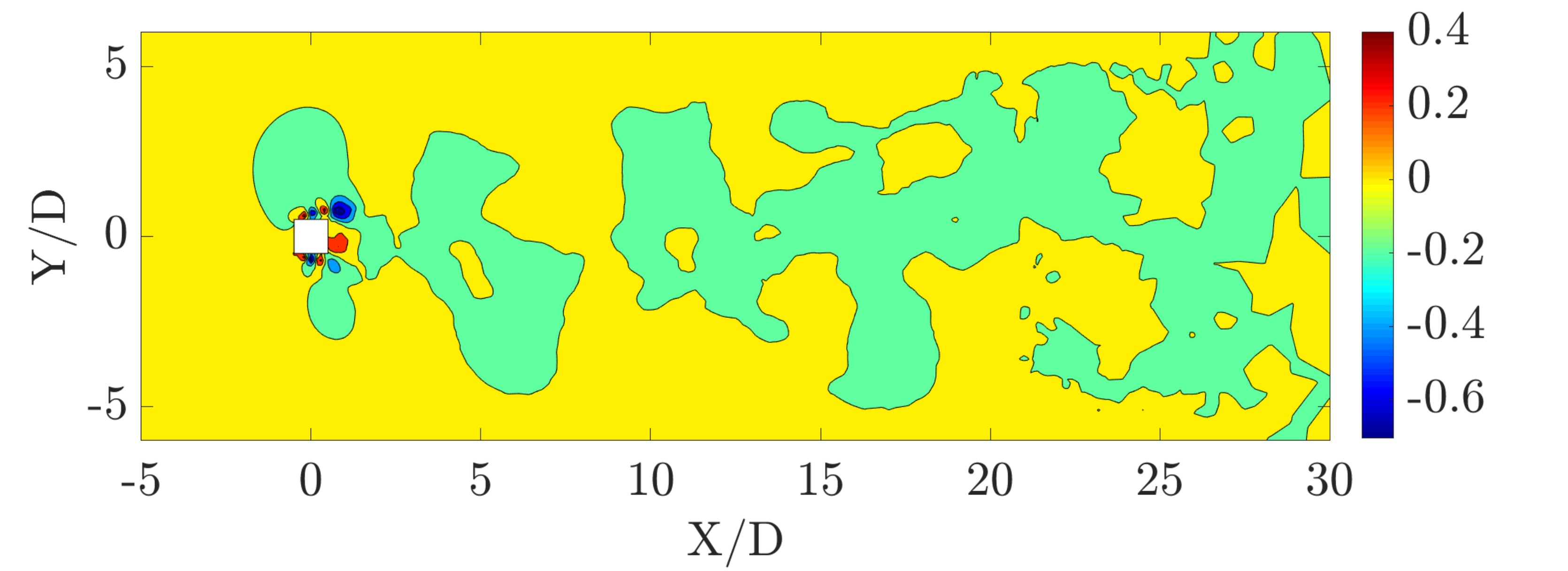}
\caption{Mode 10 (1.54\%)}
\end{subfigure}
\caption{The mean field and the first 10 significant POD modes for a oscillating square cylinder at $Re=22,000$. The energy fraction of the POD mode is mentioned in brackets. The plots are of the mid $Z$-plane.}
\label{PODModesHRe}
\end{figure}

Figure \ref{PODModesHRe} displays the mean pressure field and 10 most energetic POD modes obtained using the same snapshot technique for moderate $Re$ case. All the modes exhibit distorted and scattered patterns compared to the low $Re$ cases. However, the POD modes further portray general large-scale features. For example, the modes 1, 3, 4, 5, 6, 7 and 9 exhibit the flow patterns related to vortex shedding while the mode 2 is related to the shear layer and the modes 8 and 10 are of the near-wake phenomena. The distortions occur in each mode throughout the spatial domain due to the broadband and multiscale phenomenon of turbulence, which are not decomposed by the singular value decomposition. Turbulence distributes the modal energy across the modes which make it require significantly more modes to reconstruct the flow field and the underlying wake dynamics. Hence, the POD based reconstruction becomes computationally more expensive in turbulent flows due to the broadband and multiscale character.

\begin{figure}
\centering
\includegraphics[trim={1cm 0cm 0cm 0cm},clip,scale=0.41]{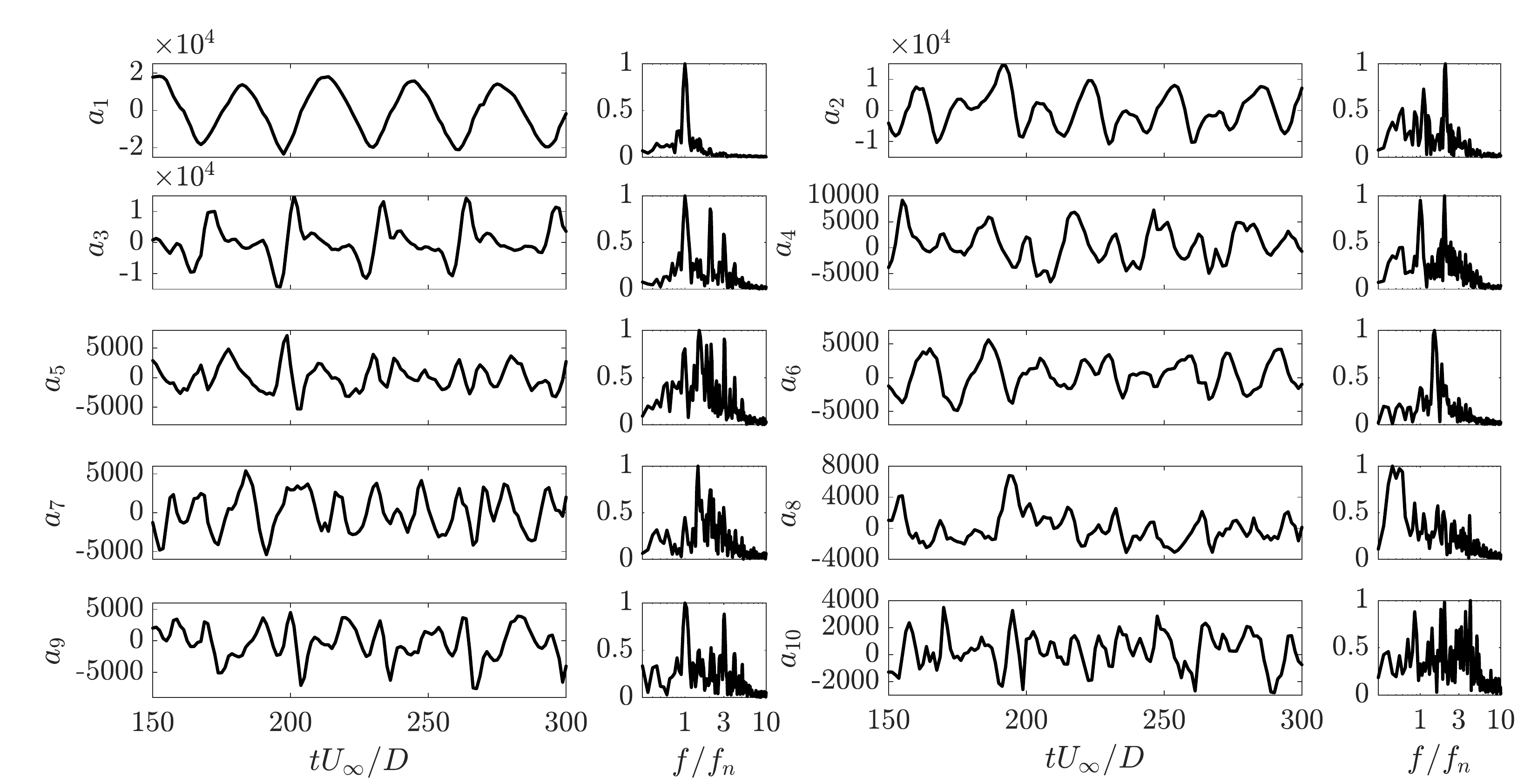}
\caption{Time dependent contributions of the 10 most energetic modes and their normalized FFT spectra for $Re=22,000$ case.}
\label{fig:TimeCoeffHRe}
\end{figure}
Similarly, figure \ref{fig:TimeCoeffHRe} demonstrates the broadband nature of turbulence in the temporal domain. Even with the multiscale spatial distortions, the first mode of the moderate $Re$ has a similar temporal contribution as the first mode of the low $Re$ case with a single dominant frequency close to the natural frequency of the system. However, the temporal coefficients of the other modes have multiple harmonics. Some of the modes exhibit predominant frequencies among the broadband FFTs. For example, the mode 2 has a dominant $2f_n$ frequency behavior, the mode 3 has $f_n, 2f_n$ and $3f_n$ harmonics and mode 4 has $f_n$ and $2f_n$ harmonics. These multiple harmonics occur due to the bombardment of turbulence on the corresponding flow features of the POD modes.

\begin{figure}
\begin{subfigure}[]{\textwidth}
\centering
\includegraphics[trim={2cm 2cm 2cm 3cm},clip,scale=0.45]{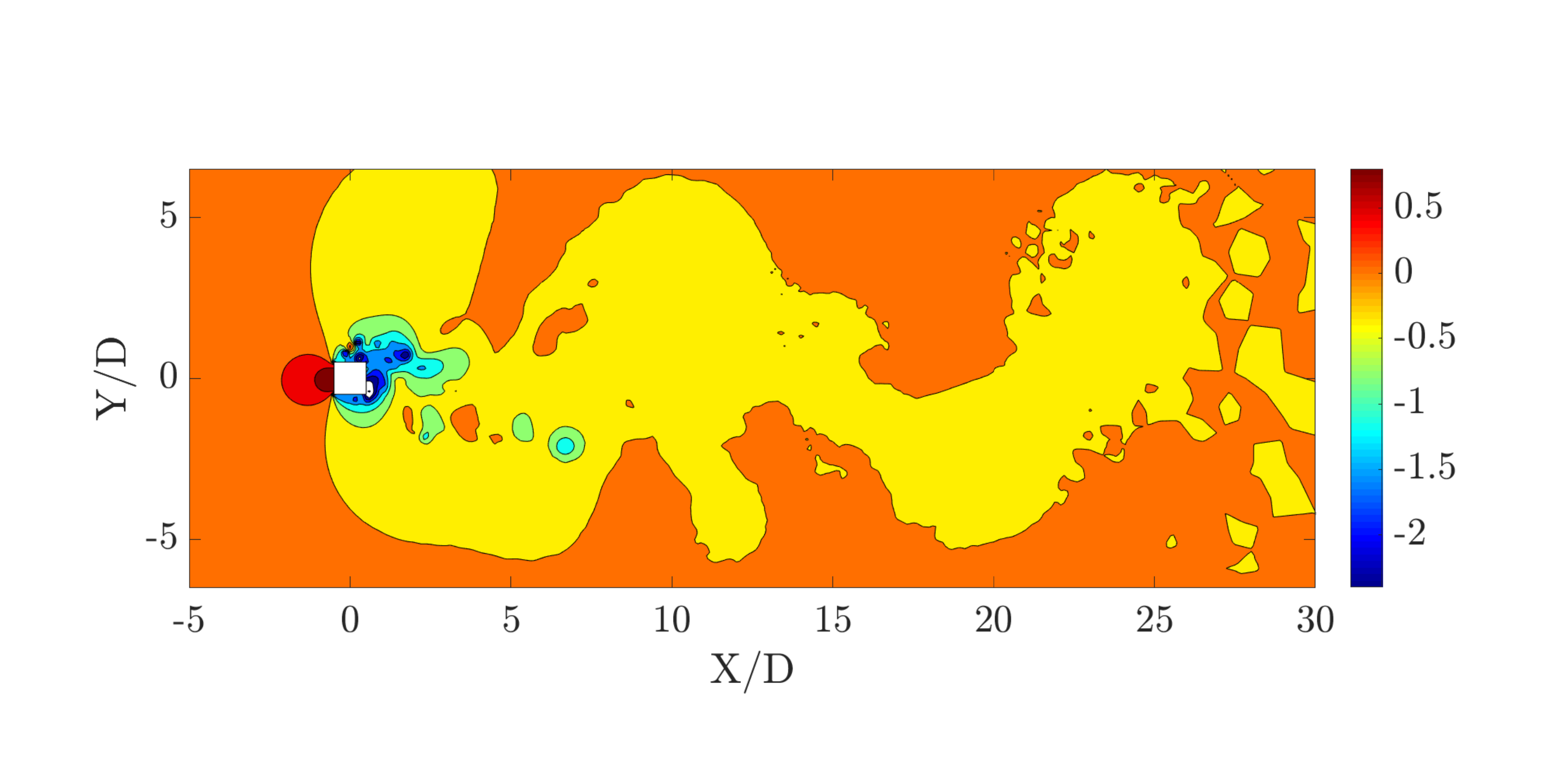}
\caption{}
\end{subfigure}
\begin{subfigure}[]{\textwidth}
\centering
\includegraphics[trim={2cm 2cm 2cm 3cm},clip,scale=0.45]{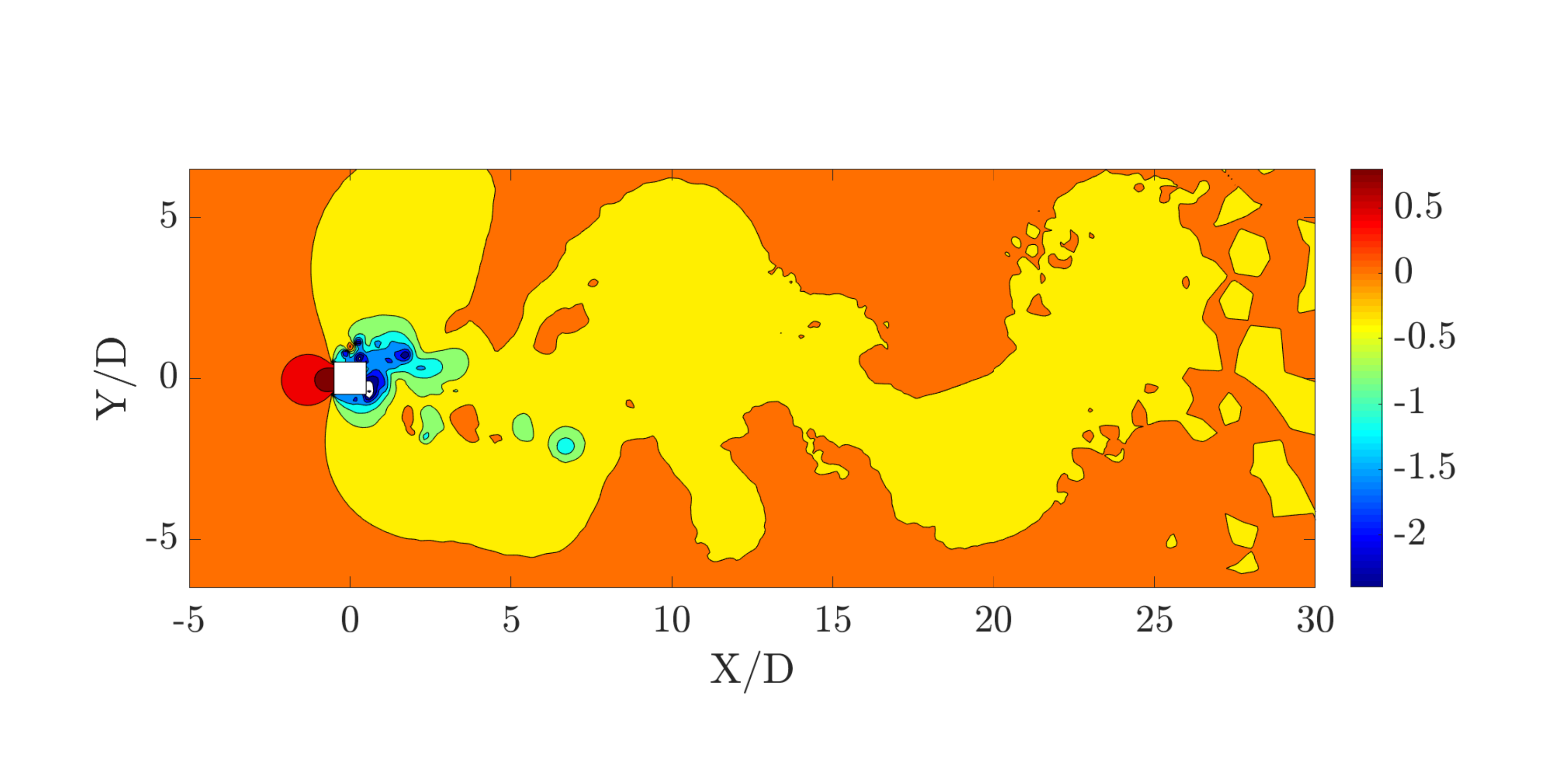}
\caption{}
\end{subfigure}
\begin{subfigure}[]{\textwidth}
\centering
\includegraphics[trim={2cm 2cm 2cm 3cm},clip,scale=0.45]{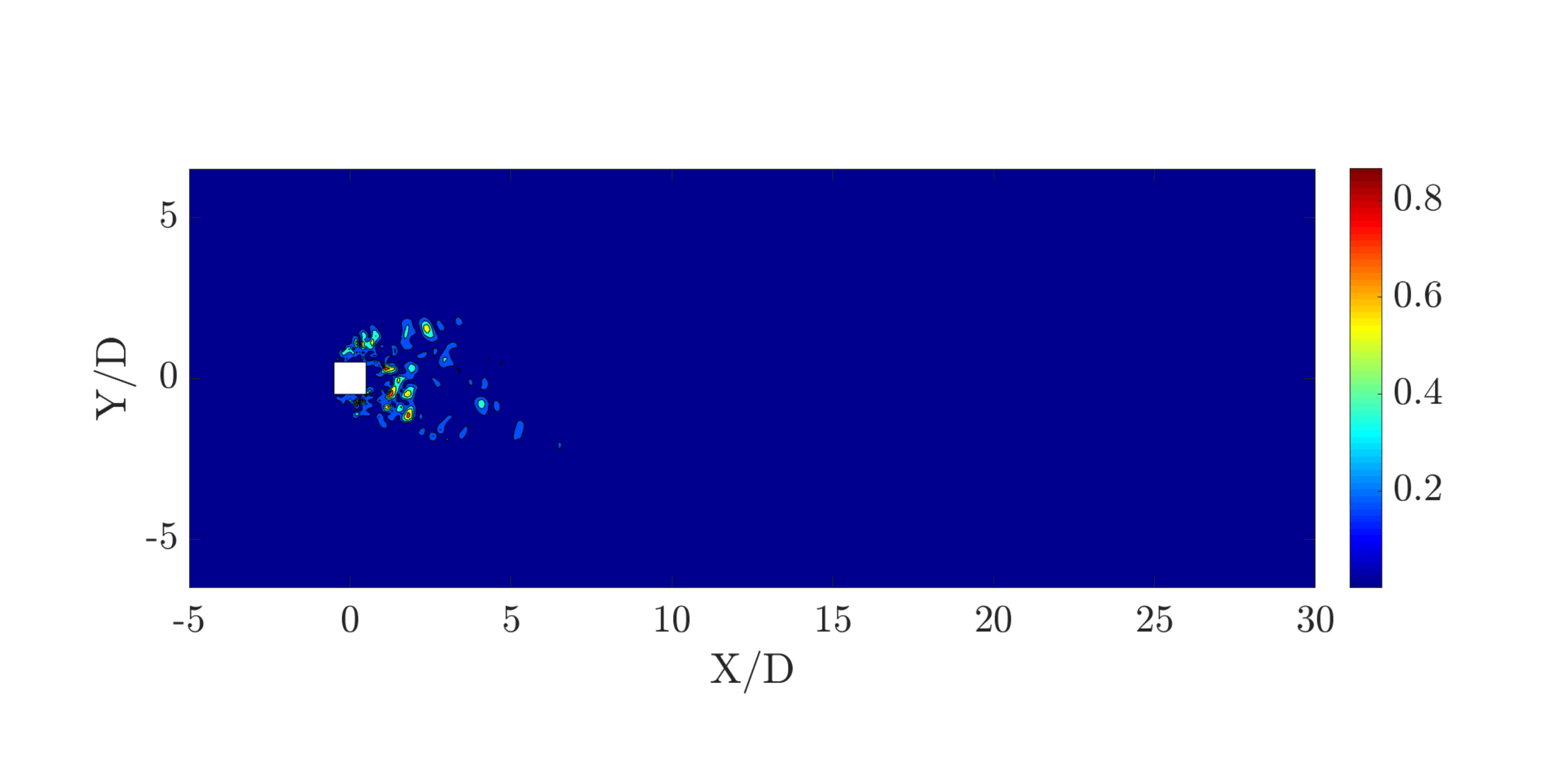}
\caption{}
\end{subfigure}
\caption{FOM and POD-DEIM reconstructed pressure field comparison for $Re=22,000$ case at $tU_{\infty}/D=100$: pressure distribution obtained by (a) full-order model and (b) 123 POD modes and 200 DEIM points and (c) the relative error (\%). The plots are of the mid $Z$-plane. The pressure values are normalized by $1/2\rho^{\mathrm{f}}U_{\infty}^2$. The flow is from left to right.}
\label{PODErrorHRe}
\end{figure}
\pgfplotsset{every tick label/.append style={font=\large}}
\pgfplotsset{every tick label/.append style={scale=1}}
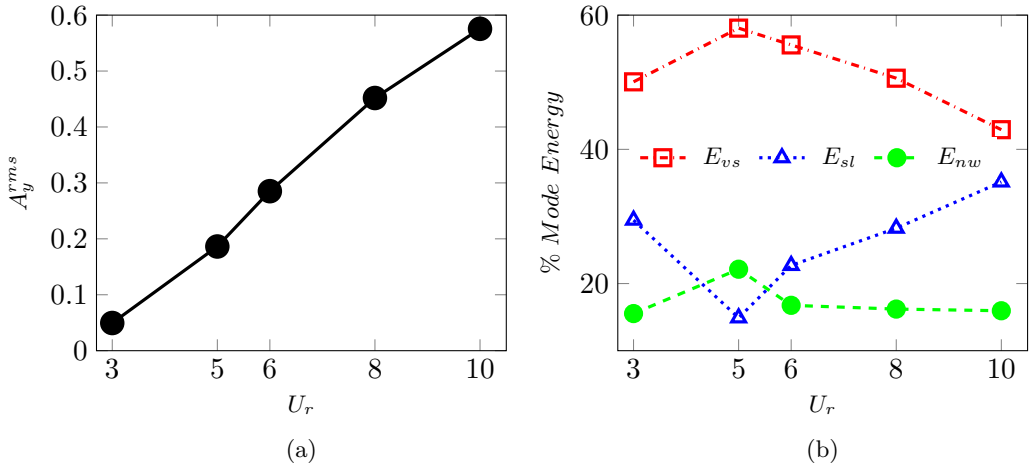
\begin{figure}
\begin{subfigure}{0.5\textwidth}
\centering
\begin{tikzpicture}[trim axis left, 
trim axis right, 
baseline]
\begin{axis}[
    xlabel={$U_r$},
    ylabel={$A_y^{rms}$},
    xmin=2.7, xmax=10.5,
    ymin=0, ymax=0.6,
    xtick={3,5,6,8,10},
    ytick={0,0.1,0.2,0.3,0.4,0.5,0.6},
    legend pos=north east,
    legend style={draw=none},
    width = 7cm,
    height = 6cm,
]

\addplot[
    color=black,
    solid,
    mark=*,
    very thick,
    mark size = 4,
    ]
    coordinates {
(3,0.0495)(5,0.1863)(6,0.2852)(8,0.4518)(10,0.5755)
};
\end{axis}
\end{tikzpicture}
\caption{}
\end{subfigure}~
\begin{subfigure}{0.5\textwidth}
\centering
\begin{tikzpicture}[trim axis left, 
trim axis right, 
baseline]
\begin{axis}[
    xlabel={$U_r$},
    ylabel={$\%$ $Mode$ $Energy$},
    xmin=2.7, xmax=10.5,
    ymin=10, ymax=60,
    xtick={3,5,6,8,10},
    ytick={0,20,40,60,80,100},
    legend style={at={(0.03,0.575)},anchor=west},
    legend columns=-1,
    legend style={draw=none,column sep=1ex},
    width = 7cm,
    height = 6cm,
]


\addplot+[
    color=red,
    dash dot,
    mark=square,
    very thick,
    mark size = 3,
    mark options=solid,
    ]
    coordinates {
(3,50.06)(5,58.06)(6,55.56)(8,50.60)(10,42.93)
};
\addlegendentry{{$E_{vs}$}}

\addplot[
    color=blue,
    dotted,
    mark=triangle,
    very thick,
    mark size = 3,
    mark options=solid,
    ]
    coordinates {
(3,29.45)(5,14.89)(6,22.70)(8,28.25)(10,35.14)
};
\addlegendentry{{$E_{sl}$}}
\addplot[
    color=green,
    dashed,
    mark=*,
    very thick,
    mark size = 3,
    mark options=solid,
    ]
    coordinates {
(3,15.53)(5,22.14)(6,16.76)(8,16.20)(10,15.95)
    };
    \addlegendentry{{$E_{nw}$}}

\end{axis}
\end{tikzpicture}
\caption{}
\end{subfigure}
\caption{Response and energy distribution for a freely vibrating square cylinder at $Re=22,000$: (a) transverse amplitude as a function of reduced velocity, (b) POD mode energy contribution of wake features. Note that $U_r=3.0$ represents the pre-synchronization, $U_r=5.0$ and $6.0$ denote the wake-body synchronization regime and $U_r=8.0$ and $10.0$ cases are in the galloping regime. For the mode energy analysis, the most energetic modes containing 95\% of the total mode energy are considered.}
\label{fig:AyRehigh}
\end{figure}

Figure \ref{PODErrorHRe} displays the pressure field reconstruction using the POD-DEIM technique. The actual instantaneous field contains some distortions and fine near-wake variations which are not completely captured by the reconstruction. However, the general large-scale variations are properly reconstructed with a maximum local error of $\approx 2\%$. 
The broadband energy distributing nature of turbulence has reduced the contribution of significant POD modes to the wake dynamics. Due to this behavior, many flow features are not captured when few of the most energetic modes are considered for the reconstruction. Hence, the inclusion of many POD modes is required for an accurate reconstruction which makes the POD reconstruction computationally expensive and time-consuming. However, this can be mitigated by selective reconstruction of few required timesteps instead of the entire time history. Using this to our advantage, we investigate the validity of the wake-body interaction cycle at this moderate $Re$.

Figure \ref{fig:AyRehigh} illustrates the response characteristics and the wake feature contributions for the mode energy at $Re=22,000$. In contrast to the low $Re$ cases, the bluff body undergoes galloping at this Reynolds number. We observe the same wake-body synchronization reported in \cite{miyanawala2018self}, i.e. 1:1 frequency synchronization of the bluff body motion and force at $U_r \in [5,6]$ and 1:3 synchronization at $U_r=10.0$. Our analysis here is focused on the 1:1 frequency lock-in where the vertical fluid forcing and the bluff body motion are in synchronization. According to figure \ref{fig:AyRehigh}(b), it is clear that the vortex shedding and near-wake modes become more energetic in the synchronized regime while the shear layer modes become less energetic. This is the same behavior observed in the low $Re$ analysis which proves that the proposed wake-body interaction cycle is valid even with the presence of turbulence.

In summary, the presence of turbulence distorts the spatially symmetric/anti-symmetric nature of POD modes and distributes the mode energy throughout many POD modes. Hence the reconstruction requires many modes and the classification of features is more complex at moderately high $Re$. However, even with this complexity, the fundamental wake-body interaction process proposed using low $Re$ analysis is observed to be valid for the three-dimensional turbulent flows. Hence we can conclude that the wake-body interaction cycle proposed in this study is a general cycle for fluid-structure interaction systems.

\section{Concluding remarks}
Despite the prevalence of SVD-based modal reduction techniques, there are very few studies on their application to fluid-structure interaction systems. In this paper, we considered the 2-DOF free vibration of a square cylinder under laminar and turbulent flows. We explored the capability of POD decomposition to interpret the most significant wake features and their contributions to the forces on the vibrating body interacting with fluid flow. 
When the linear and nonlinear POD-DEIM reconstructions are contrasted, we found that the DEIM method is faster and has a higher local accuracy since it captures the nonlinearity of the principle vortices and the near-wake region.
For the low $Re$ cases, every POD mode clearly represents one of the large-scale flow features: vortex shedding, shear layer or near-wake bubble. In these cases, we further observed that the nine most energetic modes contain $\approx 99\%$ of the energy. 
Further, we identified that the vortex shedding modes solely contribute to the transverse (lift) force while the shear layer and the near-wake modes solely contribute to the drag force. Based on these observations, we proposed a novel force decomposition for the drag and lift forces which is different from the conventional force decomposition based on the added mass and the viscous force contributions. 

We examined the POD decomposition for a range of $U_r$ values and we proposed the mechanism of the sustenance of synchronized wake-body lock-in. This further provided the explanation to the counter-intuitive observation: in the lock-in region, even though the kinetic energy is transferred from the fluid to the bluff body, the principle vortices are much more energetic than the pre- and post-lock-in regimes. 
It is seen that the bluff body motion widens the wake and it causes a vorticity transfer from the shear layer to the near-wake and vortices. We proposed the wake feature interaction cycle based on these observations. We further confirmed that this mechanism is valid for laminar $Re>Re_{cr}$ range.
For below critical $Re$ flows, we observed that the bluff body and the wake still synchronize and undergo large amplitude motion at some $U_r$ values when $Re\geq26$. Decomposition of these wakes further exhibited a similar behavior as the synchronized large amplitude motion cases at $Re>Re_{cr}$. This revealed that the flexibility of the bluff body induced the unsteadiness in the near-prwake bubble causing it to break and generate the vortices. With this observation, we can conclude that the fundamental requirements for the wake-body synchronized motion are, large enough flow inertia and appropriate flexibility of the structure. 

When the moderate $Re$ turbulent bluff body flow is decomposed, we observe that all the dominant wake modes are bombarded with different scales of turbulence. The broadband nature of turbulence resulted in a wide mode energy distribution, which required up to 123 modes to reach the $95\%$ mode energy threshold. Further analysis of the time coefficients of the modes confirmed that the large-scale features are battered by the multiple frequency turbulence. However, they generally correspond to a large-scale wake feature similar to the laminar cases.
The wake decomposition of turbulent flows for $U_r\in[3,10]$ confirmed that the wake interaction cycle proposed for laminar cases is valid for turbulent flows as well. 

\appendix
\setcounter{equation}{0} 
\setcounter{figure}{0}
\renewcommand{\theequation}{A.\arabic{equation}}
\renewcommand{\thefigure}{A.\arabic{figure}}
\section{Force decomposition based on modal contribution}
The aim of this appendix is to present a general decomposition of the fluid force exerted on a moving body in  an incompressible viscous flow. To begin, we provide some background on existing force decomposition techniques that characterize the fluid inertial and the viscous forces on a  moving body.
In one of the pioneering works, \cite{morison1950force} proposed a force decomposition for the in-line force acting on a cylindrical object which is widely used in many engineering applications. This semi-empirical decomposition can be written as a linear sum of a velocity squared-dependent drag force and an acceleration-dependent inertial force:
\begin{equation}
F(t) = \frac{1}{2}C_dD|U|U + \rho C_m \frac{\pi D^2}{4} \frac{dU}{dt},
\label{eq:Morison}
\end{equation}
where  $C_d$ and $C_m$ represent the averaged drag and inertia coefficients, 
which can be determined by experiments or numerical computations. Owing to the nonlinear dependency of these coefficients on the evolution of vorticity field, \cite{sarpkaya2001force} argued that "It does not perform uniformly well in all ranges of $K$, $\beta$ and $k/D$", where $K$ denotes the Keulegan Carpenter number, $\beta=Re/K$ and $k/D$ is the relative roughness. In \cite{lighthill1986}, a different approach is taken by the assertion that the viscous drag and the inviscid inertia force operate independently, by re-writing Eq. (\ref{eq:Morison}):
\begin{equation}
F(t) = \frac{1}{2}C_d\rho A_p U^2  + C^*_m\rho \frac{dU}{dt} V_b
\label{eq:lighthill}
\end{equation}
and for a flow defined by $U(t) = - U_m \cos{\omega t}$ it reduces to
\begin{equation}
C_F = - C_d |\cos{\omega t}| \cos{\omega t} + C^*_m\frac{\pi^2}{K} \sin{\omega t},
\end{equation}
where $A_p$ and $ V_b$ denote the projected area and the volume of the body, respectively and
$C^*_m$ is the ideal value of the inertia coefficient. Notwithstanding, many studies demonstrate that it is difficult to represent the actual force with this relation as long as a constant value $C^*_m$ is considered. In particular, \cite{sarpkaya2001force} clearly demonstrated that the viscous drag force and the inviscid inertia force are not completely independent and it is impossible to decompose the unsteady drag force to an inviscid and a vorticity-drag component. 
The decomposition of the total force into inviscid and viscous components by Lighthill's relation (Eq. \ref{eq:lighthill}) can be considered as an effort to lump the effects of the complex generation and evolution of vorticity field into mutually independent forces related to the inviscid inertia and the viscous effects. In such force decomposition techniques, the characteristics vorticity patterns and their dynamics generated during the motion of a body are not included.

In what follows, we propose an alternative force decomposition for the in-line (drag) and transverse (lift) pressure forces applied on a bluff body which extends the above decompositions to incorporate significant features of unsteady separated flow. In particular, the unsteady force is decomposed to include the nonlinear generation and evolution of vorticity field around a moving body in a fluid flow. This decomposition is based on the contributions from different POD modes to the forces and can be written as
\begin{equation}
F_j (t) = {F_j^0} + \sum_{i=1}^{n_{rj}} b_j^i (t) F_j^i,
\end{equation}
where $F_j$ is the force on a particular direction ($j=x$ for the in-line and $j=y$ for transverse). While $F_j^0$ is the time-independent contribution from the mean field, $F_j^i$ is the unsteady pressure fluctuation contribution associated with $i^{th}$ mode. Here, $b_j^i (t)$ is the time-dependent coefficient of the $i^{th}$ mode for the force in direction $j$ and $n_{rj}$ is the number of POD modes with a significant contribution for the particular force. Using the snapshot data, we can exactly determine $b_j^i(t)$ for the in-line and transverse forces by the following relation
\begin{equation}
 b_j^i (t) = \frac{F_j (t) - F_j^0}{F_j^i n_{rj}}.
\end{equation}
The magnitude of the modes, the time coefficients and the force contributions introduced in this decomposition slightly fluctuate when flow parameters and the bluff body geometry are changed. Similar to the above methods, we can create databases of $F_j^i$ for different bluff bodies. These databases can then be used to determine the total forces as well as the contribution from each flow feature to the bluff body dynamics. To further generalize the force decomposition, deep learning techniques \citep{miyanawala2017efficient} for parametric predictions can be employed.
\begin{figure}
\centering
\begin{subfigure}[]{\textwidth}
\centering
\includegraphics[trim={1.5cm 0.5cm 0.5cm 0},clip,scale=0.4]{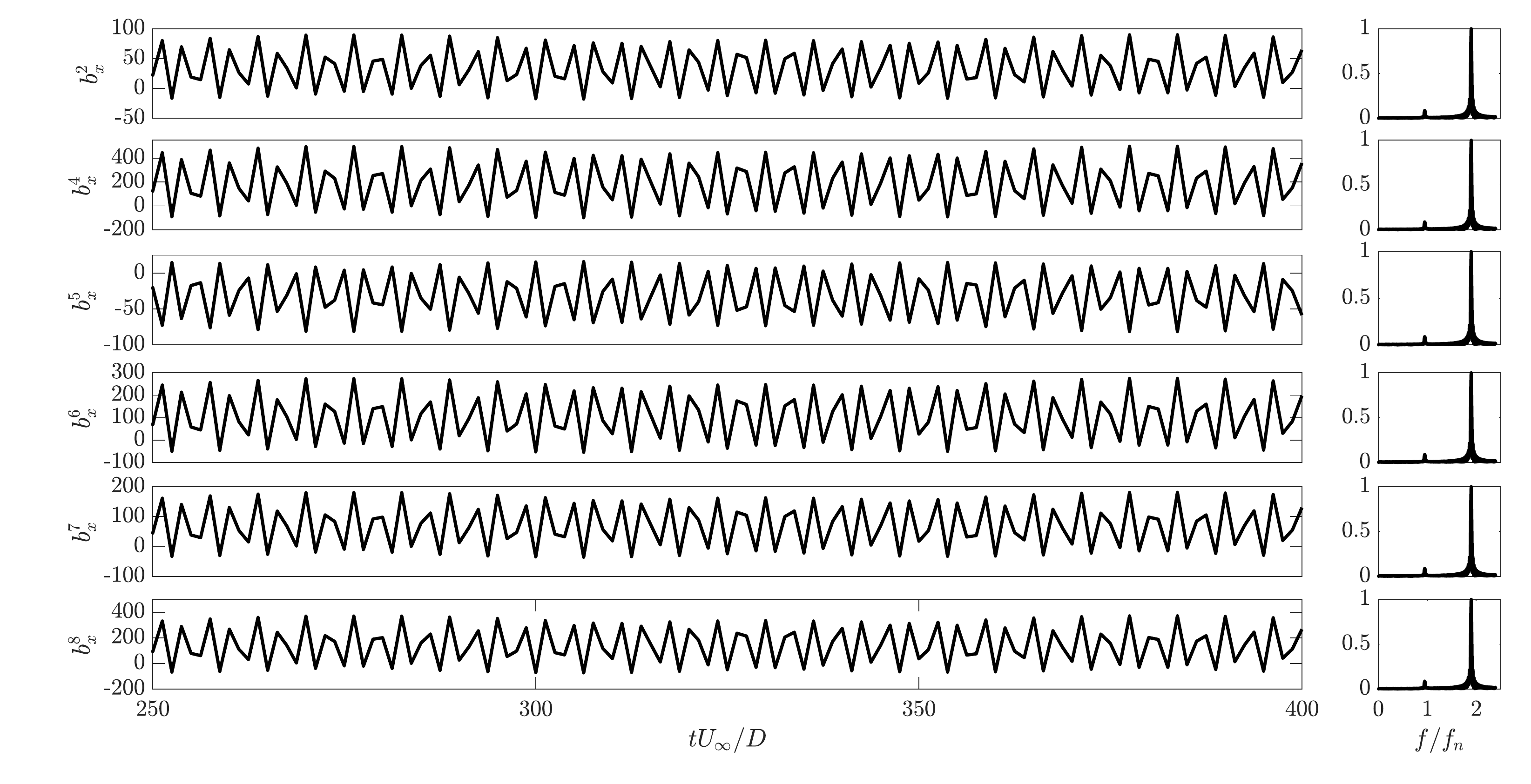}
\caption{}
\end{subfigure}
\begin{subfigure}[]{\textwidth}
\centering
\includegraphics[trim={1.5cm 0cm 0.5cm 0cm},clip,scale=0.4]{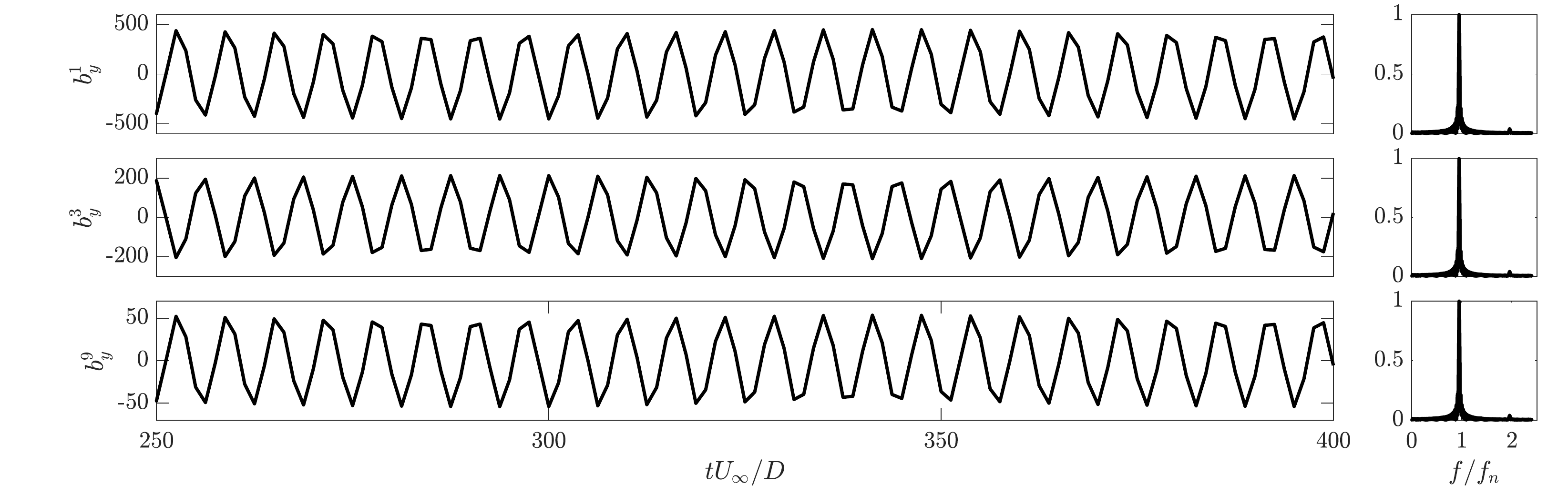}
\caption{}
\end{subfigure}
\caption{Force decomposition based on POD-based projection technique: (a) time independent force coefficient magnitudes ($|F_j^i|$) for in-line (drag) and transverse (lift) forces and time dependent coefficients ($b_j^i (t)$) for, (b) drag modes and (c) lift modes. Note that the lift modes (1, 3 and 9) entirely contribute to the lift while the drag modes entirely contribute to the drag. Moreover, note that $b_j^i (t) $ of drag modes have non-zero mean and for lift modes $\overline{b_j^i} \approx 0$.}
\label{TimeCoeff}
\end{figure} 
Figure \ref{TimeCoeff}b and \ref{TimeCoeff}c present the reconstructed values of the time dependent coefficient for drag and lift modes. Note that the relevant six out of the first 9 modes are considered for the drag (i.e. $n_{rx} = 6$ and $n_{ry} = 3$) based on the values of figure \ref{TimeCoeff}a.
In a nutshell, the force component represented by the modal decomposition implicitly characterizes the three constituent components involving an inviscid inertial force, the dynamics of vorticity field, and a skin friction force. 

\appendix 
\renewcommand{\thesection}{B}
\setcounter{equation}{0} 
\setcounter{figure}{0}
\setcounter{table}{0}
\renewcommand{\theequation}{B.\arabic{equation}}
\renewcommand{\thefigure}{B.\arabic{figure}}
\renewcommand{\thetable}{B.\arabic{table}}
\section{Performance comparison of POD reconstruction methods}
\begin{table}
\centering
\caption{Number of floating points operations (FLOPs) required for linear and nonlinear POD reconstruction. Mesh count ($m$) = 87,120, number of snapshots ($k$) = 320, number of significant modes for linear POD ($r$) = 9, number of significant modes for DEIM ($n_d$) = 7 and number of DEIM points ($n_p$) = 70. }
\label{Table:FLOPComp}
\begin{tabular}{l|c|c|c}
 & Operation & FLOPs & Order \\
\hline
Modes generation   &   & &  \\
$\mathbf{Y}^T\mathbf{Y}$ & 
Matrix multiplication    & 
$2k^2m- k^2$             & 
$2k^2m$ \\
$\mathbf{Y}^T\mathbf{Y}\bs{\mathcal{W}}=\bs{\Lambda}\bs{\mathcal{W}}$ & 
Eigenvalue solution 												  & 
$k^3$    															  & 
$k^3$ \\
$\bs{\mathcal{V}}=\mathbf{Y}\bs{\mathcal{W}}\bs{\Lambda}^{-1/2}$ & 
Matrix multiplication 											 & 
$2k^2m$             											 & 
$2k^2m$ \\
Total  & &  & $4k^2m$  \\
\hline
Linear POD & & & \\
$\mathbf{\hat{Y}}(t)= \langle \b{P}-\overline{\b{P}}, \bs{\mathcal{V}} \rangle$  &
Inner products 																	 &
$rk(2m-1)$       																 &
$2rkm$ \\
$\bs{\mathcal{V}}\mathbf{\hat{Y}} $ &  
Matrix multiplication   			& 
$2rkm-mk$          					& 
$2rkm$  \\
$\sum_{j=1}^{r} \hat{y}_j(t)\b{v}_j$ & 
Summation 							 & 
$rk^2m$          					 & 
$rk^2m$   \\
Total &  &   & $rk^2m$  \\ 
\hline
DEIM    & &  &  \\
$(\bs{\mathcal{P}}^T\bs{\mathcal{U}})\hat{\b{c}} = \bs{\mathcal{P}}^T\b{v}_i$  &
Matrix solution 															   &
$(n_p-1)n_p(n_p+6m-1)/6$   													   &
$n_p^2m$   \\
$\hat{\b{r}} = \b{v}_i - \bs{\mathcal{U}}\hat{\b{c}}$ & 
Matrix subtraction 									  & 
$(n_p-1)n_pm$            							  & 
$n_p^2m$       \\
$\bs{\mathcal{V}}_{\wp}^+$ & 
Matrix multiplication      & 
$4n_d^2n_p-n_d^2-n_dn_p$   & 
$4n_d^2n_p$       \\
$\mathbf{Y-\overline{Y}} = \bs{\mathcal{V}}\bs{\mathcal{V}}_{\wp}^+\mathbf{Y}_{\wp}$  & 
Matrix multiplication 														& 
$k(2n_dn_p+2mn_d-n_d-m)$ 															& 
$2mkn_d$    \\
Total & & & $2n_p^2m + 2mkn_d$ \\
\end{tabular}
\end{table}

Herein, we briefly compare the number of floating point operations (FLOPs) required for the linear and DEIM based POD reconstructions in Table (\ref{Table:FLOPComp}). The generation of the POD modes which is essential for both reconstructions requires $\sim O(4k^2m)$ FLOPs. In this study, we estimate this value to be $3.571\times10^{10}$. The linear POD reconstruction needs $\sim O(rk^2m)$ computations where the equal main contributions are from the multiplication between the time coefficient and the POD modes, and the summation of the multiplied POD contributions. With the use of 9 POD modes, the estimated FLOP count is $8.029\times10^{10}$. The DEIM technique consumes fewer computations than POD as it requires $\sim O(2n_p^2m + 2mkn_d)$ FLOPs. Generating the matrix solution in DEIM first step is the most expensive operation as it needs more FLOPs for last DEIM points. With the use of 7 POD modes and 70 DEIM points, the POD-DEIM reconstruction needs $1.244\times10^{9}$ FLOPs. According to this calculation, the linear POD reconstruction demands $\approx 64.54$ times more FLOPs than the DEIM reconstruction. However, the overall linear POD process needs $\approx 3.14$ times more FLOPs than the overall DEIM process.

\begin{table}
\centering
\caption{Performance comparison between linear POD and POD-DEIM for the most accurate reconstruction.}
\label{Table:PODDEIM}
\begin{tabular}{r|c|c}
\multicolumn{1}{l|}{}     & Linear POD (10 Modes) & DEIM (7 Modes  + 70 Points) \\ \hline
POD mode generation       & 35.1s  & 35.1s \\
Reconstruction            & 113.2s & 12.2s \\
Total time elapsed        & 148.3s & 37.3s \\
Speedup (reconstruction) & -      & 9.28 \\
Speedup (total)          & -      & 3.98 \\
Maximum local error       & 1.43\% & 1.24\%  \\
Cumulative domain error   & 3.85\% & 4.05\%      
\end{tabular}
\end{table}
The actual performance of the linear POD and POD-DEIM reconstruction techniques is compared in Table (\ref{Table:PODDEIM}). All the calculations are performed using the same computer. When just the reconstructions are considered, DEIM is $9.28$ times faster than linear POD. When the total processes are compared, DEIM has a speed gain of $3.98$. DEIM is more accurate in the nonlinear flow regions. However, linear POD method is more accurate when the cumulative error of the fluid domain is considered.

\section*{Acknowledgements}
The first author gratefully acknowledges the financial support from the Ministry of Education, Singapore through the National University of Singapore Research Scholarship. The high-fidelity data sets are obtained using the computational resources at High-Performance Computing (HPC) at National University of Singapore Computer Center and the National Supercomputing Center (NSCC), Singapore.

\newpage
\bibliography{PODRef}
\bibliographystyle{jfm}
\end{document}